\documentclass[manuscript, screen]{acmart}
\usepackage{multirow}
\usepackage{hhline}
\usepackage{graphicx}
\usepackage{subcaption}
\usepackage{hyperref}
\usepackage{soul}
\usepackage{booktabs} 
\AtBeginDocument{%
  }

\setcopyright{acmlicensed}
\copyrightyear{2025}
\acmYear{2025}
\acmDOI{XXXXXXX.XXXXXXX}
\acmISBN{978-1-4503-XXXX-X/2018/06}

\begin{document}

\title{Human-Centered Development of Indicators for Self-Service Learning Analytics: A Transparency through Exploration Approach}

\author{Shoeb Joarder}
\email{shoeb.joarder@uni-due.de}
\affiliation{%
  \institution{University of Duisburg-Essen}
  \city{Duisburg}
  \country{Germany}
}
\email{shoeb.joarder@uni-due.de}

\author{Mohamed Amine Chatti}
\affiliation{%
  \institution{University of Duisburg-Essen}
  \city{Duisburg}
  \country{Germany}
}
\email{mohamed.chatti@uni-due.de}







\renewcommand{\shortauthors}{Joarder \& Chatti}

\begin{abstract}
The aim of learning analytics is to turn educational data into insights, decisions, and actions to improve learning and teaching. The reasoning of the provided insights, decisions, and actions is often not transparent to the end-user, and this can lead to trust and acceptance issues when interventions, feedback, and recommendations fail. In this paper, we shed light on achieving transparent learning analytics by following a transparency through exploration approach. To this end, we present the design, implementation, and evaluation details of the Indicator Editor, which aims to support self-service learning analytics (SSLA) by empowering end-users to take control of the indicator implementation process. We systematically designed and implemented the Indicator Editor through an iterative human-centered design (HCD) approach. Further, we conducted a qualitative user study (n=15) to investigate the impact of following an SSLA approach on the users' perception of and interaction with the Indicator Editor. Our study showed qualitative evidence that supporting user interaction and providing user control in the indicator implementation process can have positive effects on different crucial aspects of learning analytics, namely transparency, trust, satisfaction, and acceptance.
\end{abstract}





\maketitle

\section{Introduction}  \label{introduction}
The learning analytics (LA) domain is maturing and often provides insights and informs decisions or actions (e.g., interventions, feedback, recommendations) regarding learning and teaching behaviors.
Before insights, decisions, and actions can impact learning, the data analysis process and outcomes must be transparent to the end-users.
This will build trust and increase acceptance of LA systems \citep{hakami2020learning, drachsler2016privacy}.
Transparency is often linked to users' understanding of a system's inner logic and supports users in building an accurate mental model of how the system works \citep{ngo2020exploring, hellmann2022development}.
Trust is a multifaceted concept, as evidenced by differences in definitions across disciplines such as psychology, sociology, and economics, as well as multiple ways to measure trust \citep{siepmann2023trust, kaur2022trustworthy, miller2022we, mcknight2002developing, yang2020visual, pu2011user, knijnenburg2012explaining}. \citet{tintarev2015explaining} conceptualize trust as ``Increase users' confidence in the system''. According to \citet{pu2011user}, trust is part of attitudes, i.e., the user's overall feeling towards the system. Trust can either be conceptualized as a personal characteristic (PC), the general tendency to trust others, or as a situational characteristic (SC), the trust in the system \citep{knijnenburg2012explaining}.

Transparency and trust are often linked, following the intuition that you will more likely trust a system that you can understand than one that is a black box to you \citep{siepmann2023trust}. 
\citet{hellmann2022development} included the effects of transparency on trust in their development of a questionnaire to measure the perceived transparency in a recommender system. The authors found a positive link between transparency and trust-related measures.
Providing transparency could enhance users' trust in the system.
For example, \citet{tintarev2015explaining} pointed out that an increase in transparency should lead to an increase in trust. Similarly, \citet{nunes2017systematic} identified transparency as one of the key factors for users to develop trust.
However, some studies found that revealing too much detail about the system's inner logic may result in information overload, confusion,
and a low level of perceived understanding, which may in turn reduce users' trust in the system \citep{ananny2018seeing, hosseini2018four, cramer2008effects}.

The evolution toward a more transparent and trustworthy LA is urgent, as recent data protection and privacy regulations like the EU General Data Protection Regulation (GDPR) and the California Consumer Privacy Act (CCPA)
stipulate that transparency is a fundamental right \citep{verbert2020xla}. 
In recent years, transparency in LA has received more attention in research from its relevance to ensuring fair, ethical, responsible, and trustworthy LA practices \citep{hakami2020learning}. 
Although the role of system transparency is well-researched in artificial intelligence (AI), machine learning (ML), and recommender systems (RS), the application of this research in LA is in its early stages. 
There is a pressing need for research on adapting and extending this work into the LA domain, where trust and acceptance are of primary importance \citep{khalil2023fairness}. 

One important aspect that may contribute to increased transparency is the degree of control users have over the system \citep{shneiderman2022human, shneiderman2020bridging, amershi2014power, he2016interactive, spinner2019explainer}. 
Recent research on human-centered learning analytics (HCLA) strengthens the need for empowering stakeholders to take control of the LA process to ensure transparency
\citep{alfredo2024human}.
Current research on HCLA demonstrates successful co-design processes for LA systems involving various stakeholders.
However, research on actively involved users in the LA indicator implementation process is lacking.
To address this research gap, in this paper, we aim to develop a better understanding of the potential of human control to increase the transparency of LA. To this end, we present the \textit{Indicator Editor} for self-service LA (SSLA), following a \textit{transparency through exploration} approach. We aim to empower the involved stakeholders (e.g., students, teachers, researchers) to create transparent LA indicators and take full control over the indicator implementation process by giving them access to the data and the analysis and visualization methods underpinning the indicators. We suggest that following an HCLA approach that involves LA stakeholders in the indicator implementation process can increase transparency, build trust, and push forward the acceptance of LA. Towards this challenge, we present the systematic design, implementation, and evaluation details of the \textit{Indicator Editor} to provide transparency in the indicator implementation process. 

We conducted a qualitative user study (n=15) based on moderated think-aloud sessions and semi-structured interviews with students and teachers to investigate the impact on users' perceptions of control, transparency, trust, and acceptance. Due to the subjective nature of these aspects, a qualitative approach seems to be the most appropriate to investigate them. This approach allows us to investigate the users' unique perspectives and expectations from an SSLA indicator implementation approach and get more in-depth insights into users' perceptions of and interaction with the \textit{Indicator Editor}. The objective of the study was to answer the following research question: 
\textit{What are the effects of an self-service LA approach that empowers users to control the LA indicator implementation process on the user's perceptions of (1) control \& personalization, (2) transparency \& trust, and (3) satisfaction \& acceptance?} \textbf{(RQ)}. 
The results of our user study show that there is qualitative evidence that having the human-in-the-loop in the indicator implementation process positively affects different vital aspects in LA, namely transparency, trust, satisfaction, and acceptance. 

To summarize, this work makes the following main contributions: First, we highlight the critical importance of taking a transparency through exploration approach to addressing the challenge of trustworthy LA.
Second, we apply the human-centered design (HCD) process to effectively design and implement the \textit{Indicator Editor}. Third, we instantiate the transparency through exploration approach by empowering the end-users to take control of the indicator implementation process using the \textit{Indicator Editor}. Fourth, we provide qualitative evidence that following this approach can positively impact the users' perception of transparency and trust, thus potentially leading to improved satisfaction and acceptance of LA.
\section{Background and Related Work}   \label{background}
\subsection{Transparent Learning Analytics}
Trust has recently gained much attention in the LA community
\citep{alfredo2024human,khalil2023fairness,drachsler2016privacy}.
The positive connection between trust and transparency is drawn in the LA literature.
Transparency is identified as one of the main factors for LA stakeholders to develop trust \citep{gedrimiene2023transparency, ahn2021co, li2019impact, hakami2020learning}.
In the LA context, transparency refers to the openness and communication of the analyzed data and the mechanisms underlying LA insights, decisions, or actions.
The transparency of the LA system is crucial to help users build a transparent mental model of the system's functioning as well as understand the decisions or actions taken by the system, thus driving forward the acceptance and adoption of LA 
\citep{clow2012learning, tsai2020privacy, hakami2020learning, cukurova2020modelling}.
Here, it is important to differentiate between objective transparency and user-perceived transparency 
\citep{gedikli2014, zhao2019users}.
In the LA domain, on the one hand, objective transparency means that the LA system reveals all the information about the data it had at its disposal (e.g., what data is collected, how this information was collected, by whom, for what purposes, who will have access to this data, and how data might be combined with other datasets (and for what purposes) \citep{prinsloo2015student} (i.e., data transparency), and how it used that information to develop indicators, including details about the used analysis and visualization methods (i.e., process transparency).
On the other hand, user-perceived transparency is based on users' subjective opinions about the extent to which they perceive such information to be available and feel they understand the meaning of the provided information.
While different aspects of the LA system, for example, the student data or the LA process, may be exposed to the user, transparency as a user-centric quality can only be assessed by measuring users' perception and understanding of those system aspects \citep{hellmann2022development}.

Several studies highlighted the importance of transparency in designing, developing, and evaluating LA systems. 
\citet{hakami2020learning} presented results of a literature review that was conducted across all the editions of the Learning Analytics \& Knowledge (LAK) conference addressing how the societal values of Fairness, Accountability, and Transparency (FAT) are being considered in LA research. According to the authors, 75\% of the reviewed papers (37 out of 49) identified transparency as one of the significant concerns in LA research. 
Related to transparency, the results of this review provide different insights into how transparency is considered in LA systems. Transparency is currently being addressed in LA research concerning how student (and staff) data can be handled transparently, including data collection, storage, and usage 
\citep{tsai2017learning, lang2018complexities, haythornthwaite2017information, drachsler2016privacy, prinsloo2015student, slade2019learning, whitelock2017students}.
Moreover, the option to opt out of student data collection can increase transparency 
\citep{prinsloo2015student, prinsloo2013evaluation, slade2013learning}. 
This research considers an LA system to become transparent if users know what, how, and why their data was collected and can opt out of the data collection processes without any consequences 
\citep{scheffel2015developing, tsai2017learning}.
The review further shows that LA researchers have presented only a few technical solutions to enhance transparency. These include Open Learner Models (OLMs) 
\citep{bodily2018open, conati2018ai},
transparent educational recommender systems 
\citep{abdi2020complementing, barria2019making, barria2019explaining},
and transparent predictive modeling approaches such as decision trees \citep{cukurova2020modelling}.

Although the LA community is concerned about ethics in LA and increasingly interested in providing transparent LA practices, transparency was mainly addressed through privacy and ethical conceptual frameworks 
\citep{drachsler2016privacy,hoel2017influence,lang2018complexities,prinsloo2013evaluation,swenson2014establishing,tsai2017learning}
and issues of transparency about tracking learners' data have been a cornerstone in these frameworks 
\citep{hakami2020learning,tsai2017learning}.
The view of transparency that centers on student data (i.e., data transparency) is one possible way to achieve transparency. Another way to enhance transparency is to help users understand the entire process behind LA outcomes (i.e., process transparency). To build effective LA interventions, LA systems need to focus on more than just students' concerns about the use of data. They need to help students, e.g., understand how indicators are generated. The view of process transparency is increasingly adopted in recent studies on transparent LA 
\citep{duan2024towards, conijn2023effects, pozdniakov2022question, tsai2020empowering}.
For example, 
\citet{duan2024towards} highlighted the need for transparent AI algorithms that align with the needs of educational stakeholders. \citet{conijn2023effects} explored the extent to which transparency regarding an AI grading system impacted both trust and motivation in students.
In the study conducted in \citep{pozdniakov2022question}, teachers raised concerns regarding the lack of transparency of the provided LA visual interface. Particularly, teachers wondered how phrases in the LA questions might be connected with the algorithm based on which a visualization was created. While these studies reflect concerns about the lack of transparency that centers on the LA process, concrete solutions regarding how to provide students and teachers with insights into the design decisions behind LA systems in order to achieve transparency are under-explored. 

Users' perception of a system's transparency may be influenced by several factors \citep{hellmann2022development}. 
Providing explanations is one important factor, and some studies have shown that providing explanations can enhance transparency \citep{khosravi2022explainable, duan2024towards, tintarev2015explaining, pu2011user, cramer2008effects}.
Another significant factor that may contribute to increased transparency is related to users' control over the system \citep{shneiderman2022human, shneiderman2020bridging, amershi2014power, he2016interactive, spinner2019explainer, knijnenburg2012explaining}. 
A positive influence of interaction possibilities as well as perceived control on the perceived transparency of the system was reported by \citet{pu2011user}. 
Based on these two factors, the AI literature distinguishes between two main approaches for achieving transparency, namely \textit{transparency through explanation} and \textit{transparency through exploration} \citep{siepmann2023trust}. In this paper, we focus on the exploration approach for achieving transparency that centers on the LA process.
Human control can contribute to increased transparency in decision-making systems. Researchers recognized the potential for human control in different application domains, including human-centered AI \citep{shneiderman2022human}, interactive machine learning \citep{amershi2014power}, and interactive recommendation 
\citep{jugovac2017interacting, he2016interactive, tsai2017providing, verbert2013visualizing},
and visual analytics \citep{keim2006challenges, spinner2019explainer}. 
Recently, \citet{shneiderman2022human} suggested a shift from dependence on retrospective explanations (aka post-hoc explanations
\citep{du2020techniques}
) that describe the local or global process of an AI system to prospective user interfaces that are interactive, visual, and exploratory, with the new goal to give users a better understanding of how the system works so that they can prevent confusion and surprise that lead to the need for explanations. 
According to \citet{shneiderman2022human}, exploration works best when the user's inputs are actionable, allowing users to control and modify them. Exploratory user interfaces can potentially guide users incrementally toward their goals and increase user control of AI processes. To describe the process underlying exploratory user interfaces, \citet{shneiderman2022human} further proposed an updated mantra of `Preview first, select \& initiate, then show execution' so that users can monitor the process and decide whether they want to change it along the way. This asserts the centrality of human control over AI, thus achieving more transparent and trustworthy AI systems.

LA following \textit{transparency through exploration} approach suggests that human control is crucial for providing transparent insights, decisions, and actions into the LA process. Human control can, however, only reach its full potential if the LA system is understandable, easy to use, and intuitive for all stakeholders involved, including lay users. While attention to human control is also growing within the domain of LA, as discussed in the next section, theoretically- and technically sound solutions that empower end-users to steer the LA indicator implementation process are lacking. To close this research gap, in this paper, we aim to achieve transparent LA by following a transparency through exploration approach. Concretely, our goal is to provide a self-service LA (SSLA) tool that can support students and teachers with prior knowledge in data analysis and visualization in creating transparent LA indicators and take complete control over the indicator implementation process by giving them access to the data and the analysis and visualization methods underpinning the indicators.


\subsection{Human-Centered Learning Analytics}
Recent research strengthens the need for engaging stakeholders in the LA process to ensure transparency, leading to an emerging LA sub-field referred to as human-centered learning analytics (HCLA) \citep{buckingham2019human, chatti2019perla}. HCLA is an approach that emphasizes the human factors in LA. It seeks to include humans in designing, developing, and evaluating the LA process to serve the needs of various LA stakeholders with their multiple goals effectively \citep{buckingham2019human, chatti2020design}. HCLA can be achieved by bringing Human-Computer Interaction (HCI) approaches (e.g., design thinking, Human-Centered Design, participatory design, co-design, and value-sensitive design) that prioritize human needs, values, and perspectives into the field of LA \citep{sarmiento2022participatory, lang2023learning,viberg2023designing,dimitriadis2021human}.
Recently, a growing body of research has provided mature examples of how these HCI approaches can be applied to LA to overcome the challenge of designing LA tools that lack the voice of the 2-users \citep{alfredo2024human, topali2025designing}.
Most of these works mainly present case studies that target teachers \citep{lawrence2024teachers, campos2024leveraging, wiley2024human, hutchins2024co, dollinger2018co, holstein2019co, martinez2015latux, ahn2019designing, rehrey2019engaging, chen2019towards, dimitriadis2021human} 
or students \citep{hilliger2024curriculum, prieto2018co, alvarez2020deck, de2019student, hilliger2020learners, sarmiento2020engaging} as stakeholders. 
The involvement of educational stakeholders in the design of HCLA systems is essential to meet their needs and preferences. 
This involvement can take passive or active forms \citep{lang2023learning, sarmiento2022participatory}. 
In their recent review of educational stakeholder involvement in the design phases of current HCLA systems, \citet{alfredo2024human} found that students and teachers were the most involved stakeholders (71\% and 59\% of 108 surveyed papers, respectively), with students having the highest representation of passive involvement (52\%) and teachers having the highest representation of active involvement (39\%).
The study results further revealed a relatively low representation of student and teacher active involvement, at 19\% and 39\%, respectively. 
Using the five phases of the HCD process \citep{hanington2019universal}, the authors pointed out that most active stakeholder involvement has been observed in phase 2 (exploration, synthesis, and design implications – 32\%) and phase 3 (concept generation and early prototype iteration – 33\%). Furthermore, the authors noted that active stakeholder involvement in phase 4 (evaluation, refinement, and production – 25\%) and phase 5 (launch and monitor – 25\%) was relatively low. This indicates that active stakeholder involvement in the post-production phase of HCLA systems is still limited.
These findings suggest that there is a pressing need to increase student and teacher active involvement throughout the various phases of the design process, from early conception and requirements analysis, to implementation and
evaluation, through to deployment and adoption \citep{alfredo2024human,buckingham2024human,martinez2023human}. In particular, it is crucial to include end-users' voices after the LA system is finalized and deployed in real-world settings to re-align the system more effectively with their educational goals. To address this challenge, in this paper, we aim to actively involve students and teachers in the post-production phase of HCLA systems by giving them agency in implementing LA indicators that can meet their real-world educational needs and preferences. This ensures HCLA system flexibility and adaptability according to changing requirements and contexts.

Recent systematic literature reviews about HCLA  highlight that most of the HCLA studies focus on standalone or embedded LA dashboards (LADs) and the identification of relevant LA indicators for different purposes \citep{topali2025designing,alfredo2024human}. 
Before HCLA, students and teachers rarely had a say in how LADs are designed and what information they receive through LADs \citep{jivet2021quantum}. HCLA allowed to move away from a one-size-fits-all model, whereby the same indicators are provided to each end-user to a personalized approach tailored to the goals, needs, and contexts of different users. While current research on HCLA has demonstrated successful participatory designs of LADs that incorporate the perspectives of both students and teachers, most LADs in the HCLA literature provide manually pre-configured indicators resulting from co-design sessions and often lack adaptability when faced with new end-user requirements after deployment \citep{alfredo2024human}. Additionally, in the studies surveyed in \citet{alfredo2024human} and \citet{topali2025designing}, active stakeholder involvement in the systematic implementation of LA indicators is under-explored. In fact, the participation of stakeholders was primarily limited to the LA indicator design process, e.g., contributing to ideation and prototyping and assisting in testing. In contrast, stakeholders were not actively involved in the implementation of high-fidelity LA indicators, which is done by developers or researchers. 
Furthermore, while investigating the connection between the stakeholders who participated in the process of co-designing the HCLA systems and the end-users aiming to take advantage of the final systems, \citet{topali2025designing} noted that in many cases (14.90\%) the involved stakeholders were different from the targeted ones. The authors further highlighted that focusing on addressing the specific needs of the co-design participants does not guarantee covering the diverse needs of the larger target audience \citep{topali2025designing}.
Therefore, the challenge is how to generalize LA indicators to meet the requirements of different stakeholders. In particular, how to provide the right indicators for new stakeholders who were not involved in the LA indicator co-design process but may have different needs and preferences. 
To address this challenge, previous works identified the need to adapt the LADs to the learning context and target group. \citet{oliver2022adapting} and \citet{teasley2017student} pointed out that LADs should be adapted to the learning context and/or be adaptable by
the students. This aligns with the findings of \citet{roberts2017give}, indicating that students would like to customize their dashboard to include only information they perceive as benefiting them. Similarly, \citet{bennett2019four} concluded that LADs need to be customizable by the student to respond to their individual needs and in this way helps to support a student’s sense of empowerment and agency.
As a concrete solution to meet the need for personalized LADs, \citet{jivet2021quantum} have built a customizable LAD on which students can set goals and decide what indicators they wish to see as part of self-regulating their learning towards achieving their goals. And,
\citet{shreiner2022information} proposed a data visualization inquiry tool that requires students to control and customize visualizations through pull-down menus.  
However, in many cases, what new stakeholders want to see and use on a dashboard differs from what LADs provide. Therefore, in addition to engaging stakeholders (i.e., students and teachers) in customizing LADs, putting them in the driver's seat and empowering them to adapt existing indicators according to their needs or develop new indicators beyond the pool of existing ones can further support their agency and better meet their individual goals, needs, and learning context.
To this end, in this paper, we aim to include students' and teachers' voices in the post-production phase by empowering them to take control of the LA indicator implementation process.
Implementing LA indicators requires a significant investment of time. Moreover, it can be expensive, requiring sophisticated technology, resources, and technical expertise \citep{buckingham2024human}. This may lead to heavy reliance on developers and researchers with programming experience. To fill this gap, in this paper, we present the \textit{Indicator Editor}, a no-code, low-cost, interactive LA tool that enables educational stakeholders with knowledge of data analysis and visualization to implement LA indicators with minimal effort. 
We argue that ensuring active stakeholder engagement in the LA indicator implementation process can foster a sense of ownership, enhance transparency, build trust, and drive forward the acceptance and adoption of HCLA systems.
\section{Indicator Editor}  \label{indicator-editor}
The primary goal of \textit{Indicator Editor} is to achieve transparent and human-centered LA. Following a transparency-through-exploration approach, \textit{Indicator Editor} empowers LA stakeholders to steer the LA indicator development process to bring new insights and gather knowledge through exploratory data analysis and visualization. Designed for students and teachers with data analytics and visualization knowledge, \textit{Indicator Editor} adopts a human-in-the-loop approach to bring these stakeholders into the LA development process to turn educational data into value. The stakeholders can gain insights into educational data by visually exploring data, iteratively refining hypotheses, and getting answers to their questions.

\subsection{User Scenarios}
We present two user scenarios for how LA stakeholders (teachers and students) can use the \textit{Indicator Editor} to develop LA indicators based on their needs and goals.
\subsubsection{Teacher Scenario:}
Aryan, a professor at XYZ University, uses a MOOC platform to manage his course, which hosts hundreds of students. To effectively oversee his course, he relies on the personalized dashboard in CourseMapper
, which provides various predefined indicators. These include insights into student participation in lectures, activity in discussion forums, and assignment progress, giving him a high-level overview of the course's general health. However, he has specific pedagogical interests that go beyond these predefined metrics. He is keen on understanding how frequently students engage with his learning materials and wants to identify how many students consistently access these resources. Unfortunately, none of the existing indicators on CourseMapper fully capture this detailed information. This is where the \textit{Indicator Editor} comes into play. Using the \textit{Indicator Editor}, he creates two custom indicators that better align with his instructional goals. The first indicator he created tracked the total access to the learning materials across all students. This allowed him to gauge the overall level of student interaction with his content. For the second indicator, he wanted to get a more granular view by measuring how many individual students are accessing the materials, helping him understand the breadth of engagement among the student body. 
Aryan configures these indicators through the \textit{Indicator Editor} by selecting relevant parameters and setting them to automatically perform statistical analyses, such as counting total accesses and calculating the number of unique student views. Once these indicators are generated, he embeds their code snippets into his personalized CourseMapper dashboard. Now, alongside the standard indicators, he has access to his customized metrics, which give him a deeper understanding of how students interact with the course materials.
\begin{figure}[!ht]
    \centering
    \includegraphics[width=\linewidth]{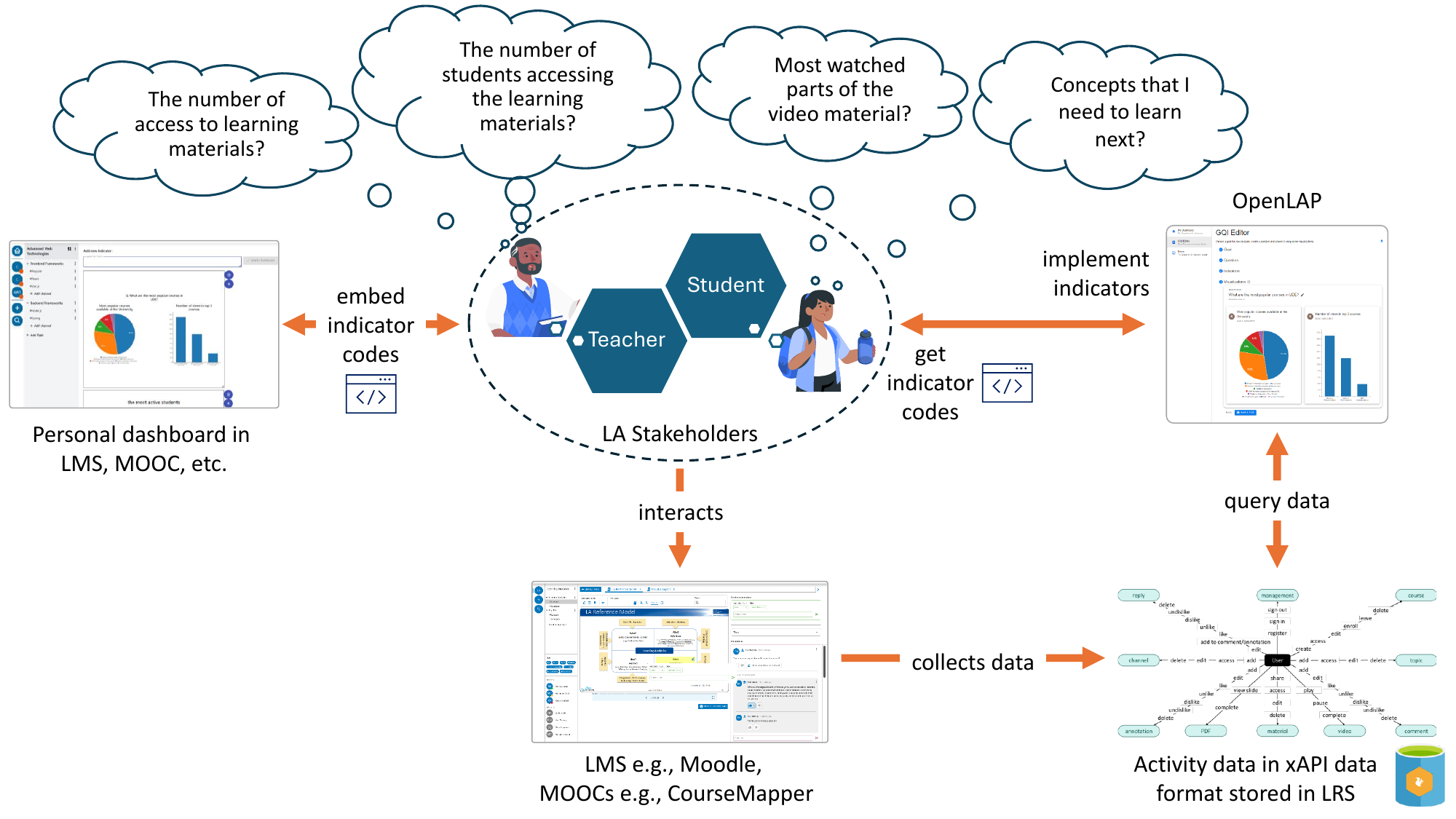}
    \caption{User scenario - Teachers/students use \textit{Indicator Editor} to create indicators for their CourseMapper dashboard}
    \label{fig:openlap-teacher-scenario}
\end{figure}

\subsubsection{Student Scenario:}
Qintha is a student in Aryan's course at XYZ University, delivered through CourseMapper. As she navigates the course, she wants to understand better which parts of the course materials are most popular among her peers and receive personalized recommendations on what concepts to focus on next. While using CourseMapper, she encounters predefined indicators, such as her overall progress in the course and her grades on assignments. However, these metrics do not fully satisfy her curiosity about how other students interact with the course or which concepts she should prioritize. To address this, she turns to the \textit{Indicator Editor} to create two personalized indicators that give her the insights she needs. The first indicator she creates identifies the most-watched parts of the course videos. She is interested in knowing which sections of the video lectures attract the most attention from her classmates. This helps her identify which topics or explanations are perceived as important or challenging by others, giving her a sense of what might require extra attention. Using the \textit{Indicator Editor}, she configures a custom metric that analyzes student activity within the videos and pinpoints the specific timestamps or sections with the highest view counts or replay rates. With this indicator embedded in her dashboard, she can easily see which parts of the videos are the most popular among her peers, helping her align her studies with the broader class focus.

Qintha created another indicator that aimed at helping her understand which concepts she needed to learn next. As the course material becomes increasingly complex, she wants personalized recommendations on what concepts to focus on to stay on track. Using the \textit{Indicator Editor}, she creates another indicator that evaluates her quizzes, assignments, and video engagement performance. Based on this data, the indicator highlights key concepts she should review or study further. It pulls information from the course concept map and her activity to suggest the most relevant concepts to learn next. By embedding these custom indicators into her personalized CourseMapper dashboard, she can track which parts of the video lectures are popular among her peers and receive actionable recommendations on what to focus on next.
\subsection{Design Goals}
We investigated the LA literature to extract requirements
and determine design goals (DGs)
for the \textit{Indicator Editor}. We analyzed the existing works, considering the steps needed to create an LA indicator. LA is an iterative and cyclical process, generally carried out in different steps. For example, data-gathering and preprocessing, analysis, graphical visualization, interpretation, and reflection \citep{dyckhoff2012design}, data collection and pre-processing, analytics, and actions, post-processing \citep{chatti2012reference}, learning activities, data collection, data storage and processing, analysis, visualization, and action \citep{chatti2019perla, chatti2020lava}, data gathering, information processing, and knowledge application \citep{elias2011learning}, awareness, (self-)reflection, sensemaking, and \citep{verbert2013learning}, learners, data, metrics and interventions \citep{clow2012learning}.
The cyclical nature of the LA process suggests that the development of LA indicators, which is at the core of the LA process, is a user-centered, iterative, and interactive process that aims to help users generate hypotheses and questions, interact with visualizations, explore insights, and take actions based on gained knowledge.
Based on this, we determine as a first design goal for the \textit{Indicator Editor} \textbf{(DG1) Interactive and exploratory user interface:} Users should be able to develop indicators that meet their needs and goals. The process should be iterative and exploratory, allowing users to continue making changes until they find satisfactory answers to their questions. This process should simplify each step while allowing users to go back and change their previous decisions by modifying different analysis parameters or directly interacting with the visualization.

From a technical perspective, the steps of the LA process further indicate that \textit{data}, \textit{analysis}, and \textit{visualization} are the building blocks of the LA development process. LA indicators can be described as ``specific calculators with corresponding visualizations, tied to a specific question'' \cite[p. 60]{dyckhoff2012design}. The Goal-Question-Indicator (GQI) approach proposed in \citep{muslim2017goal} suggests a progressive step-by-step process, which can guide users incrementally toward their goals by formulating questions and defining indicators to answer them. The GQI approach includes three main steps. The first step is for users to define a specific goal for LA. The second step is to formulate a particular LA question based on the goal defined in the previous step. Lastly, users can select existing or defined indicators to answer the LA question specified in the previous step. Users must follow four steps when defining new indicators: explore the datasets, apply filters, and choose the analysis and visualization methods. The users can also select the appropriate parameters for the analysis and visualization. 
Using these four steps, we determine the following design goals for the implementation of LA indicators using the \textit{Indicator Editor}  
\textbf{(DG2) Datasets:} Users should be able to explore learning activity data to define the required dataset for the indicator. This includes, for instance, selecting data sources, identifying which platforms’ data to include, and choosing the types of activities to focus on. 
\textbf{(DG3) Data filters:} Users should be able to apply additional filters to refine the dataset, select a specific time, define which users to include, and other relevant filtering criteria. 
\textbf{(DG4) Analysis:} Users should be able to choose the analysis method to apply to the filtered dataset. This could involve selecting the appropriate analytical technique and adjusting its parameters. 
\textbf{(DG5) Visualization:} Users should be able to define how the analyzed results will be visualized. This might involve choosing the visualization library, selecting the chart type, and specifying the data to be displayed in the visualization. 
\subsection{Iterative Design}
The \textit{Indicator Editor} allows users to create personalized basic indicators that use simple statistics to create visualizations. More complex indicators, e.g., based on machine learning algorithms, classification, or prediction, are out of the scope of this paper. The \textit{Indicator Editor} was iteratively designed in three iterations by using low- and high-fidelity prototypes. We systematically followed the Human-Centered Design (HCD) approach and usability design principles to create an intuitive UI \citep{norman2013design}.
\subsubsection{First iteration, initial design, low-fidelity prototypes:}
\begin{figure}[!ht]
    \begin{center}
        \begin{subfigure}[normla]{0.24\linewidth}
            \includegraphics[width=\linewidth]{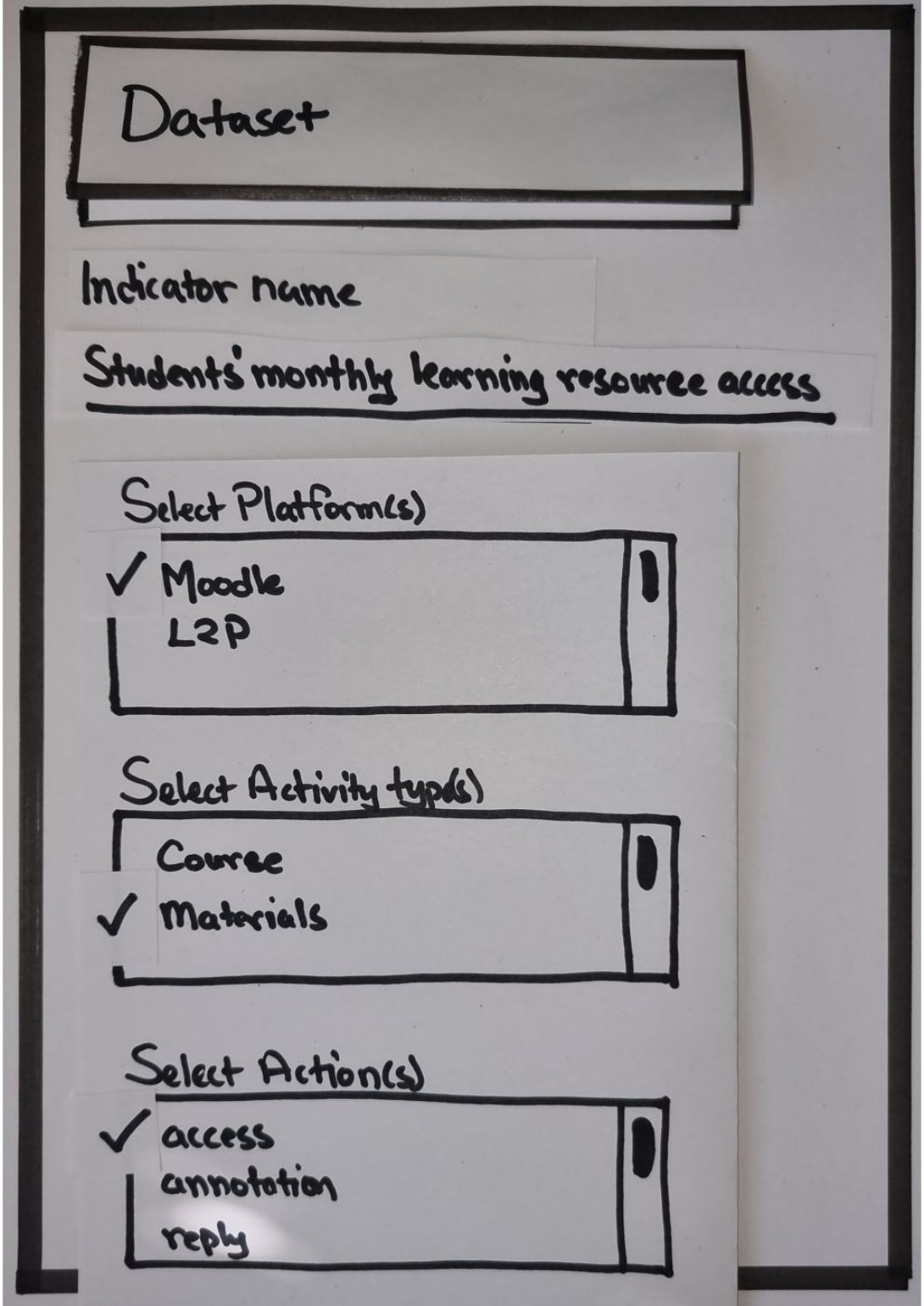}
            \caption{}
            \label{subfig:low-select-dataset}
        \end{subfigure}
        ~
        \begin{subfigure}[normla]{0.24\linewidth}
            \includegraphics[width=\linewidth]{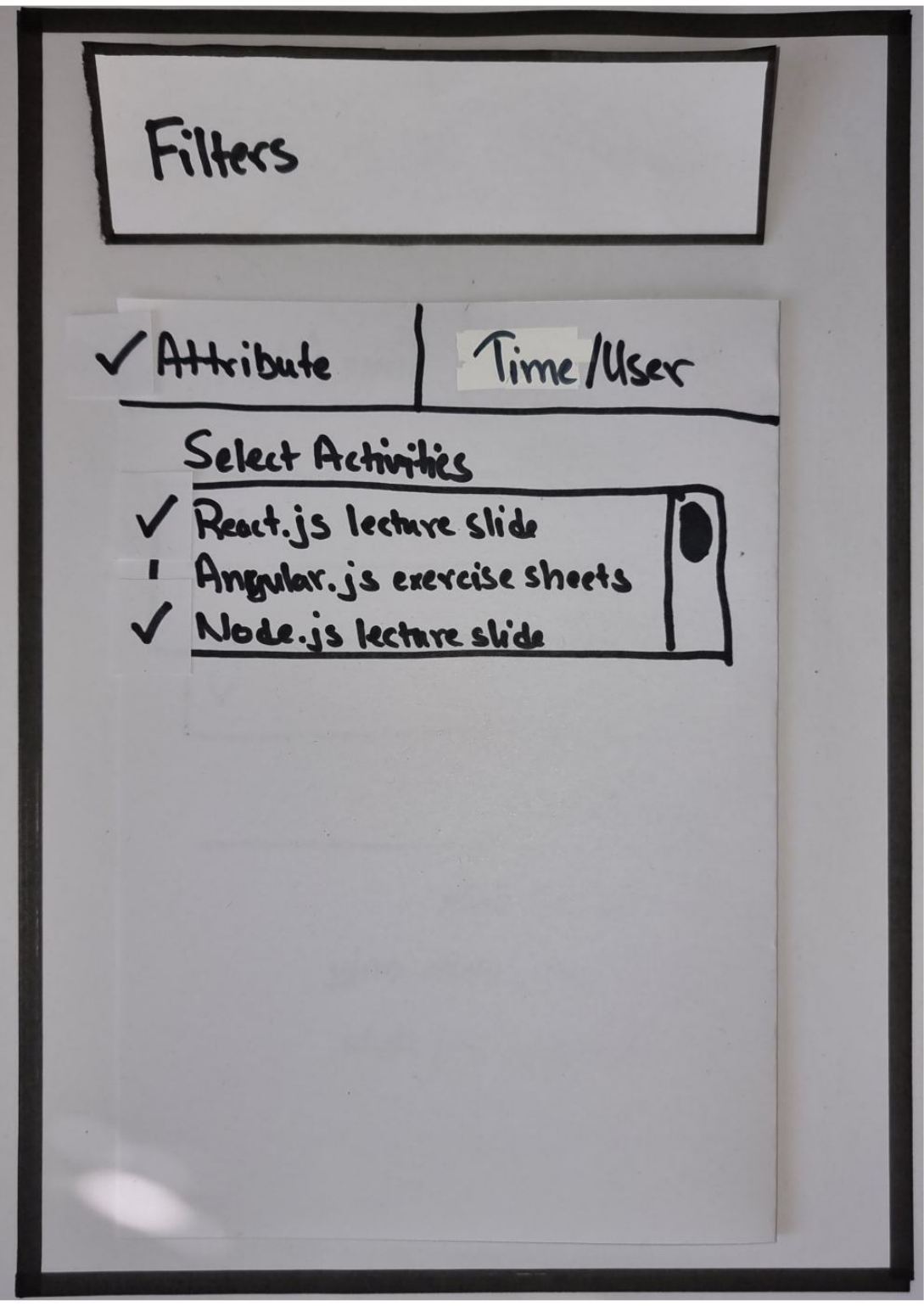}
            \caption{}
            \label{subfig:low-select-filters-activities}
        \end{subfigure}
        ~
        \begin{subfigure}[normla]{0.24\linewidth}
            \includegraphics[width=\linewidth]{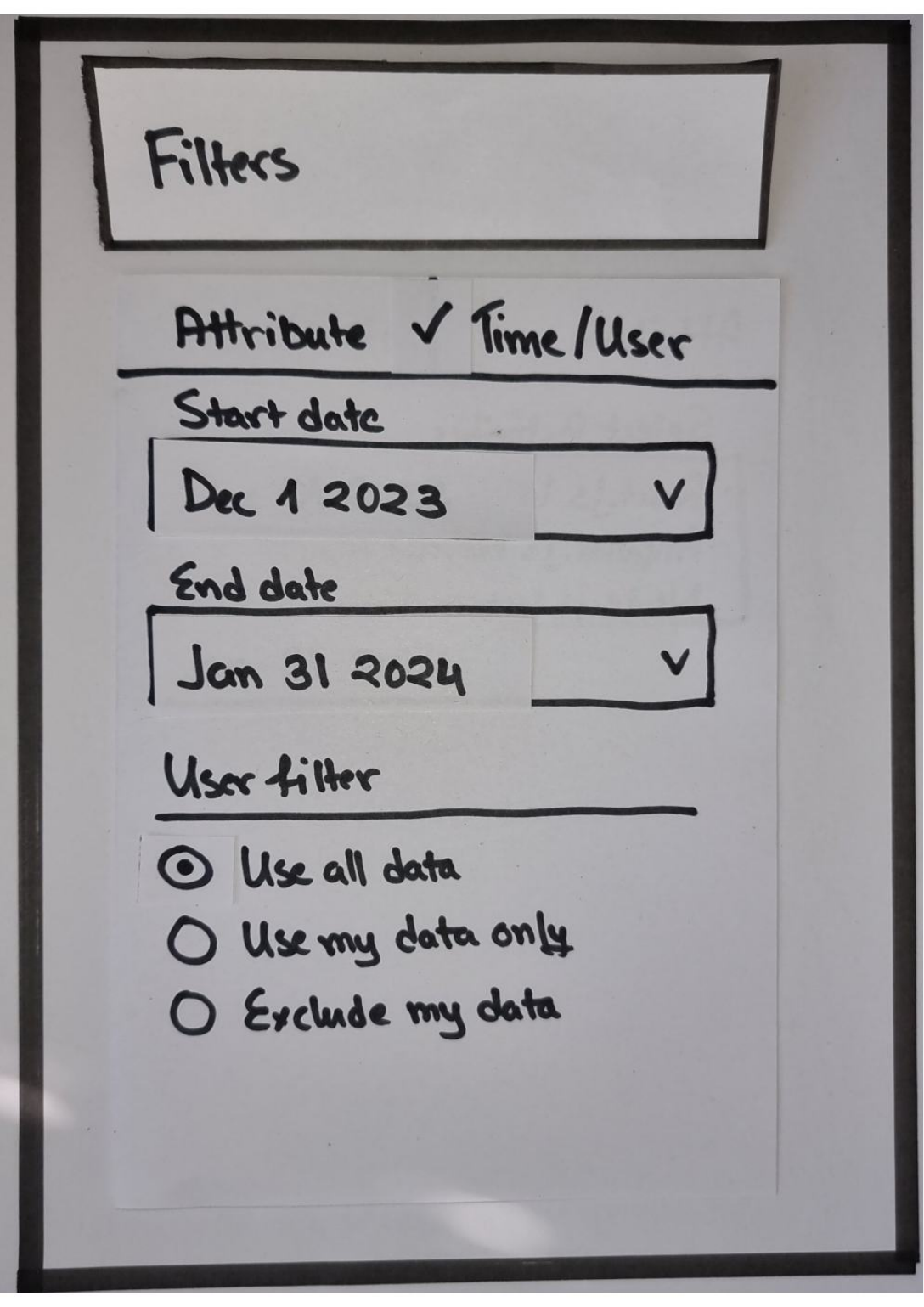}
            \caption{}
            \label{subfig:low-select-filters-time-user}
        \end{subfigure} \\
        ~
        \begin{subfigure}[normla]{0.24\linewidth}
            \includegraphics[width=\linewidth]{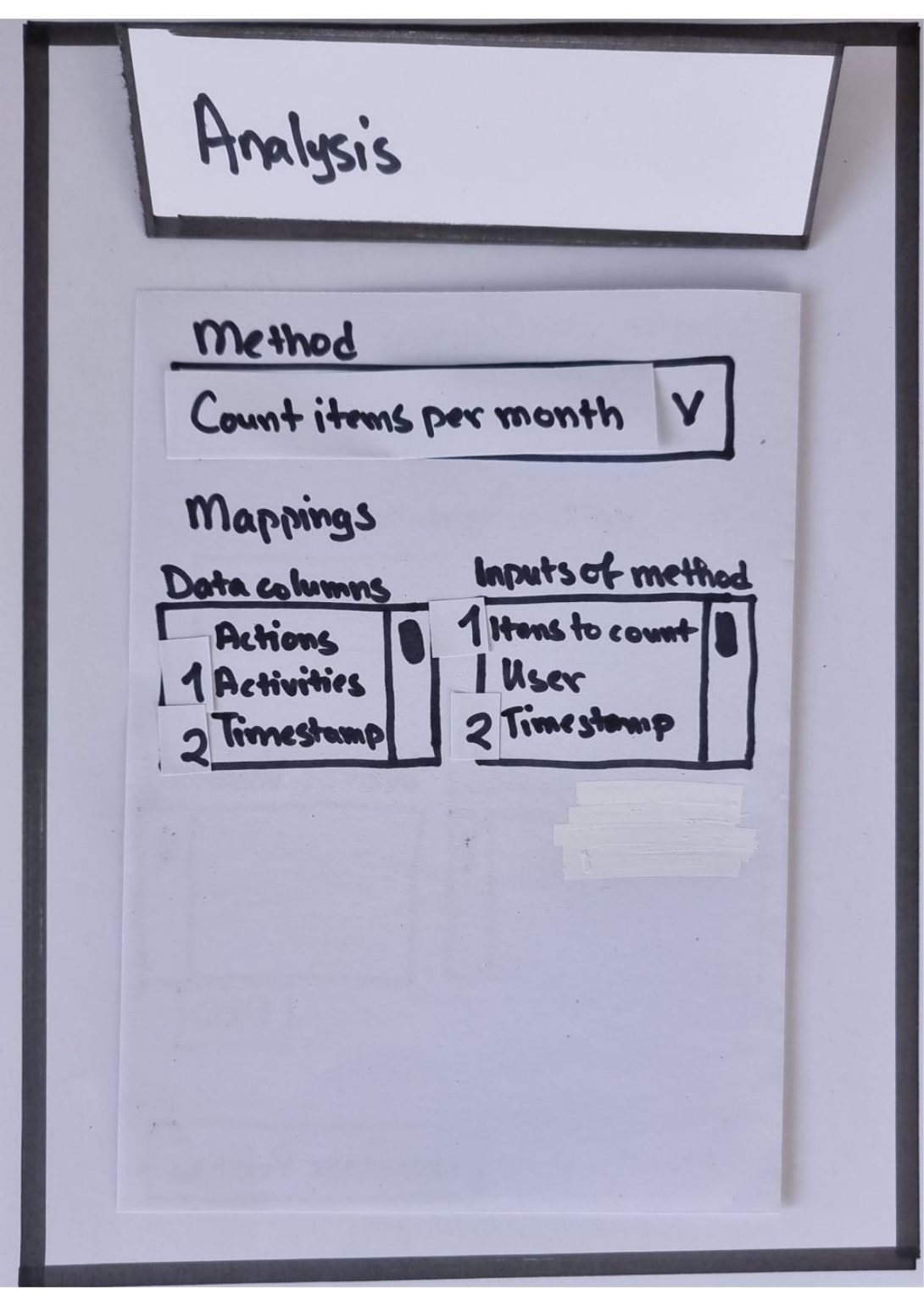}
            \caption{}
            \label{subfig:low-select-analysis}
        \end{subfigure}
        ~
        \begin{subfigure}[normla]{0.24\linewidth}
            \includegraphics[width=\linewidth]{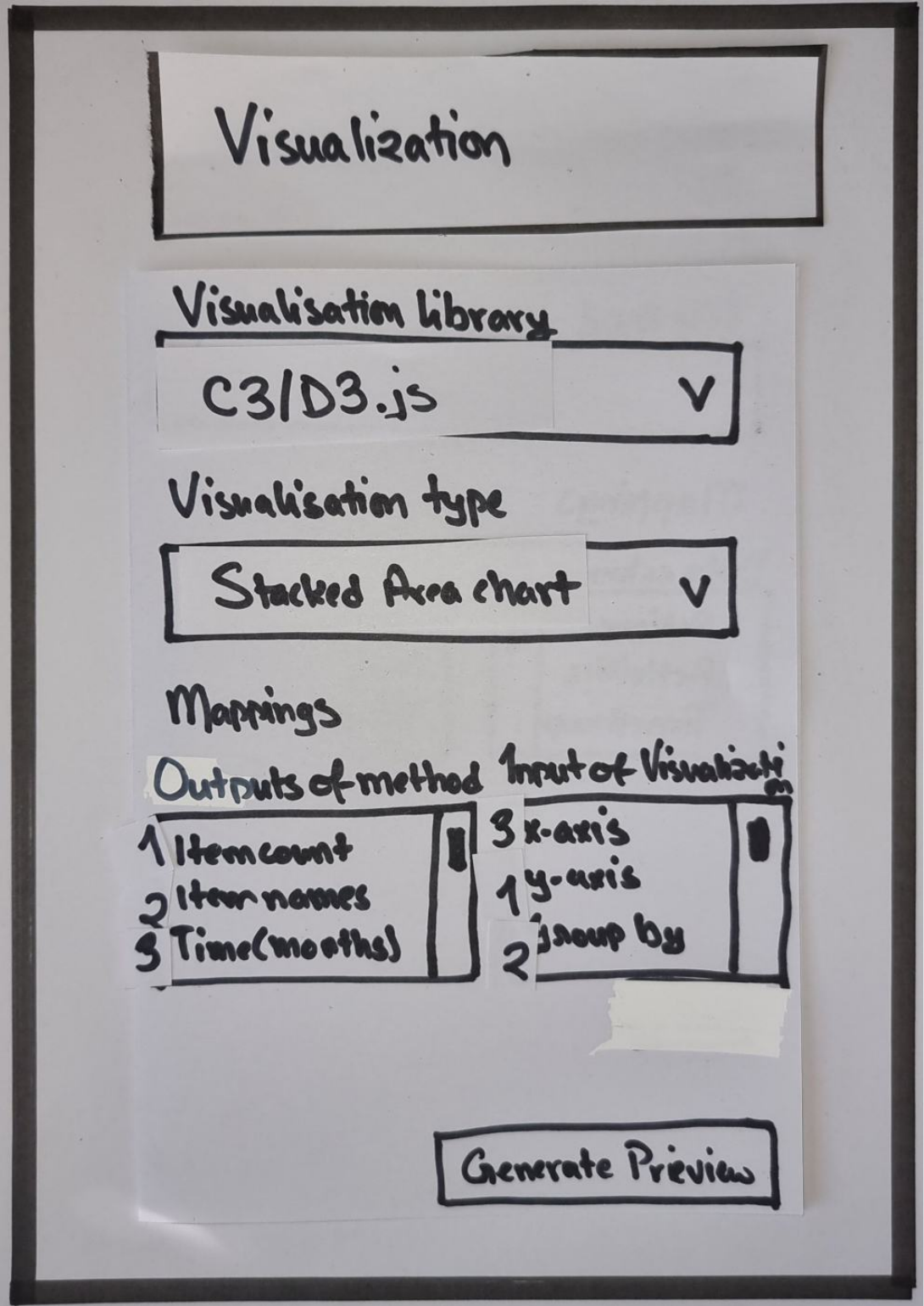}
            \caption{}
            \label{subfig:low-select-visualization}
        \end{subfigure}
        ~
        \begin{subfigure}[normla]{0.24\linewidth}
            \includegraphics[width=\linewidth]{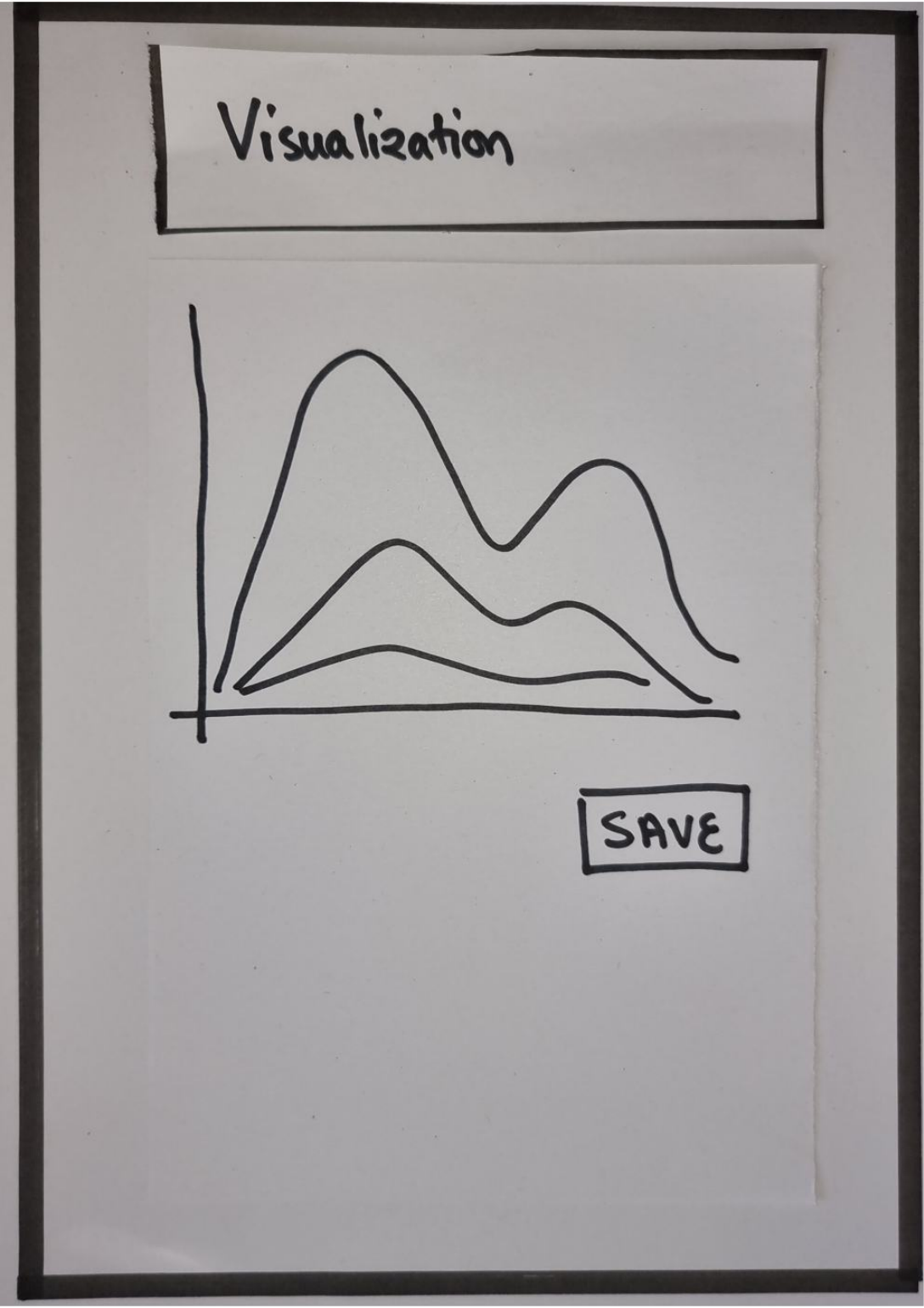}
            \caption{}
            \label{subfig:low-preview-indicator}
        \end{subfigure} \\
        ~
        \begin{subfigure}[normla]{0.24\linewidth}
            \includegraphics[width=\linewidth]{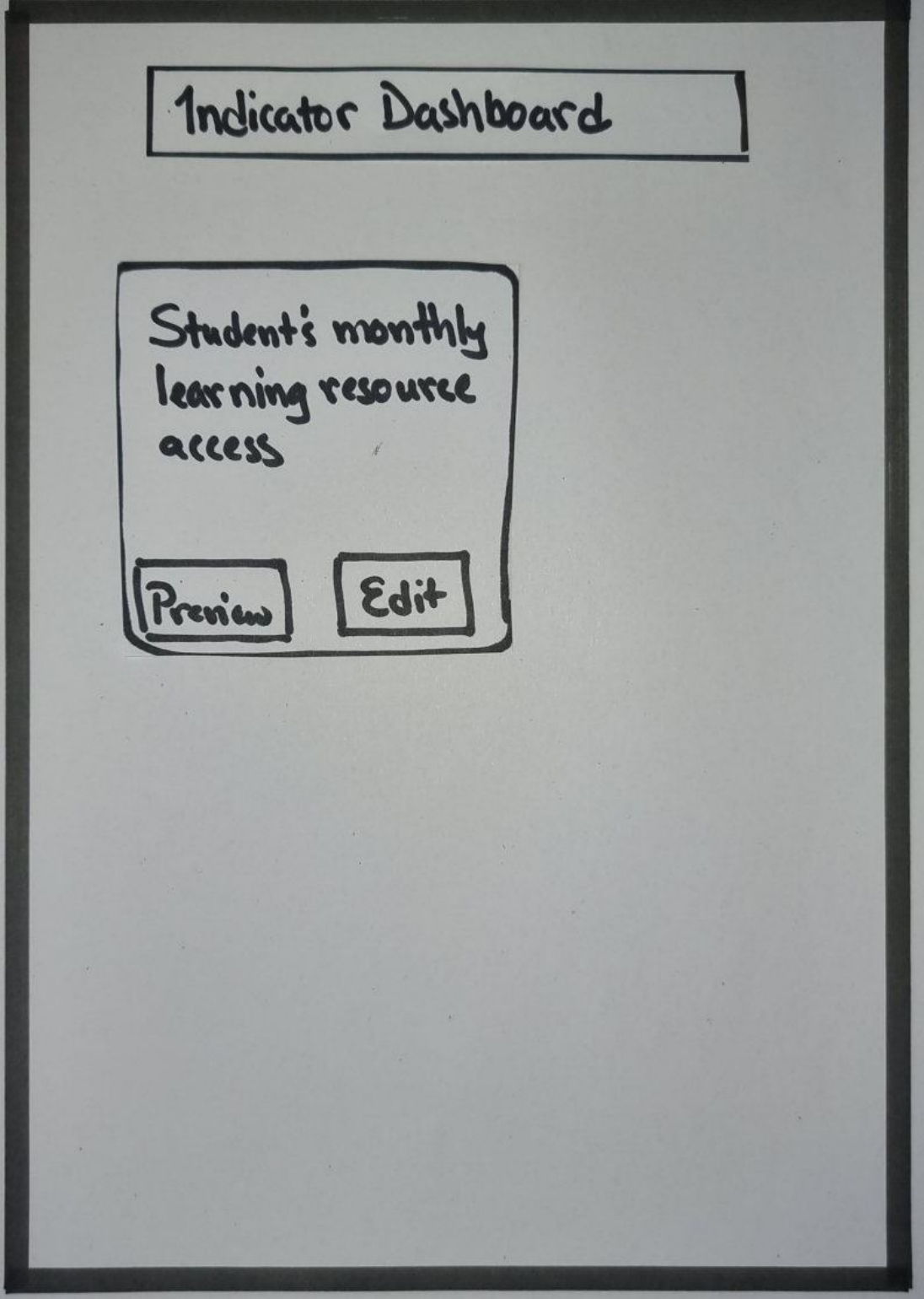}
            \caption{}
            \label{subfig:low-dashboard}
        \end{subfigure}
    \end{center}
    \caption{Paper Prototypes of the \textit{Indicator Editor}}
    \label{fig:low-fil-prototypes}
\end{figure}

We generated preliminary design ideas using low-fidelity prototypes, starting with paper prototypes due to their simplicity and convenience. As shown in Figure \ref{fig:low-fil-prototypes}, users can begin the indicator creation process by selecting the source of the data \textbf{(DG2)} (Figure \ref{subfig:low-select-dataset}), filtering the data based on activities \textbf{(DG3)} (Figure \ref{subfig:low-select-filters-activities}) and timestamp or user \textbf{(DG3)} (Figure \ref{subfig:low-select-filters-time-user}), selecting the analysis method and defining the mappings between the data and the inputs of the analytics method \textbf{(DG4)} (Figure \ref{subfig:low-select-analysis}), selecting the visualization technique and defining the mappings between the outputs of the analysis method and the inputs of the visualization technique \textbf{(DG5)} (Figure \ref{subfig:low-select-visualization}), previewing the indicator before saving it \textbf{(DG5)} (Figure \ref{subfig:low-preview-indicator}), viewing the indicator in a dashboard, and editing the indicator, if needed \textbf{(DG1}) (Figure \ref{subfig:low-dashboard}).

To evaluate the low-fidelity prototypes, we conducted interview sessions with five participants (three females and two males) from the local university, including two master's and two bachelor's students and one teaching assistant, aged between 22 and 30. They were studying Engineering and Business Intelligence programs. All participants were familiar with data analytics and visualization tools. Before the interviews, we prepared a list of core activities for participants to perform using the \textit{Indicator Editor}, such as selecting a dataset, applying filters, choosing analytics methods, and selecting a visualization. 
Overall, participants found the \textit{Indicator Editor}'s UI elements clear and the interactions intuitive. However, three participants noted that the abundance of options presented in the UI from the beginning was overwhelming, leading to confusion about where to start. Specifically, they preferred a step-by-step guide that would reveal options progressively as they advanced through the steps. Moreover, they wanted to see the number of steps required to create the indicator. Most participants (n=4) further commented that the selections in the steps of the dataset (platforms, activity types, actions) and in filters (activities) using checkboxes were not intuitive. Moreover, they found the selections in the mappings using numbers confusing. We used this feedback in the subsequent design iteration to enhance the low-fidelity prototypes.
\subsubsection{Second iteration, high-fidelity prototypes}
\begin{figure}[!ht]
    \begin{center}
        \begin{subfigure}[normla]{0.4\linewidth}
            \includegraphics[width=1\linewidth]{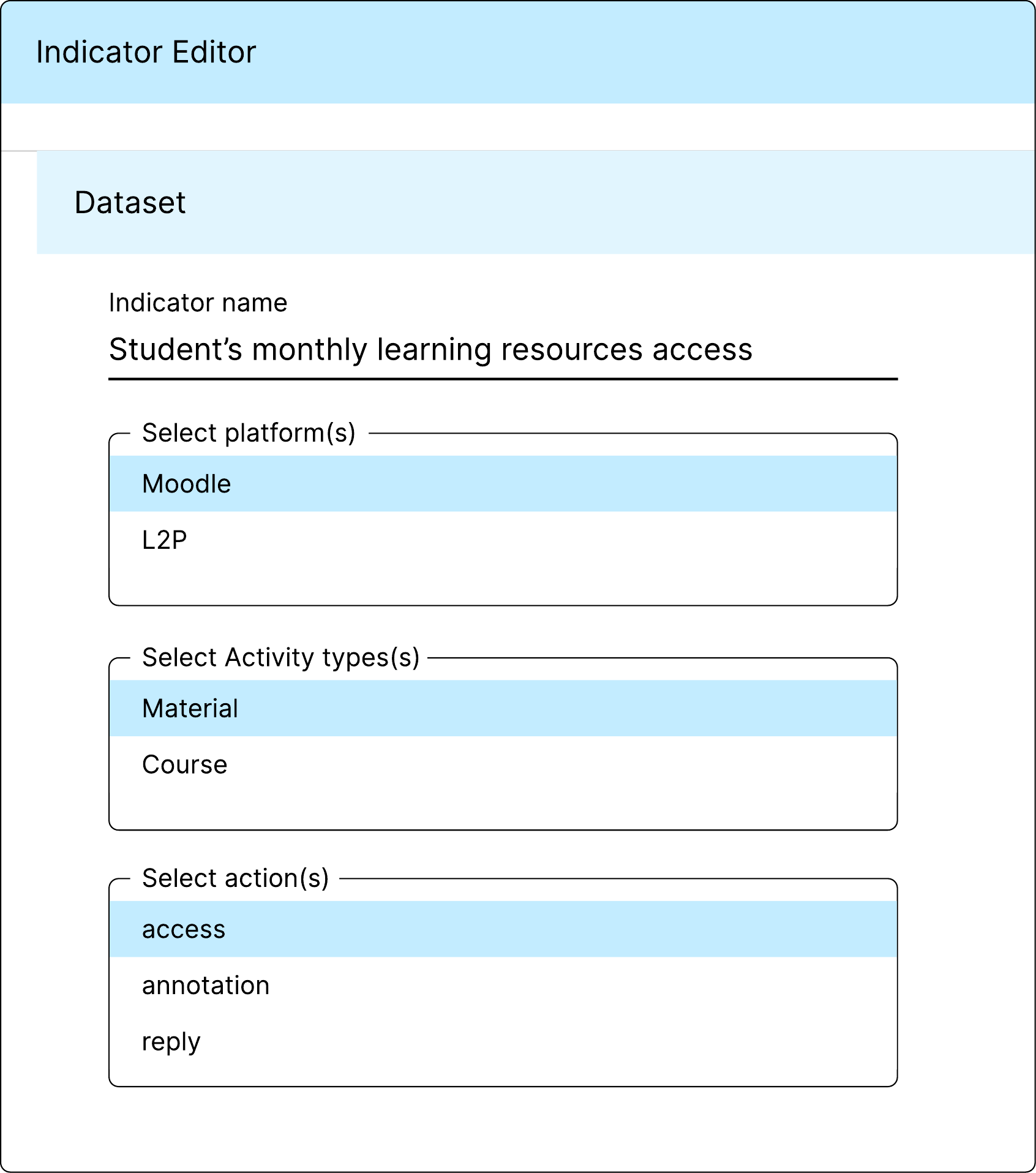}
            \caption{}
            \label{subfig:high-select-dataset}
        \end{subfigure}
        ~
        \begin{subfigure}[normla]{0.4\linewidth}
            \includegraphics[width=1\linewidth]{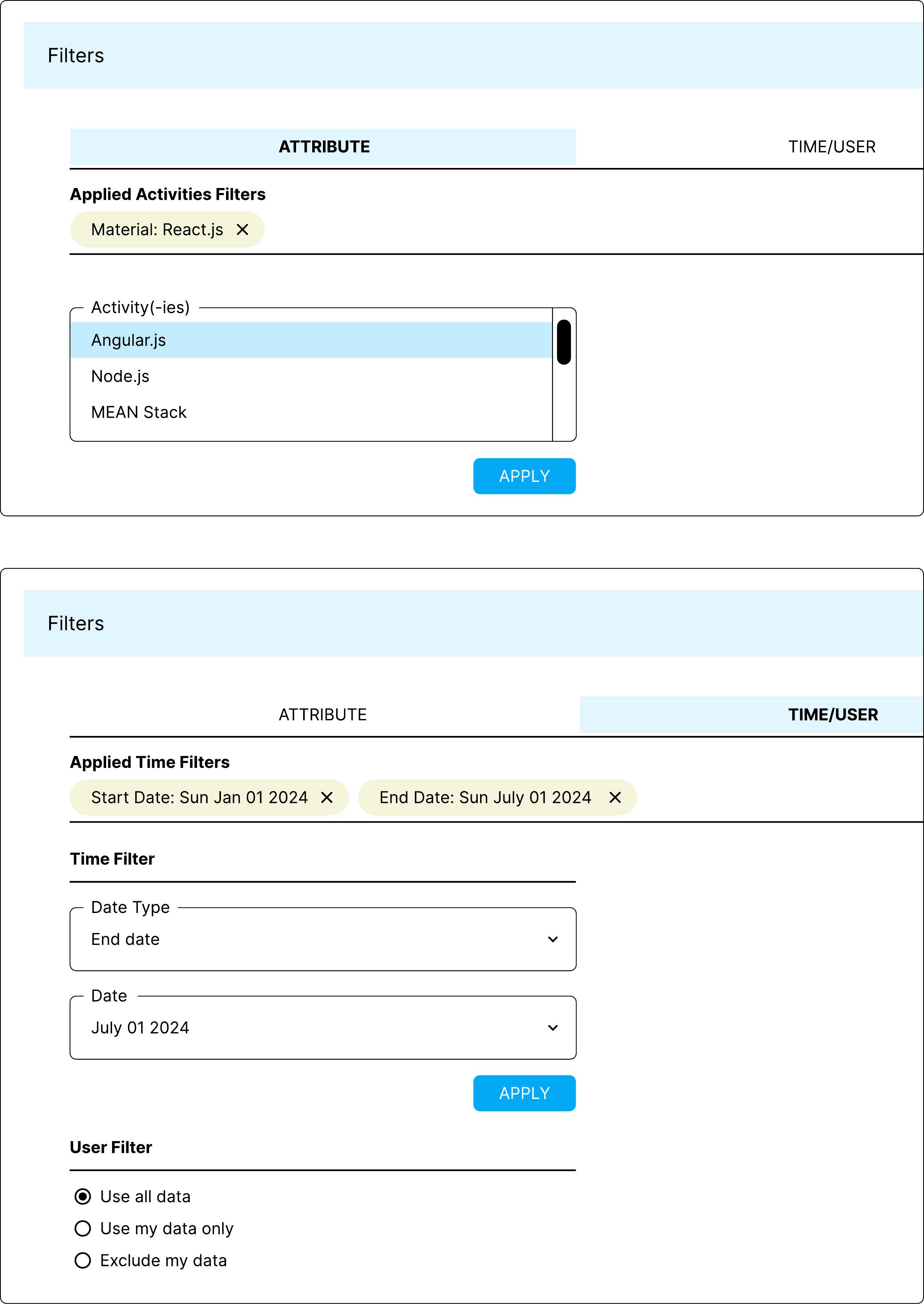}
            \caption{}
            \label{subfig:high-select-filters-activities}
        \end{subfigure}
    \end{center}
    \caption{High-fidelity Prototypes of the  \textit{Indicator Editor}: Dataset and Filters}
    \label{fig:high-fil-prototypes-1}
\end{figure}

\begin{figure}[!ht]
    \begin{center}
        \begin{subfigure}[normla]{0.4\linewidth}
            \includegraphics[width=1.0\linewidth]{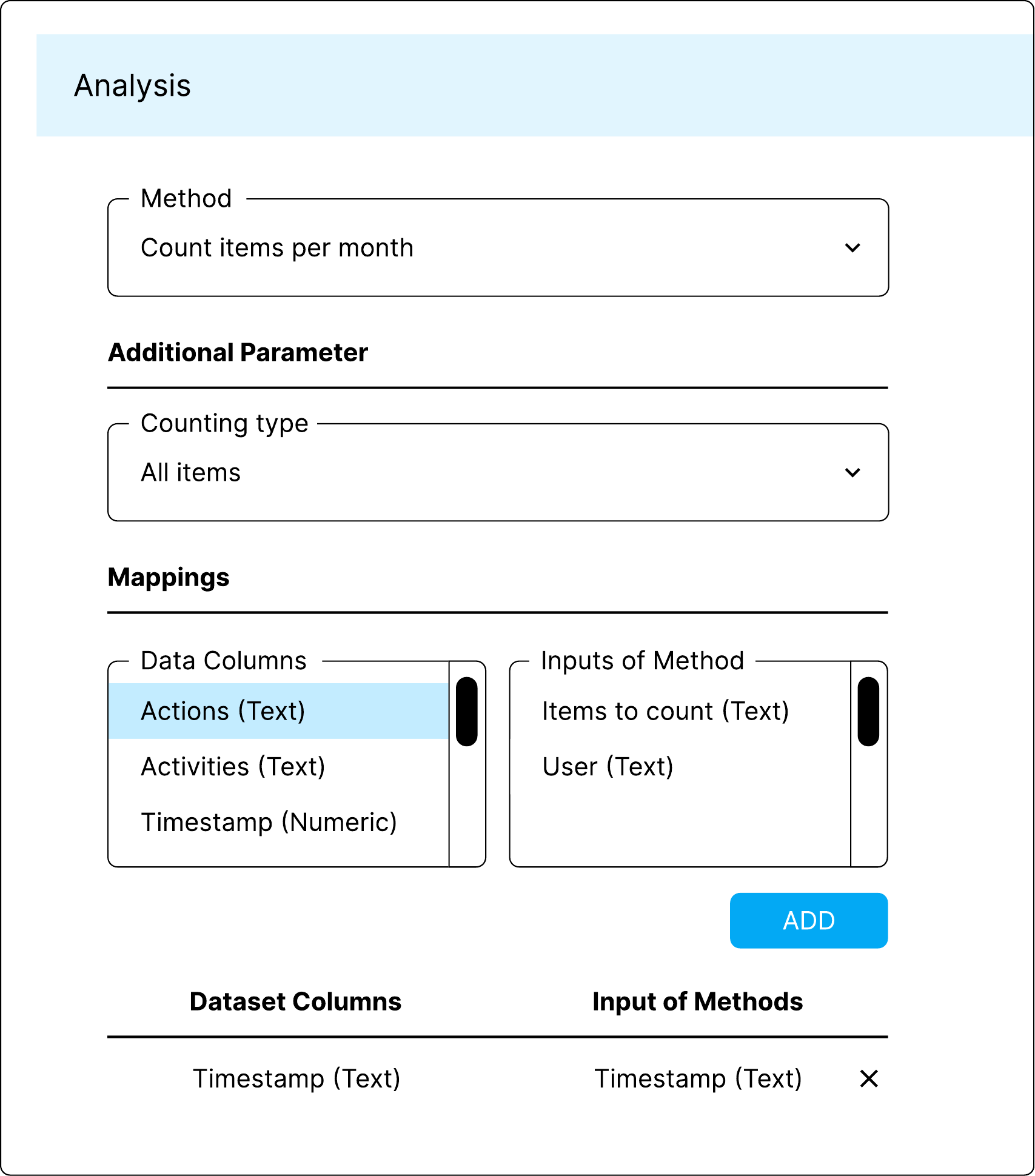}
            \caption{}
            \label{subfig:high-select-analysis}
        \end{subfigure}
        ~
        \begin{subfigure}[normla]{0.4\linewidth}
            \includegraphics[width=1.0\linewidth]{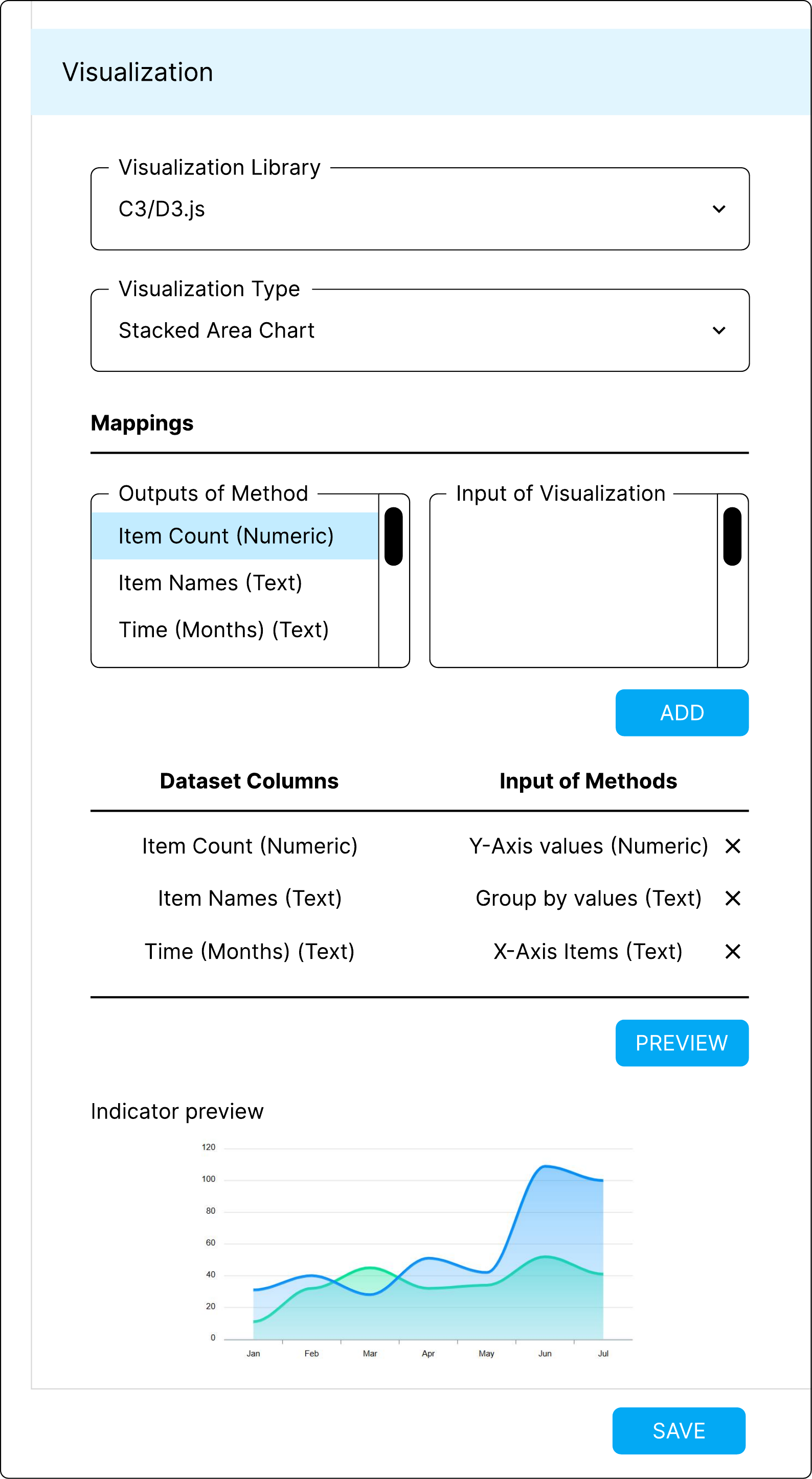}
            \caption{}
            \label{subfig:high-select-visualization}
        \end{subfigure} \\
        ~
        \begin{subfigure}[normla]{0.4\linewidth}
            \includegraphics[width=1.0\linewidth]{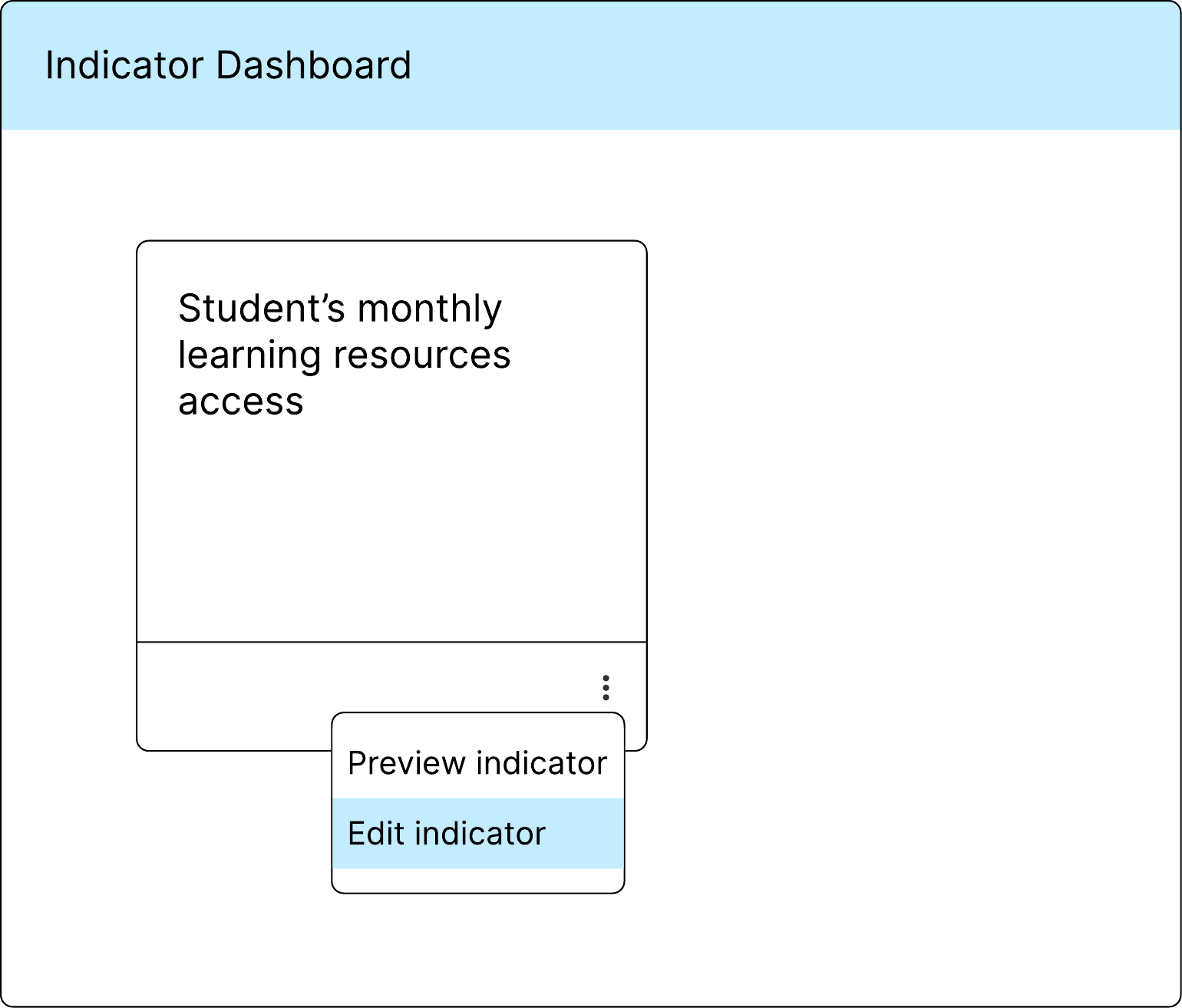}
            \caption{}
            \label{subfig:high-preview-indicator}
        \end{subfigure}
    
    \end{center}
    \caption{High-fidelity Prototypes of the  \textit{Indicator Editor}: Analysis, Visualization, and Dashboard}
    \label{fig:high-fil-prototypes-2}
\end{figure}

In this iteration, we created high-fidelity prototypes using Figma. One of the key issues raised by users in the previous iteration was their desire to see a clear progression of the steps required to create an indicator. As shown in Figure \ref{fig:high-fil-prototypes-1} and Figure \ref{fig:high-fil-prototypes-2}, the updated prototype of the \textit{Indicator Editor} incorporated a form-like structure that provides users with an overview of these steps. This prototype allows for greater interactivity, such as clicking to reveal the next set of options after each step \textbf{(DG1)}. 
Additionally, the selections were improved by adding new sections in the UI, such as direct selection and highlighting dataset options (i.e., platforms, activity types, actions) \textbf{(DG2)} (Figure \ref{subfig:high-select-dataset}). Furthermore, `Applied Activities Filters' and `Applied Time Filters' were added in filters, which show the selected filtered items, and `APPLY' buttons to perform these selections \textbf{(DG3)} (Figure \ref{subfig:high-select-filters-activities}).
Furthermore, the selections for mappings between the data and the input for the analysis method as well as between the output of the analysis method and the visualization inputs have been enhanced by displaying two lists side by side. Users can select a mapping and then click the `ADD' button. The chosen mapping will be shown below the lists, and users can also deselect it \textbf{(DG4, DG5)} (Figures \ref{subfig:high-select-analysis} \& \ref{subfig:high-select-visualization}).

We gathered feedback on the high-fidelity prototypes from five new participants from the local university: one PhD student, three master students, and one bachelor student, aged between 20 and 32, studying Computer Science and Engineering degrees. All participants were familiar with data analytics and visualization tools. We asked them to complete a task that involved creating an indicator to monitor students' monthly learning resource access. Throughout the task, we encouraged them to think aloud. Overall, the participants responded positively to the prototypes and provided constructive criticism and suggestions for improvements. 
Specifically, they recommended making the indicator chart more prominent, such as spanning the entire width or displaying it in full-screen. Additionally, three participants expressed confusion about mapping the data columns to the method's inputs and mapping the method outputs to the visualization inputs. They suggested tooltips to help them choose suitable options. Moreover, all participants expressed dissatisfaction with the form-like structure of the \textit{Indicator Editor}. While the interface displayed the upcoming steps, further options were unresponsive due to the incompletion of previous steps, which led to frustration and caused users to click around aimlessly. Although they appreciated knowing the number of steps required to complete the process, they preferred a more guided approach where options for future steps would only appear once the current step had been completed.
\subsection{Final prototypes and implementation}
\begin{figure}[!ht]
    \begin{center}
        \begin{subfigure}[normla]{0.32\linewidth}
            \includegraphics[width=1\linewidth]{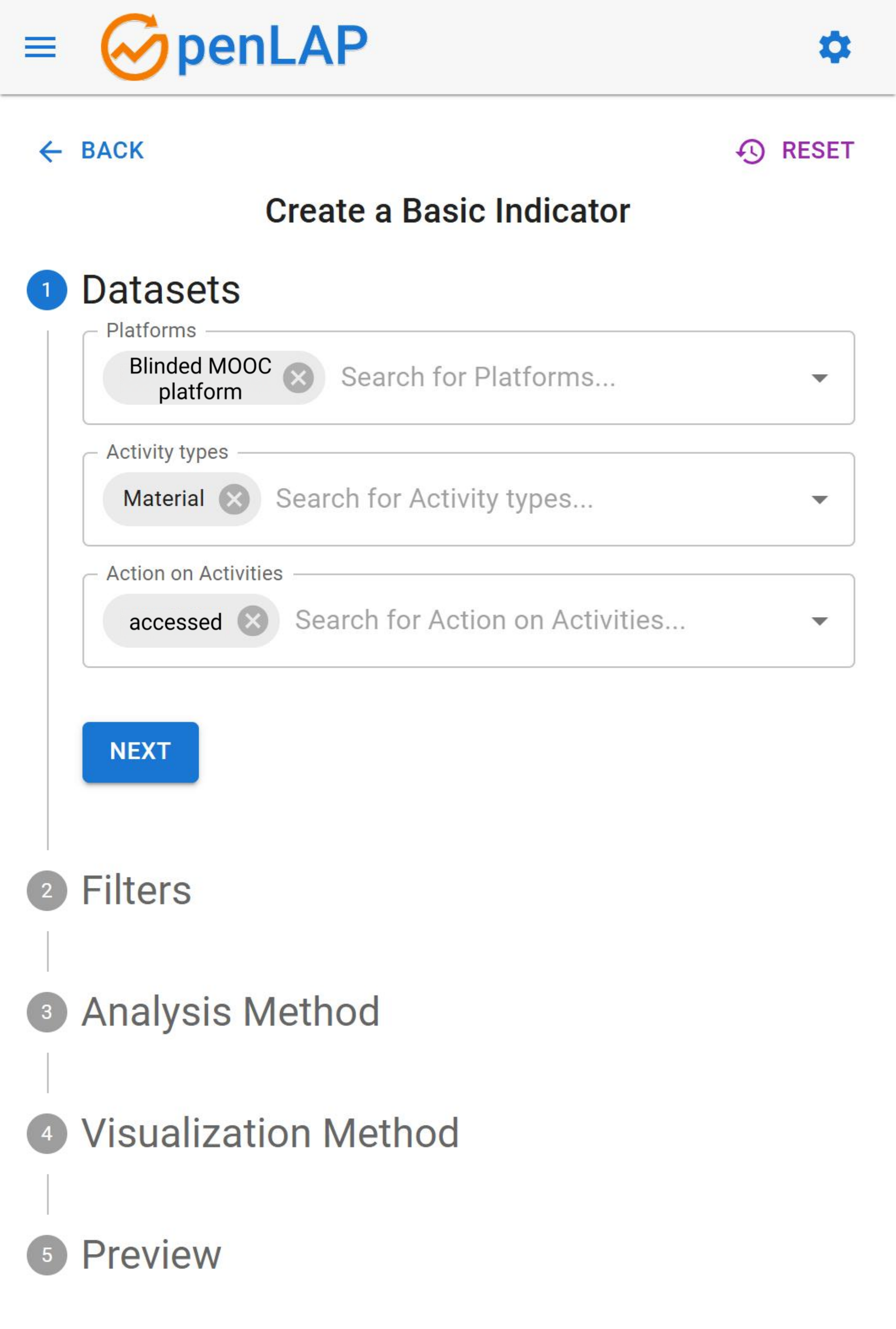}
            \caption{Select Dataset}
            \label{subfig:select-dataset}
        \end{subfigure}
        ~
        \begin{subfigure}[normla]{0.32\linewidth}
            \includegraphics[width=1\linewidth]{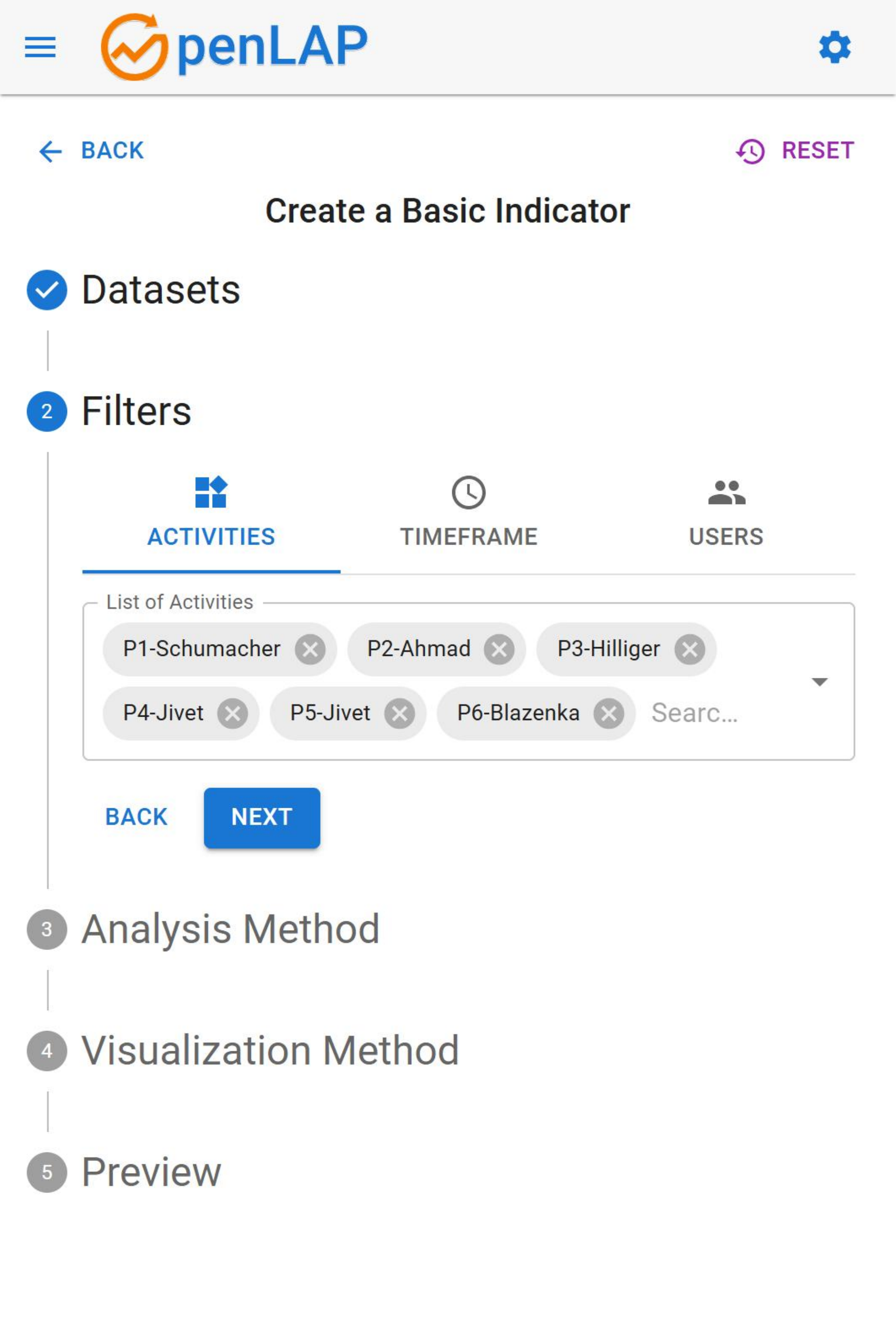}
            \caption{Select Filters: Activities}
            \label{subfig:select-filters-activities}
        \end{subfigure}
        ~
        \begin{subfigure}[normla]{0.32\linewidth}
            \includegraphics[width=1.0\linewidth]{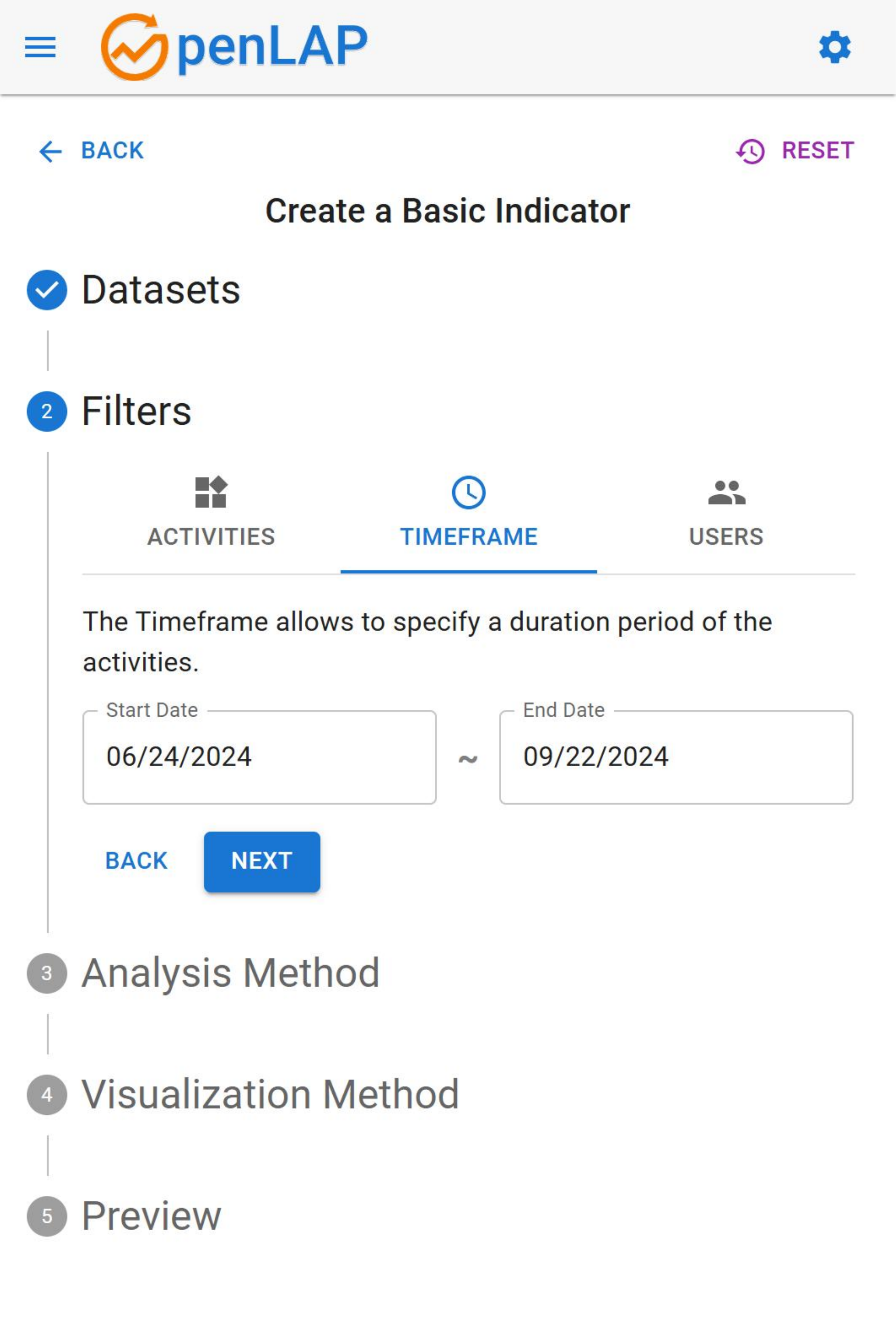}
            \caption{Select Filters: Time}
            \label{subfig:select-filters-time}
        \end{subfigure} \\
        ~
        \begin{subfigure}[normla]{0.32\linewidth}
            \includegraphics[width=1.0\linewidth]{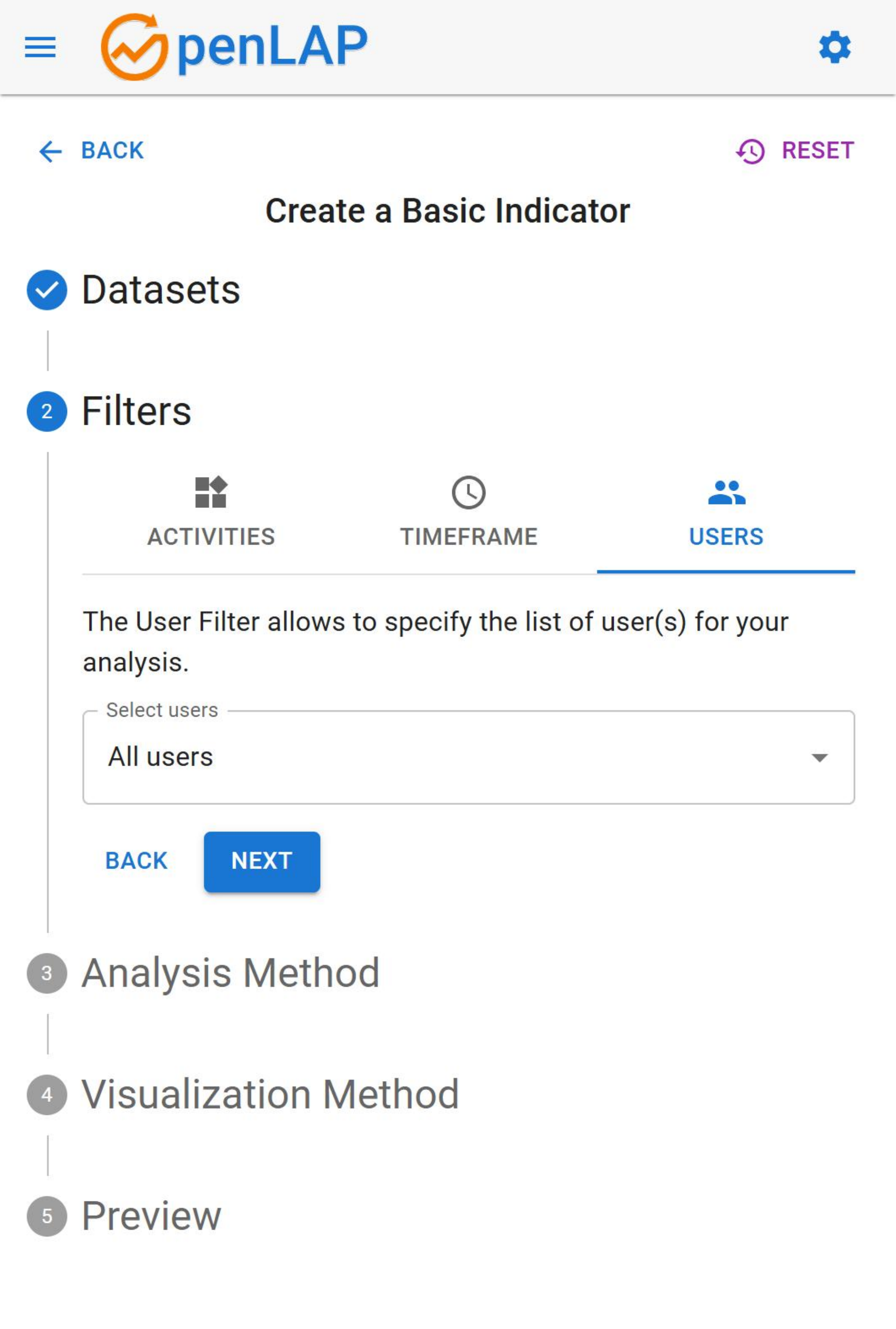}
            \caption{Select Filters: User}
            \label{subfig:select-filters-user}
        \end{subfigure}
        \begin{subfigure}[normla]{0.32\linewidth}
            \includegraphics[width=1.0\linewidth]{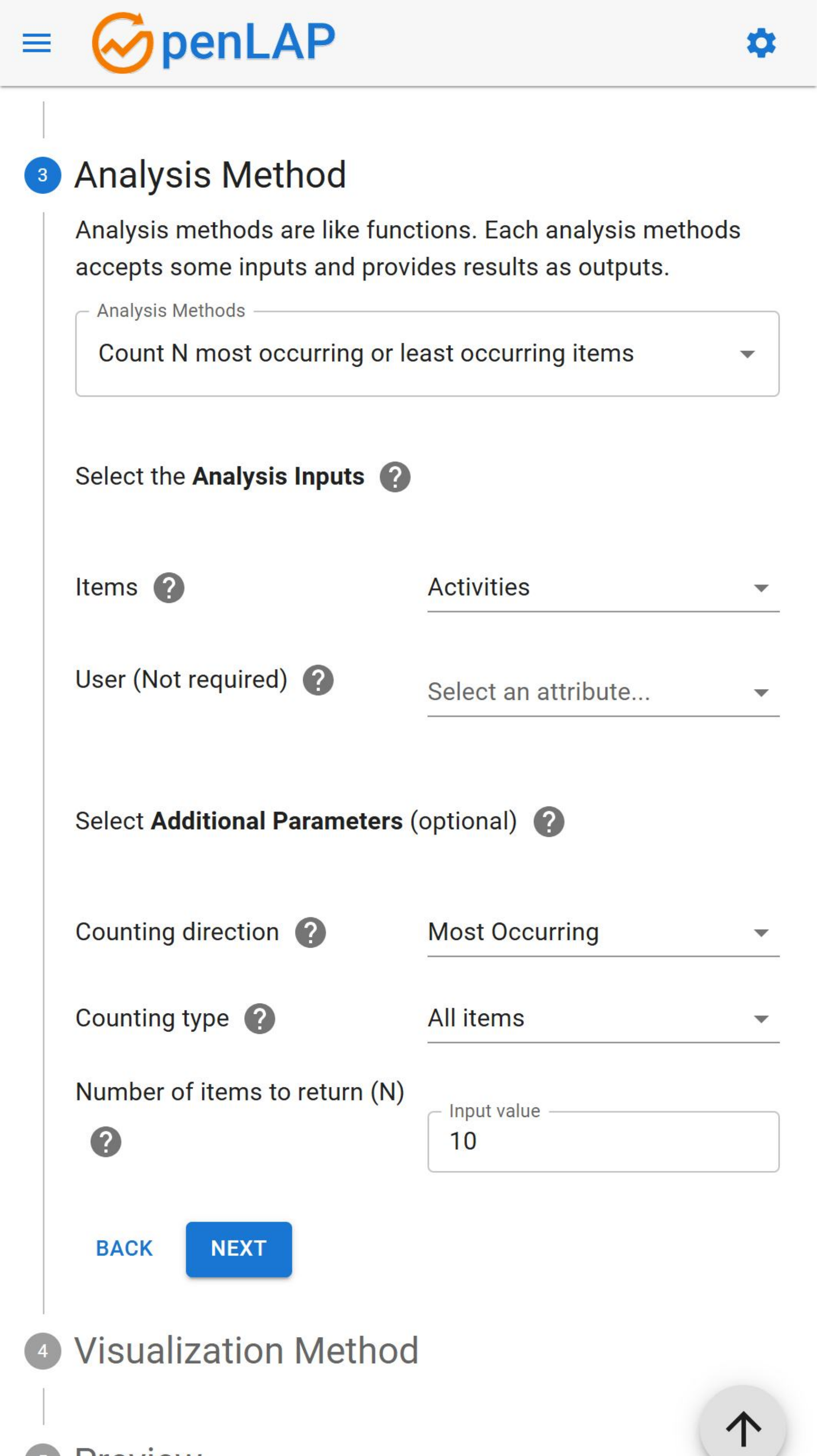}
            \caption{Select analysis method}
            \label{subfig:select-analysis}
        \end{subfigure}
        ~
        \begin{subfigure}[normla]{0.32\linewidth}
            \includegraphics[width=1.0\linewidth]{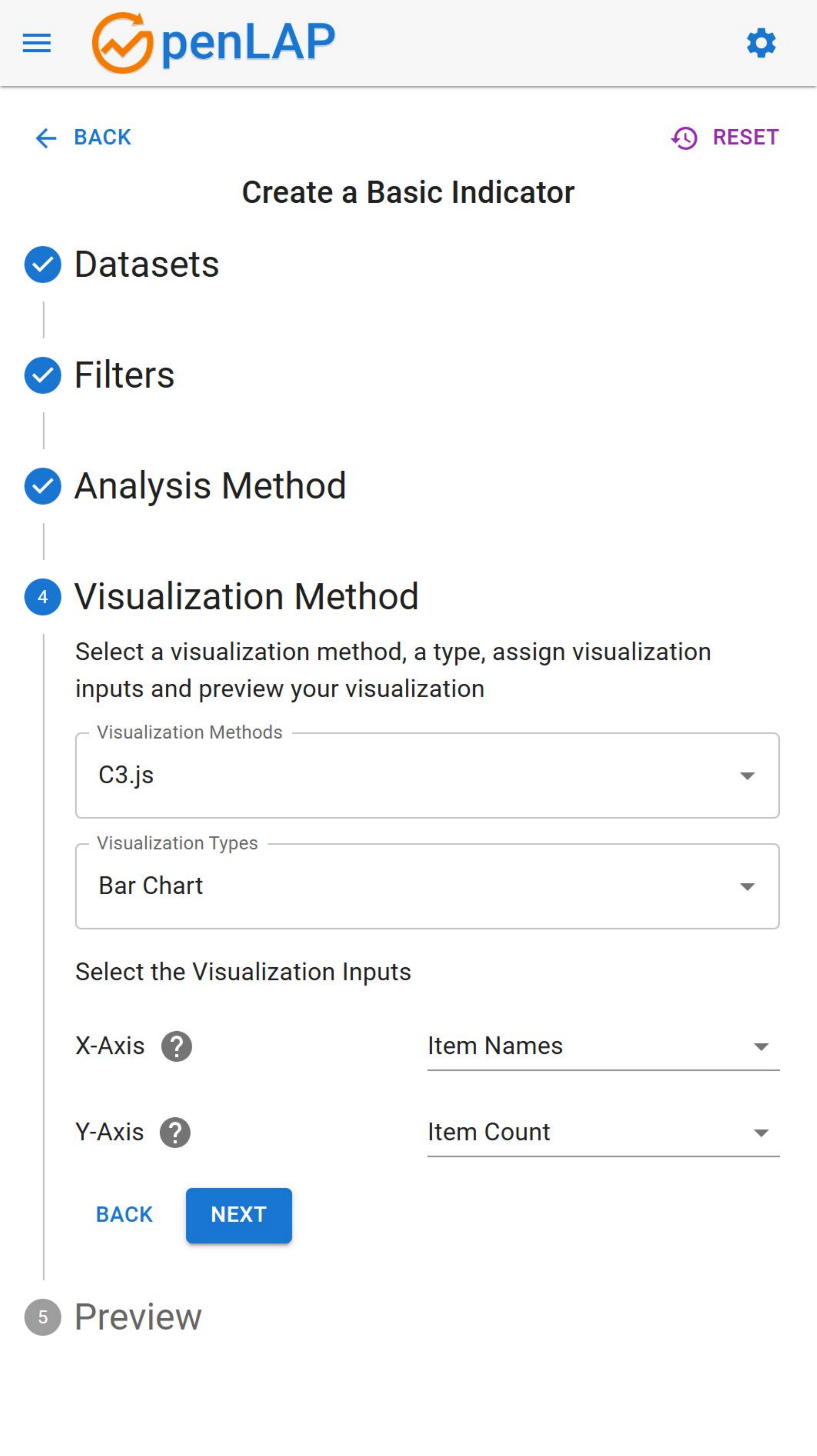}
            \caption{Select Visualization}
            \label{subfig:select-visualization}
        \end{subfigure}
    \end{center}
    \caption{Final Prototypes of the \textit{Indicator Editor}}
    \label{fig:final-prototypes-1}
\end{figure}

\begin{figure}[!ht]
    \begin{center}
        \begin{subfigure}[normla]{0.32\linewidth}
            \includegraphics[width=1.0\linewidth]{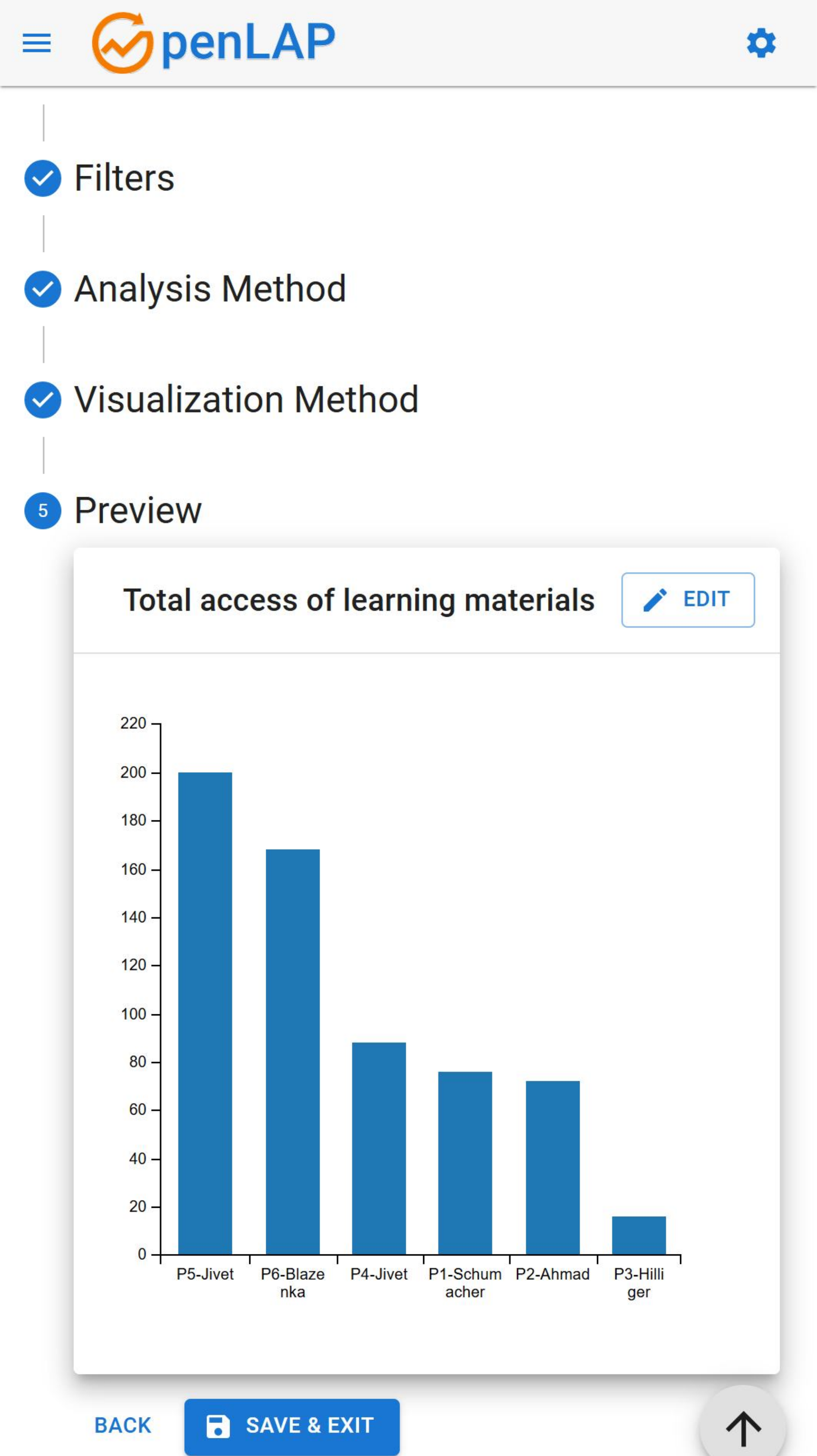}
            \caption{Preview indicator}
            \label{subfig:preview-indicator}
        \end{subfigure}
        ~
        \begin{subfigure}[normla]{0.32\linewidth}
            \includegraphics[width=1.0\linewidth]{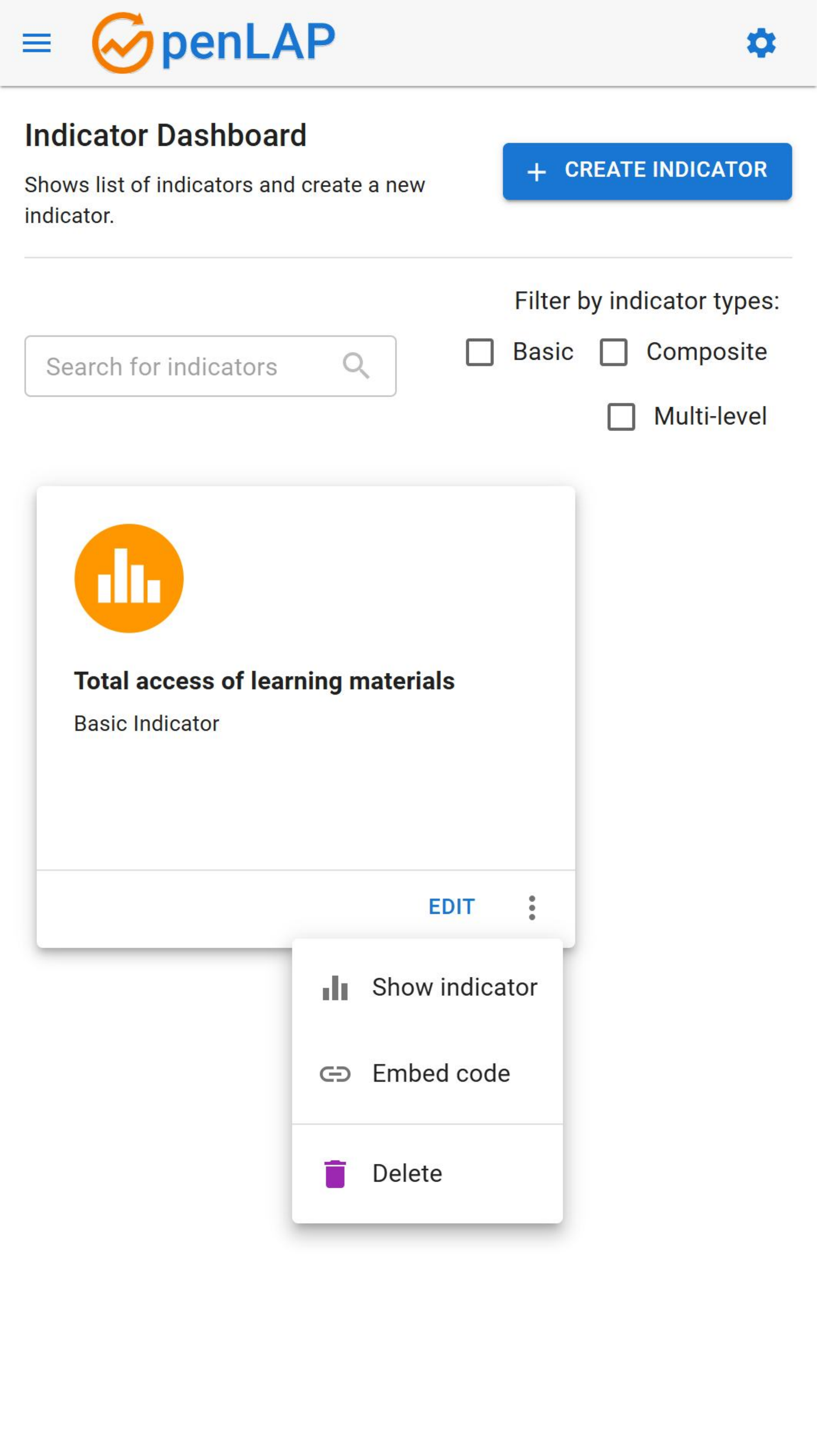}
            \caption{Indicator Dashboard}
            \label{subfig:dashboard}
        \end{subfigure}
    \end{center}
    \caption{Final Prototypes of the \textit{Indicator Editor}: Preview and Dashboard}
    \label{fig:final-prototypes-2}
\end{figure}

After incorporating feedback from previous iterations, we improved the high-fidelity prototypes and implemented the \textit{Indicator Editor} using ReactJS and Material Design. The final implementation details of the high-fidelity prototypes are described using the example from the user scenario, where a teacher wants to find out the most frequently accessed learning materials in their course provided in the CourseMapper. As shown in Figure \ref{subfig:dashboard}, the teacher can create a new indicator by clicking the `CREATE INDICATOR' button from the dashboard to direct them to the \textit{Indicator Editor} page. One of the critical issues from the previous iteration was the participants' dissatisfaction with the form-like structure of the \textit{Indicator Editor} UI. Therefore, as shown in Figure \ref{fig:final-prototypes-1} and Figure \ref{fig:final-prototypes-2}, we introduced a shopping cart checkout-like progression feature with accordion components, allowing a teacher to interact with options step-by-step while displaying the steps required to create an indicator \textbf{(DG1)}. Subsequent steps remain disabled until the previous ones are completed. Moreover, the teacher can move back and forth between completed steps, if needed \textbf{(DG1)}. The details of the four steps involved in creating an indicator are discussed in the following sections.


\subsubsection{Datasets:} 
The initial step involves selecting a suitable data source \textbf{(DG2)}. The teacher can choose from a list of platforms (e.g., Moodle platform (LMS), CourseMapper (MOOC)), activity types (e.g., Course, Material, Annotation, Reply, PDF), and actions performed on activities (e.g., accessed, viewed, edited, replied, enrolled, played). These multi-select dropdown menus appear one after another, as each option depends on the previous selections. As illustrated in Figure \ref{subfig:select-dataset}, the teacher selects `CourseMapper' as the data source, then `Materials' as the chosen activity type, and specified `accessed' as the action performed.
\subsubsection{Filters:} 
The teacher refines the selected dataset by applying various filters in the next step \textbf{(DG3)}. These filters were divided into three categories that are separated by tab component, namely: `Activities' (to specify the relevant activity data from the dataset) (Figure \ref{subfig:select-filters-activities}), `Timeframe' (to set the start and end dates) (Figure \ref{subfig:select-filters-time}), and `Users' (to choose either all available user data or specific user data) (Figure \ref{subfig:select-filters-user}). Clicking on these tabs will switch the UI and show the options for each category, respectively. In this case, the teacher selects the relevant materials from their course, set a timeframe of approximately six months, and included data from all enrolled users in the course.
\subsubsection{Analysis:} 
In this step, the teacher chooses an analysis method and then defines the mappings between the filtered dataset and the inputs required for the selected analysis method \textbf{(DG4)}. Examples of analysis methods included: `Count items based on a specified column,' `Calculate the average of items,' `Calculate the median of items,' and `Calculate the average and median of items over time.' The list of available analysis methods is displayed using a dropdown menu. After selecting an analysis method, the inputs of the analysis method and additional parameters will become available. The additional parameters are often pre-set with default values. However, the inputs for the analysis methods must be set manually. Some inputs can be optional and marked as `Not required'. In our example, the teacher applies the analysis method `Count N most occurring or least occurring items', as shown in Figure \ref{subfig:select-analysis}. This method requires inputs such as `Items' and `User'. The teacher selects `Activities' as the `Items' to count and decided to keep the default values for additional parameters, such as `Counting direction' to `Most occurring,' `Counting type' to `All items,' and the `Number of items to return (N)' to 10. 
\subsubsection{Visualization:}
In this step, the teacher chooses a visualization method and selects a visualization type \textbf{(DG5)}. The visualization method shows a list of visualization libraries, such as D3/C3.js, Apexcharts.js, Google Charts, etc., and the visualization types show a list of various chart types supported by the visualization method, such as bar chart, pie chart, etc. The teacher chooses a bar chart as depicted in Figure \ref{subfig:select-visualization}. Upon selecting a chart type, inputs for the visualization become available, which needs to be set manually. In our example, the teacher chooses the `Item Names' as the x-axis input of the visualization and `Item Count' as the y-axis input. Once selected, the teacher clicks the `NEXT' button to preview the visualization. Next, the teacher can finalize the indicator by assigning a name to the indicator, such as `Total access of learning materials,' as shown in Figure \ref{subfig:preview-indicator}. The last step is for the teacher to save the indicator and return to the dashboard. 
\begin{figure}[!ht]
    \begin{center}
        \begin{subfigure}[normla]{0.47\linewidth}
            \includegraphics[width=1\linewidth]{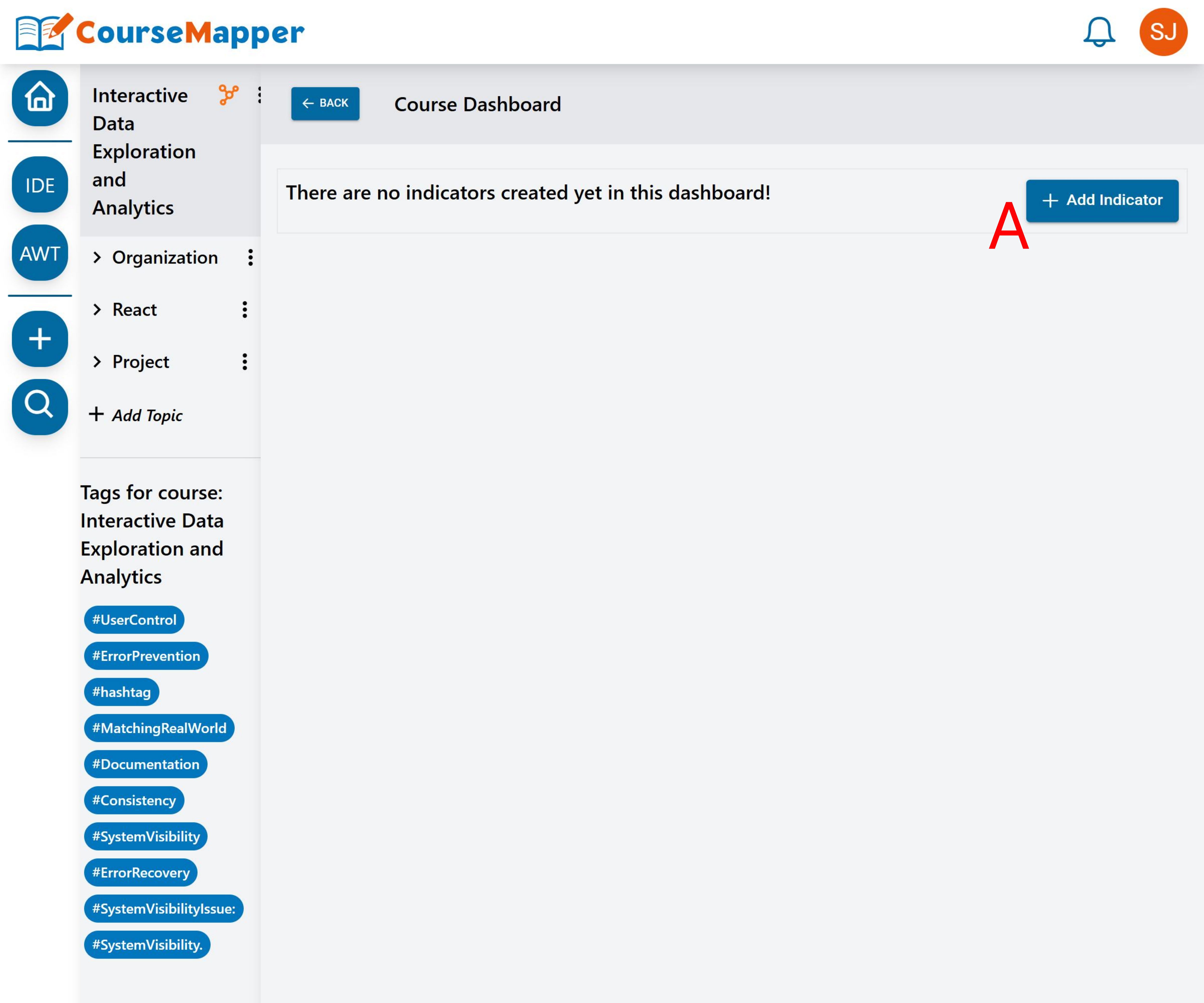}
            \caption{Click `Add indicator' button}
            \label{subfig:mooc-add-indicator}
        \end{subfigure}
        ~
        \begin{subfigure}[normla]{0.47\linewidth}
            \includegraphics[width=1\linewidth]{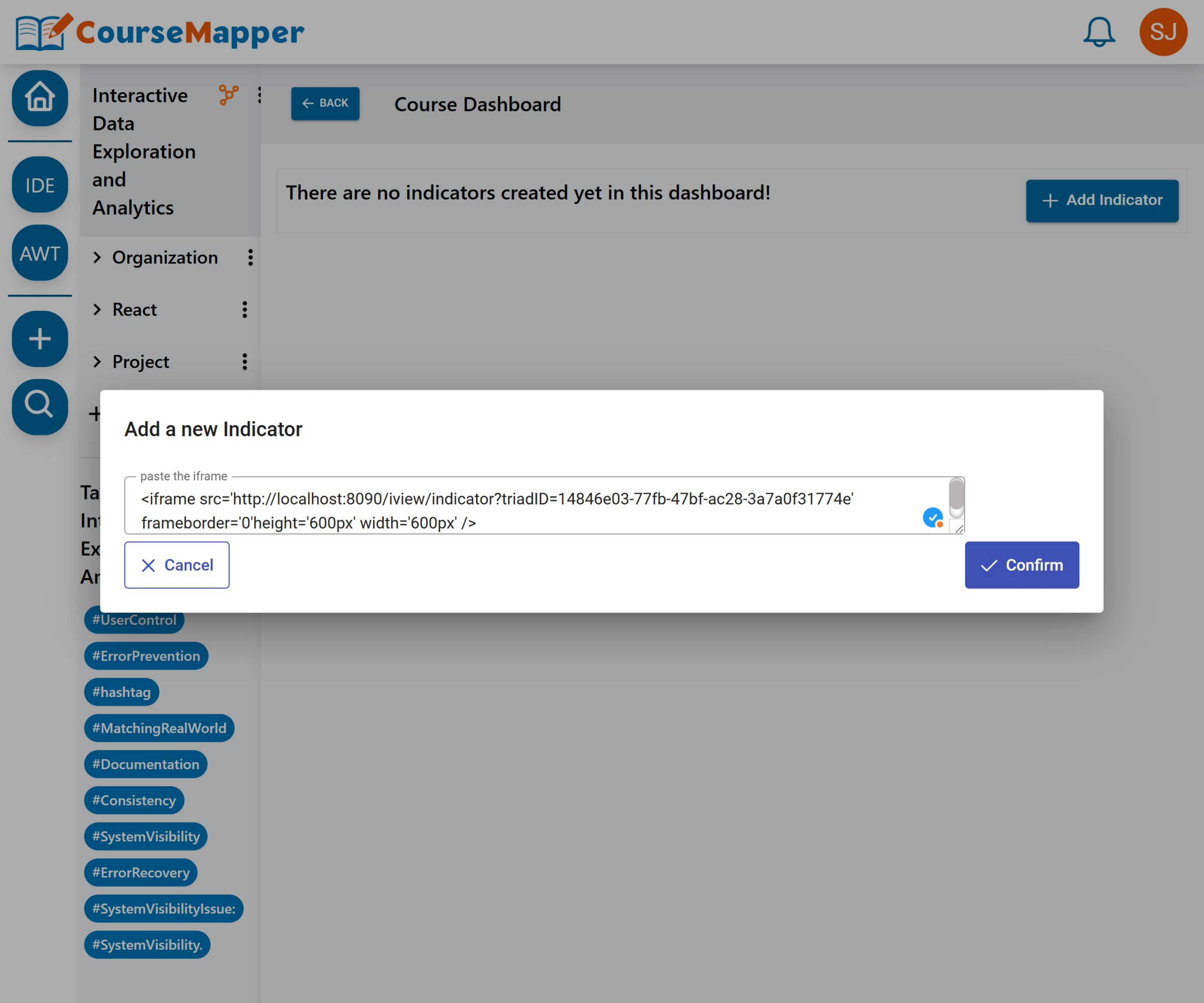}
            \caption{Insert the IIC code}
            \label{subfig:mooc-iic-code}
        \end{subfigure} \\
        ~
        \begin{subfigure}[normla]{0.47\linewidth}
            \includegraphics[width=1.0\linewidth]{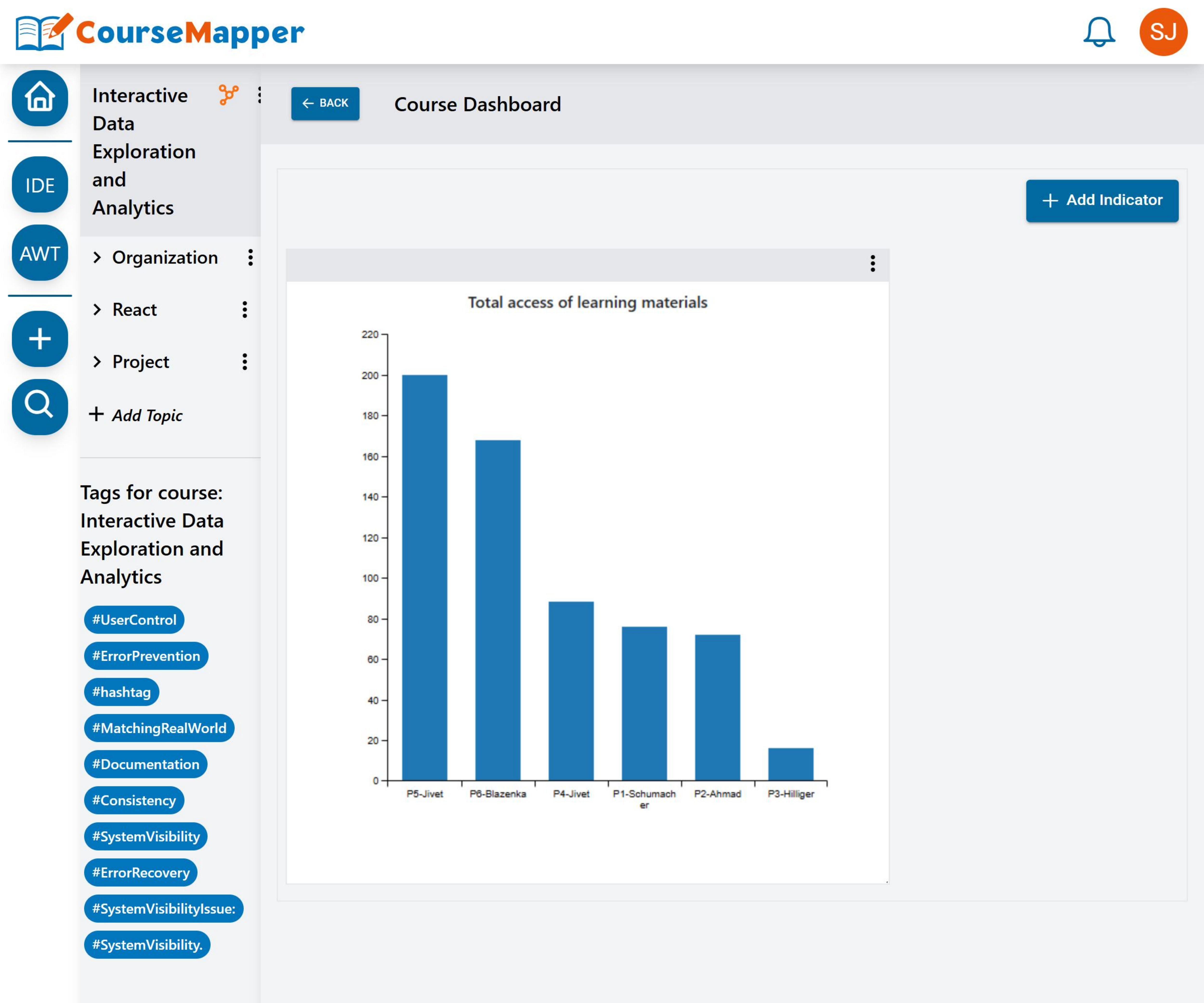}
            \caption{Preview the IIC}
            \label{subfig:mooc-preview-iic}
        \end{subfigure}
    \end{center}
    \caption{Interactive Indicator Code (IIC) in CourseMapper}
    \label{fig:mooc-share}
\end{figure}

\subsubsection{Interactive Indicator Code:}
As shown in Figure \ref{subfig:dashboard}, the teacher can modify the selections made during the creation of the indicator by clicking the `Edit' button \textbf{(DG1)}. This action prepopulates all the teacher's previous selections to be available during the editing process. The `Show Indicator' button lets the teacher preview the indicator without entering the editing mode. Additionally, the teacher can delete the created indicator by clicking the `Delete' button. The most important button is the `Embed code' button, which copies the Interactive Indicator Code (IIC). This code can be embedded in any website that supports iFrame code embedding. As shown in Figure \ref{subfig:mooc-add-indicator} marked with a red letter \textit{A}, the teacher can click the `Add Indicator' button in the CourseMapper to open a dialog box where the IIC code can be inserted, as shown in Figure \ref{subfig:mooc-iic-code}. After clicking the `Confirm' button, the teacher can preview the IIC in the course dashboard in the CourseMapper, as illustrated in Figure \ref{subfig:mooc-preview-iic}.
\section{Evaluation} \label{evaluation}
After systematically designing the \textit{Indicator Editor} and implementing it in OpenLAP, we conducted an online qualitative user study to explore the users' perceptions, expectations, and attitudes towards an indicator generation process in terms of control \& personalization, transparency \& trust, and satisfaction and acceptance. Since users' perspectives on self-service LA (SSLA) are inherently subjective and individual, a quantitative approach would be insufficient for this goal, as this approach aims at analyzing empirical data for predetermined hypotheses. Thus, we deem a qualitative approach more appropriate to explore individual expectations from an SSLA approach. We do not aim to generalize our conclusions, but to deeply understand the potential of human control to increase LA's transparency and inform the future design of SSLA systems.
\subsection{Participants}
We conducted a user study with n=15 participants (13 students and two teachers). The target group of our study was students and teachers with at least basic knowledge of data analytics and visualization. About 94\% and 87\% of the participants mentioned that they were aware of the concepts of data analytics and learning analytics, respectively, and all were familiar with visualizations. We recruited them via email, word-of-mouth, and social media to ensure a diverse sample across different nationalities, educational levels, and study backgrounds. Nine males and six females aged between 18 and 44 completed the study. Most of the students (n=11) were international students from nine different countries residing in Germany, and the rest of the participants were locals in Germany. All participants had sufficient English language skills, and the highest level of education reported by most participants was a bachelor's degree (60\%). About 53\% and 33\% of the participants had study backgrounds in Computer Science and Business and Economics, respectively.
\subsection{Study Design}
We conducted a qualitative user study to gather in-depth feedback on the usage and attitudes towards the \textit{Indicator Editor}. All participants gave informed consent to participate in the study. Participants were first presented with an online survey via Google Forms, where they answered a questionnaire about their demographics and familiarity with the concepts of data analytics, learning analytics, and visualization. Next, they were introduced to the goals and concepts used in the \textit{Indicator Editor}, with concrete examples provided for better understanding. Afterward, we conducted moderated think-aloud sessions where participants were asked to perform two tasks: (1) ``Imagine you are a student who wants to track your engagement on your study platform. Design an indicator that shows which learning material you have annotated the most. This indicator should help you understand your level of interaction with different resources and identify which materials you engage with most frequently''. (2) ``Imagine you are a teacher in a Learning Analytics course. Create an indicator that shows the total number of times students have accessed the learning materials. This indicator will help you track how often course resources are used, giving insight into student engagement with the content''. 
Following the think-aloud method, participants were encouraged to think aloud about anything that came to mind during each interaction. Afterward, we conducted semi-structured interviews to gather in-depth feedback. These interviews were conducted online, lasted approximately 30 to 45 minutes, and were recorded with the participants' consent. During the interviews, participants were asked the following open-ended questions: 
\textbf{(1)} \textit{What do you like the most about the current state of the Indicator Editor?} \textbf{(2)} \textit{What do you like the least about the current state of the Indicator Editor?} \textbf{(3)} \textit{Do you have a sense of control when you interact with the Indicator Editor? How?} \textbf{(4)} \textit{Does the interaction/controllability of the Indicator Editor influence the transparency of the application? How?} \textbf{(4.1)} \textit{Which parts or features of the Indicator Editor give you a sense of transparency of the Indicator Editor? Why?} \textbf{(5)} \textit{Does the interaction/controllability of the Indicator Editor influence your trust in the application? How?} \textbf{(5.1)} \textit{Which parts or features of the Indicator Editor give you a sense of trust in the Indicator Editor? Why?} \textbf{(6)} \textit{Does the interaction/controllability of the Indicator Editor influence your satisfaction with the application? How?} \textbf{(7)} \textit{Do you have any suggestions to improve the system?}.

After the semi-structured interviews, participants were also asked to fill out a questionnaire containing questions based on seven constructs, namely 
\textbf{(1)} \textit{Perceived Usefulness} \citep{davis1989perceived}, 
\textbf{(2)} \textit{Perceived Ease of Use} \citep{davis1989perceived, pu2011user}, 
\textbf{(3)} \textit{Intention To Use} \citep{venkatesh2003user} to evaluate the user acceptance of the \textit{Indicator Editor} based on the Technology Acceptance Model (TAM) \citep{davis1989perceived, venkatesh2003user}, 
\textbf{(4)} \textit{Control \& Personalization} \citep{pu2011user}, 
\textbf{(5)} \textit{Transparency} \citep{pu2011user, hellmann2022development}, 
\textbf{(6)} \textit{Trust} \citep{gefen2003trust, carter2005utilization}, and \textbf{(7)} 
\textit{Satisfaction} \citep{pu2011user}, as shown in Table \ref{fig:constructs}. 
For each construct, answers were given on a 5-point Likert scale, ranging from 1 (``strongly disagree'') to 5 (``strongly agree''). To note that by using these constructs, we are not aiming at conducting a quantitative evaluation and generalizing our conclusions, but rather to use participants' answers to the questionnaire as a starting point to collect their opinions towards the \textit{Indicator Editor}, which are then explored in-depth through our qualitative study.
\subsection{Analysis and Results}
We qualitatively analyzed the moderated think-aloud sessions and the semi-structured interviews. Notes and transcripts of the interview recordings were made for the analysis. 
We followed the thematic analysis instruction proposed by \citet{braun2006using}. 
The analysis involved the two authors of this paper.
Firstly, we familiarized ourselves with the depth and breadth of the qualitative data. Next, we collected all the answers from the interview transcripts in a spreadsheet and coded them systematically to identify common patterns. 

\begin{figure}[!ht]
    \centering
    \begin{minipage} {0.51\linewidth}
        \begin{subfigure}[t]{1.0\linewidth}
            \includegraphics[width=1.0\linewidth]{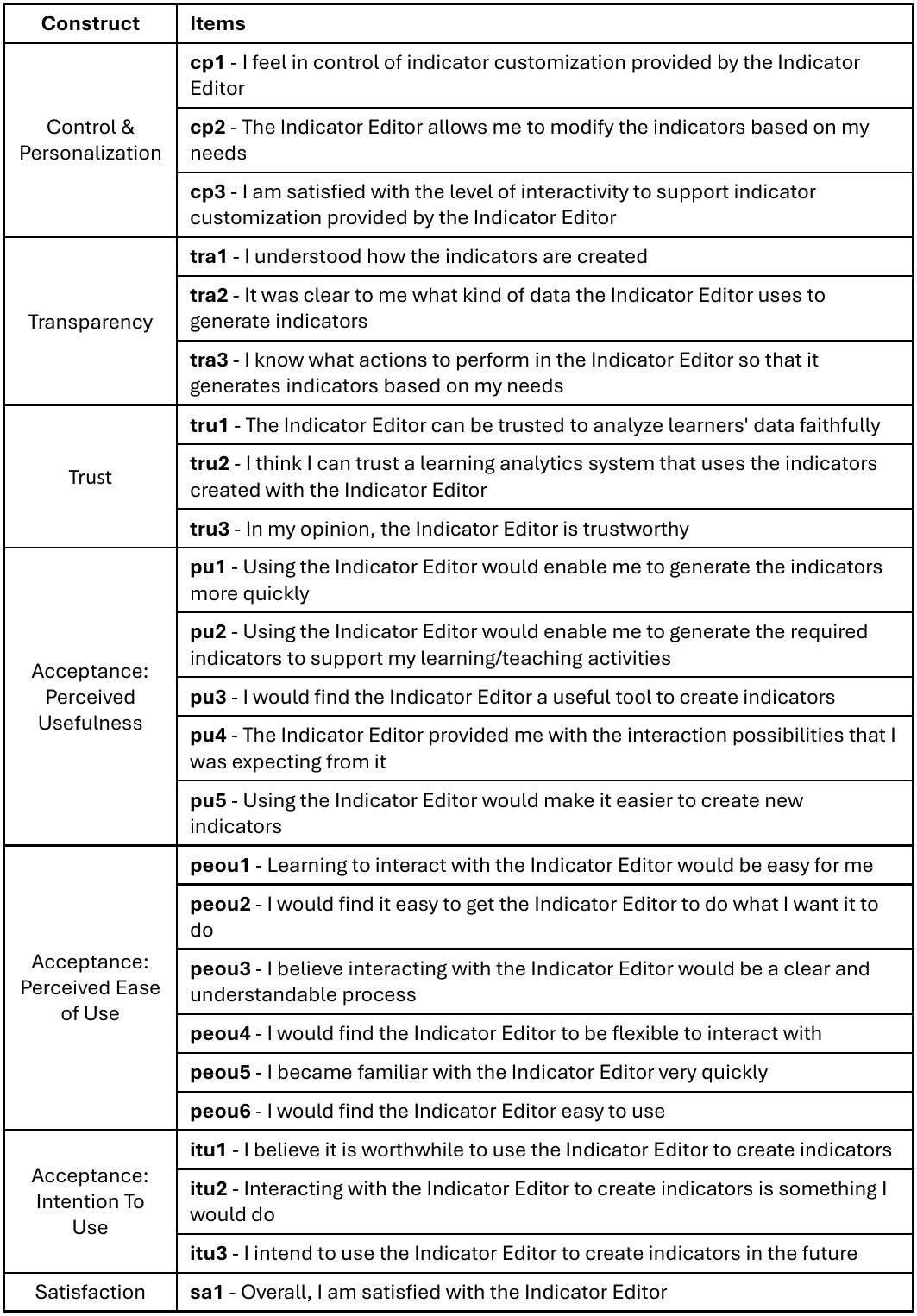}
        	\caption{Constructs and items}
        	\label{fig:constructs}
        \end{subfigure}
    \end{minipage}
    \begin{minipage} {0.45\linewidth}
        \begin{subfigure}[t]{1.0\linewidth}
            \includegraphics[width=1.0\linewidth]{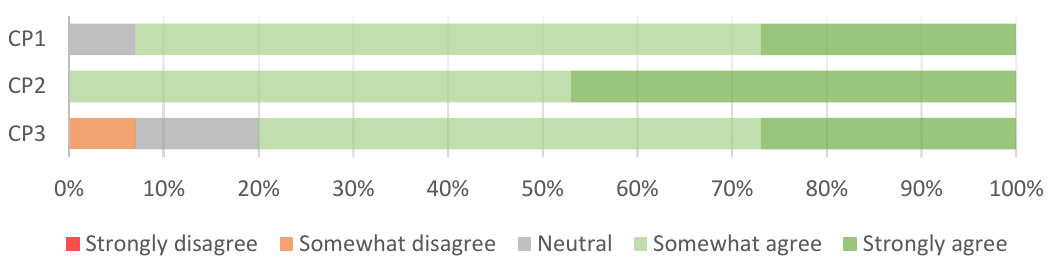}
            \caption{Control and Personalization}
            \label{subfig:control-personalization}
        \end{subfigure}
        
        \begin{subfigure}[b]{1.0\linewidth}
            \includegraphics[width=1.0\linewidth]{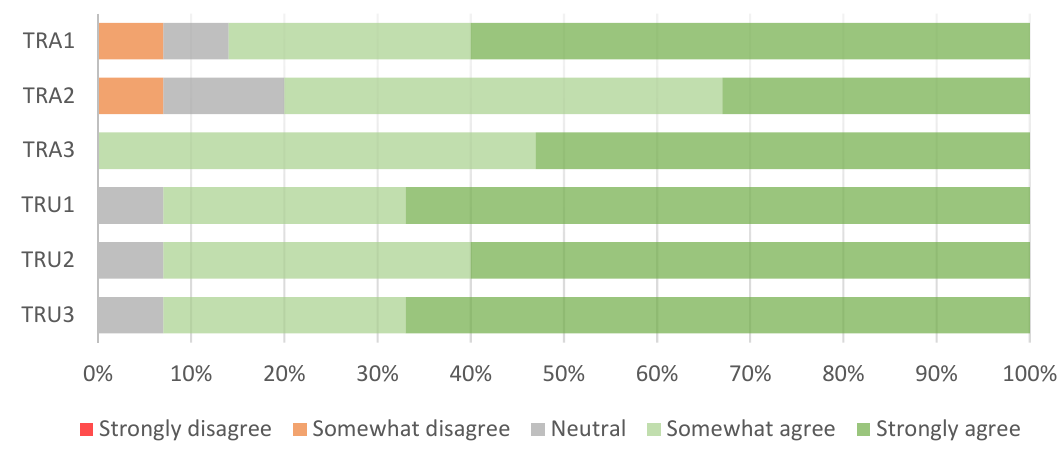}
            \caption{Transparency and Trust}
            \label{subfig:transparency-trust}
        \end{subfigure}
        \begin{subfigure}[b]{1.0\linewidth}
            \includegraphics[width=1.0\linewidth]{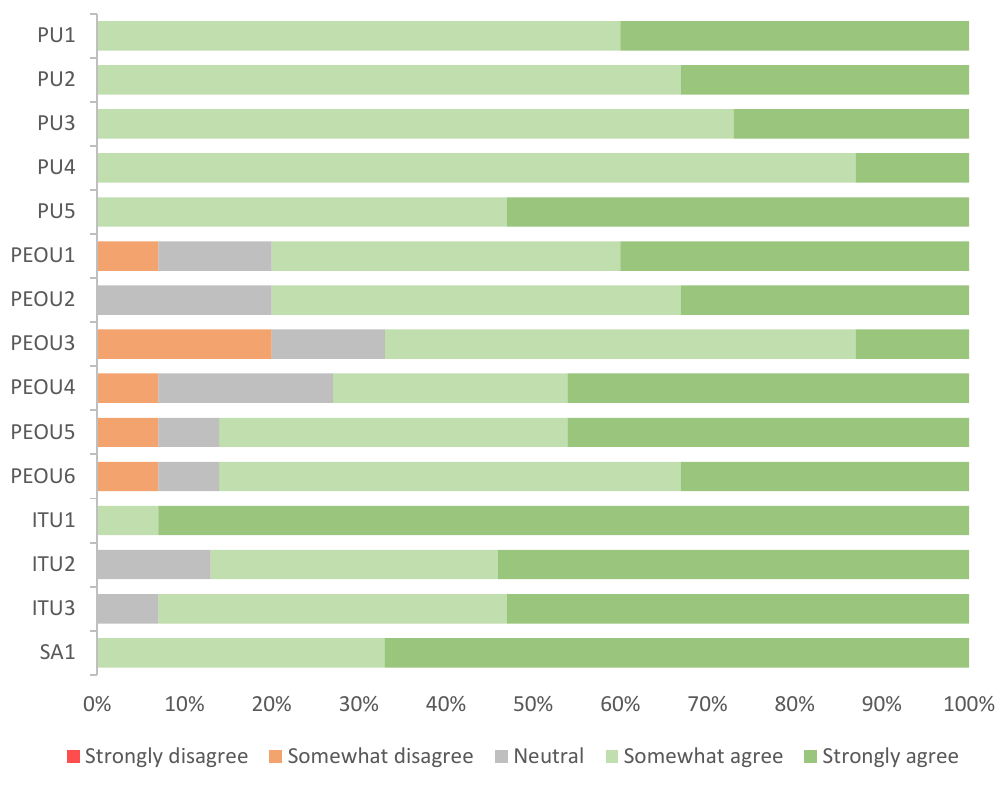}
            \caption{Satisfaction and Acceptance}
            \label{subfig:satisfaction-acceptance}
        \end{subfigure}
    \end{minipage}
    \caption{Constructs and evaluation results}
    \label{fig:evaluation-results}
\end{figure}

The coding scheme was developed based on our research question, employing predetermined themes established
prior to data analysis, namely \textit{Control \& Personalization}, \textit{Transparency \& Trust}, and \textit{Satisfaction \& Acceptance}.  
The decision on whether to include/exclude
dubious cases was solved through a joint discussion between the two authors.
The analysis was rather deductive as we aimed to find additional explanations to address our research question.
In contrast to inductive (i.e., bottom-up) thematic analysis, which is the data-driven process of coding the data without trying to fit it into a pre-existing coding frame, deductive (i.e., top-down) thematic analysis is an analyst-driven process of coding for a specific research question. This form of thematic analysis tends to provide a detailed description of the data overall and a more detailed analysis of some aspects of the data \citep{braun2006using}. Following a deductive thematic analysis approach, we present the results of the evaluation organized by the three themes, as shown in Figure \ref{fig:evaluation-results}.

\subsubsection{Control \& Personalization}
User control is defined as the degree to which users feel in control in their interaction with the system \citep{pu2011user}. Personalization refers to the capability to customize a system according to the user's needs and preferences. 
The participants indicated a positive sense of control when interacting with the \textit{Indicator Editor}, as shown in Figure \ref{subfig:control-personalization}. 
Specifically, the \textit{Indicator Editor} allowed them to personalize the indicators according to their needs. For example, one participant commented: \textit{\textbf{P10 (teacher)}: ``I was able to change it [the indicator], edit to customize it. There was a lot of customization, so I felt fully in control of the system''.} Similarly, another participant stated: \textit{\textbf{P3}: ``I felt like I was in control of the application. I was directing [the \textit{Indicator Editor}] on what kind of indicator I needed, whether it was a pie chart or a bar chart, and what I wanted on the x-axis or y-axis''.} 
Participants also appreciated the system's clear flow, noting that they could easily navigate between steps and found the hints and feedback helpful for correcting mistakes. They liked that the tool allowed them to customize existing indicators or create new ones based on their needs, and they valued the different selection options provided by the \textit{Indicator Editor}. However, some participants mentioned that certain selection options were difficult to understand: \textit{\textbf{P12}: ``I knew my end goal, but the labels were extremely confusing, particularly in selecting the analysis method. This made me feel like I wasn't fully in control''.}
Another participant expressed a similar concern: \textit{\textbf{P5}: ``I think the labels and tips are enough to guide me [in creating indicators], but in some places, I felt stuck. The analysis method was confusing, and I couldn't grasp the idea or the parameters needed to complete my task''.}
Additionally, a few students (n=3) found the \textit{Indicator Editor} overwhelming: \textit{\textbf{P13}: ``The application takes some time to get used to because there are many editing options''.} 
Another participant added: \textit{\textbf{P11}: ``Choosing from the selection options was quite difficult''.} Lastly, a participant noted: \textit{\textbf{P9}: ``There are too many controls required to create an indicator''.}
\subsubsection{Transparency \& Trust}
The term transparency is defined as the degree to which a system allows users to understand its inner logic, in our context, how the indicators are created \citep{pu2011user}. 
Participants agreed that the \textit{Indicator Editor} provided valuable insights into the design decisions behind the creation of indicators, thereby enhancing the system's transparency. For instance, \textbf{P14} noted, \textit{``I think interacting with the charts and getting insights into them gave me a feeling that the system is trying to be transparent''.} 
Most participants (n=11), as depicted in Figure \ref{subfig:transparency-trust}, felt that the ability to customize and select various options in the \textit{Indicator Editor}, such as choosing the data source and defining actions for creating indicators, was sufficient to ensure system transparency. Participants expressed a positive view regarding the influence of interaction and control on perceived transparency. 
As one participant explained, \textit{\textbf{P10 (teacher)}: ``I can see what kind of data I am selecting and what kind of visualization I am applying. I know what I have in front of me is what I created myself''.} 
Similarly, another participant commented, \textit{\textbf{P12}: ``Especially in my interaction with the visualization, I could see how it was made''.} \textbf{P11} stated, \textit{``I expected how the visualization should look because I chose everything''.}
However, a few participants mentioned being confused about selecting the appropriate analysis method and its inputs to generate the indicator, as the labels were sometimes perceived as too technical or unclear: \textit{\textbf{P12}: ``If I had understood what inputs I was selecting, the system would have felt completely transparent''.} 
One participant suggested that the application could increase transparency by allowing more interaction with the visualizations, such as through zoom or filter options. 
Additionally, some participants requested more details about the data used to generate the indicators, specifically how the data was collected, to make the indicator creation process more transparent.

Trust indicates whether the user finds the whole system trustworthy \citep{pu2011user, gefen2003trust}. 
The results were also mostly positive in terms of users' perception of trust, as shown in Figure \ref{subfig:transparency-trust}.
The feedback provided by the \textit{Indicator Editor} and its stability positively influenced participants' trust in the system. 
\textbf{P3} noted, \textit{``The application didn't hang or freeze, which was good because it retained all the inputs I had entered. It gave me appropriate feedback when needed''.} 
Furthermore, participants highlighted that their sense of control contributed to building trust in the system. As \textbf{P4} mentioned, \textit{``I started to trust the application because it gave me the flexibility to move between steps and allowed me to keep everything under control''.} 
Participants also expressed confidence in the accuracy of their indicators since they were based entirely on their selected inputs: \textit{\textbf{P11}: ``I think I can trust it because I chose all these inputs and analysis methods. I know the output is correct, with no external influence''.}
Although participants indicated that they used trustworthy data sources, such as Moodle, some expressed concerns about the lack of transparency regarding data collection from such sources. \textbf{P12} stated, \textit{``I can configure everything correctly and see the results, and the application performs as expected. However, trust is ultimately based on the dataset and the credibility of the sources. The configuration of the indicator or its purpose is secondary''.} 
Additionally, a few participants raised privacy concerns that affected their trust in the system. \textbf{P14} noted, \textit{``I'm unsure about the platform because indicators are shareable with others. I can't verify whether the platform shares user data with other applications, which could lead to privacy issues''.}

\subsubsection{Satisfaction \& Acceptance}
Evaluating overall satisfaction determines users' thoughts and feelings using the \textit{Indicator Editor} \citep{pu2011user}. 
As shown in Figure \ref{subfig:satisfaction-acceptance}, participants were generally satisfied with the indicators because they could select data, charts, and their options effectively. 
For instance, \textit{\textbf{P12}: ``It does what it's expected to do because it helps me complete the task or achieve what I want''.}
Additionally, participants noted that their satisfaction was significantly influenced by the interaction and user control mechanisms provided by the system. Specifically, they appreciated the flexible selections and controls, the ability to customize and reuse indicators, and the overall system's flow for generating indicators according to their preferences. For example, \textbf{P2} stated: \textit{``This application provides the interaction and customization I need, making me feel in control of everything''.}

To evaluate the user acceptance of the \textit{Indicator Editor}, we used the Technology Acceptance Model (TAM). The two most influential factors that describe users' intention to use a system are perceived usefulness (PU) and perceived ease of use (PEOU) \citep{davis1989perceived}. 
Perceived usefulness is defined as the degree to which people believe using a particular system would enhance their job performance \citep{davis1989perceived}. 
As shown in Figure \ref{subfig:satisfaction-acceptance}, all the participants agreed that they could generate indicators quickly and that they could support their learning/teaching activities. 
All participants agreed that the system offered the expected interaction capabilities. They felt comfortable with the \textit{Indicator Editor} and considered it a helpful tool.
For instance, \textbf{P12}: \textit{``It does what it's expected to do because it helps me complete the task or achieve what I want''}
and \textit{\textbf{P13}: ``I like that I can edit existing indicators with minimal effort, such as converting a bar chart to a pie chart. I believe this entire indicator generation process will be highly beneficial for both students and teachers''.}
Participants recognized the benefits of the \textit{Indicator Editor} for using data from various sources, analyzing it, and embedding indicators into any web application via shareable code. They agreed that the indicator development approach is convenient and the step-by-step process is logical and intuitive. 
However, \textbf{P13} noted that using a small dataset that did not include their data might hinder understanding the application's full potential for supporting learning and teaching activities.
To improve the indicator generation process, participants suggested adding features for searching and recommending indicators, filtering based on selected goals or questions, finding similar questions, supporting multiple languages, providing chatbots for assistance, and offering additional documentation and video tutorials.

The perceived ease of use refers to the degree to which a person believes that using a particular system would be effortless \citep{davis1989perceived}. 
As shown in Figure \ref{subfig:satisfaction-acceptance}, the participants generally found the \textit{Indicator Editor} intuitive. \textit{\textbf{P13}: ``I like that I can edit existing indicators with minimal effort, such as converting a bar chart to a pie chart. I believe this entire indicator generation process will be highly beneficial for both students and teachers''.} However, they initially faced difficulty understanding the system, as it appeared technical. They often relied on feedback and hints from the system to avoid making mistakes in their selections. They found the system easier to use once they became familiar with how the selection process worked. 
For example, \textit{\textbf{P6}: ``Everything is self-explanatory. It became easier to relate things in the second trial. The first time, I needed some effort. It's easy for me to go through everything now. The system mostly helps me with solving issues. Everything became easier to grasp once I understood how the options work''.}
\textit{\textbf{P13}: ``Although the interaction with the system is not overly complicated, it takes some time to get used to it. I think it is designed for users with more experience in data analysis''.}
However, there was less consensus on the system's overall ease of use. Many participants found the analysis method step in the indicator development process challenging. Specifically, they struggled with understanding the input selections and parameters and the technical terms used in the labels, such as ``group by columns'' and how the analyzed dataset looks like, \textit{\textbf{P8 (teacher)}: ``I didn't like the generic technical naming of the analysis methods because I couldn't understand them. The purpose of these methods was unclear to me. Moreover, I wish to see a preview of the analyzed dataset''.}
The intention to use is the degree to which a person has the behavioral intention to adopt the technology 
\citep{venkatesh2003user}. 
In general, the participants found the idea of the \textit{Indicator Editor} to be very beneficial, as shown in Figure \ref{subfig:satisfaction-acceptance}. 
\section{Discussion}    \label{discussion}
The primary research question we address in this work is \textbf{RQ}: \textit{What are the effects of a self-service LA approach that empowers users to control the LA indicator implementation process on the user's perceptions of (1) control \& personalization, (2) transparency \& trust, and (3) satisfaction \& acceptance?}

We aimed to provide an interactive \textit{Indicator Editor} to help users control and personalize the indicator implementation process based on their needs and preferences. Most participants could control and personalize their indicators by interacting with the \textit{Indicator Editor}. However, although all participants in our study had experience with data analysis and visualization, some of them found the controls in the \textit{Indicator Editor} overwhelming. This indicates that while allowing users to control and personalize the implementation of their indicators is beneficial in any LA system, it is essential to provide different levels of controllability to meet the needs of stakeholders with diverse experiences. This confirms the findings in other domains, such as interactive recommendations, showing that providing additional controls in interactive systems also increases cognitive load and that different users have different needs for control 
\citep{jin2018effects, andjelkovic2016mood}, and that the user preference for interaction methods depends on several personal characteristics such as domain knowledge \citep{knijnenburg2011each}.
Thus, recognizing that the effectiveness of interactive LA interfaces is likely to be affected by individual differences, it is essential to go beyond a one-size-fits-all approach with the same controls for all users and develop interactive LA interfaces that offer user controls that are adapted to the needs of different LA stakeholders to ensure acceptable cognitive load. 
Related to the \textit{Indicator Editor}, it is necessary to provide different versions that offer varying levels of controls. In addition to the current version of the \textit{Indicator Editor}, which is more suitable for advanced users with LA domain knowledge and experience in data analysis and visualization, there is also a need to offer a simpler version with fewer controls to support non-expert LA stakeholders in designing custom LA indicators. A promising approach is to use card-based approaches in the early stages of the design process to co-design low-fidelity LA indicator prototypes and then proceed to a more detailed design of high-fidelity LA indicators using the \textit{Indicator Editor}. For example, \citet{alvarez2020deck} proposed a Learning Analytics Design deck of cards (LA-DECK) to support stakeholders in co-designing LA systems and provide a common basis for understanding and communication in an LA team. More recently, \citet{joarder2025isccreatorhumancentereddesign} proposed the Indicator Specification Cards (ISC) Creator as an HCLA tool that allows the low-cost design of low-fidelity LA indicator prototypes systematically. Further research is required to (1) investigate the level of control needed for different stakeholders depending on their context, needs, goals, experience, and motivation, and (2) examine the benefits of providing an \textit{Indicator Editor} with different levels of controllability (i.e., basic and advanced modes) on the interaction with and perception of the \textit{Indicator Editor} in terms of transparency, trust, user satisfaction, and acceptance.

Most participants in our study showed a favorable opinion towards the transparency of the \textit{Indicator Editor} and liked that the system helped them to gain insights into the indicator implementation process. 
The majority of participants agreed that the possibility to control the \textit{Indicator Editor} and steer the indicator implementation process positively impacted their perceived system transparency. This indicates that human control of the indicator implementation process can contribute to increased transparency of LA systems. 
This is consistent with findings by \citet{alfredo2024human} who noted that most of the surveyed transparency strategies in HCLA systems
are built around human control \citep[e.g.,][]{shute2021design,ahn2021co,dollinger2019working}.
This is also in line with studies from different other application domains, including 
human-centered AI 
\citep{shneiderman2022human,shneiderman2020bridging},
interactive machine learning 
\citep{amershi2014power, dudley2018review, jiang2019recent},
recommender systems 
\citep{tsai2021effects,tintarev2015explaining,pu2011user}, and visual analytics \citep{spinner2019explainer}
which also showed that human control and transparency are interdependent. 
However, in our study, some participants (students and teachers), had difficulty understanding the steps required to develop indicators and found the overall process rather technical and unclear. These users perceived the \textit{Indicator Editor} as less transparent, indicating a potential relationship between perceived transparency on the one hand and users' background knowledge and experience on the other hand. 
Our results also indicated that some users felt that the amount of information provided in the \textit{Indicator Editor}, such as collected student activity data, guidance on selecting data, analysis methods, and visualizations, was sufficient for making the system perceived as transparent. This indicates that user-perceived transparency can be high, even though the system's objective transparency is partial, as the \textit{Indicator Editor} does not release all information about the data it had at its disposal (e.g., how student activity data was collected, by whom, and who else has access to this data).
On the other hand, some students suggested that including additional details, such as information on the data collection, would enhance their perceived system's transparency. This feedback indicates that perceptions of transparency in an LA system can vary depending on users' perspectives, highlighting the need for tailored information to meet diverse user transparency needs.
Overall, our study highlights that, for assessing transparency in LA systems, it is necessary to view transparency as a multi-dimensional concept that includes both data transparency (what data does the system use? how is the data collected?) and process transparency (how is the indicator generated?). Therefore, future research efforts should compare the effect of the scope of transparency, between a process-centric transparent LA system that enables users to have control over the indicator implementation process (i.e., process transparency) \citep{pozdniakov2022question,duan2024towards,conijn2023effects,khosravi2022explainable}  and a data-centric transparent LA system that provides transparency regarding the data it had at its disposal (i.e., data transparency) \citep{ahn2021co,drachsler2016privacy,tsai2017learning}. 
Moreover, it is important to view transparency as a user-centric quality and consequently focus on the assessment of subjective perceived transparency, rather than on measuring the system's objective transparency \citep{hellmann2022development}.

Most participants found the \textit{Indicator Editor} trustworthy. Some participants mentioned that the stability, reliability, and feedback provided by the \textit{Indicator Editor} positively impact their trust in the system. This suggests that usability aspects might influence trust in LA. Moreover, the majority of participants agreed that their feeling of control over the indicator implementation process gave them a sense of trust in the system. This indicates that human control can improve the trustworthiness of the LA implementation process, which aligns with discussions on the trustworthiness of LA, stating that incorporating human control is essential to make LA more trustworthy 
\citep{usmani2023human, hakami2020learning, slade2019learning, shibani2019contextualizable, drachsler2016privacy}
and that the lack of involvement of students and teachers in the design and development of LA systems can potentially lead to a lack of trust in the system \citep{alzahrani2023teaching,sarmiento2022participatory,shibani2022questioning,dollinger2019working}.  
While our study mainly showed a positive relationship between human control and trust, \citet{alfredo2024human} reported mixed findings regarding stakeholders' perceived trust when evaluating trust of HCLA systems. In particular, the authors pointed out that giving end-users a mechanism to control an HCLA system does not necessarily foster trust. For example, \citet{ooge2023steering} argued that plainly giving learners a mechanism to control a recommender system does not necessarily increase their trust in the system.
Offering human control of the HCLA process can be influential to trust, but also introduces the risk of overwhelming users and increasing cognitive load. Thus, it is essential to vary the level of human control in an HCLA system depending on who the end-user is. 
These results provide important insights into further investigating the impact of personal characteristics (e.g., domain knowledge, technical expertise, cognitive capabilities) on the stakeholders’ perception of trust in HCLA
systems. Other HCLA studies further highlighted that, in addition to human control, trust can also depend on the system's ability to provide accurate and useful information. For example, \citet{ma2022glancee} noted that teachers were
inclined to trust the system when the displayed data matched their expectations. In their study on the importance of
trustworthiness and transparency in a tool that provides guidance and recommendations for career paths, \citet{gedrimiene2023transparency} showed that accuracy had a
significant impact on the users' perceived trust in the tool.
Providing explanations is another possible factor that could foster stakeholders’ trust in the system.  
For example, \citet{khosravi2022explainable} reported that educating students about how their knowledge status is computed by AI and opening up the black-box AI models could ultimately increase their trust in the system. Similarly, \citet{wilson2021automated} noted that balancing
the presented information’s accuracy with explainable information could foster learners’ trust in the system.   

Furthermore, participants in our study raised concerns related to privacy and its impact on trust. This confirms that, besides human control, awareness of how the data is collected, and privacy are crucial factors in users' sense of trust in LA systems, as stressed in different LA ethical studies and conceptual frameworks 
\citep{drachsler2016privacy, tsai2017learning, khalil2023fairness, tsai2020privacy, hoel2017influence, pardo2014ethical,prinsloo2015student}.
Overall, the different perspectives on trust in the HCLA literature confirm that trust is a multifaceted concept, as also highlighted in previous studies on explainable AI (XAI) and recommender systems \citep[e.g.,][]{ kaur2022trustworthy, miller2022we,siepmann2023trust}. 
Further investigation is required to explore and compare the effects of usability, human control, perceived accuracy and usefulness, explanation, and privacy on users’ perceptions of trust in HCLA systems.

The broad agreement among participants that their interaction with and control of the \textit{Indicator Editor} positively impacted their satisfaction with and acceptance of the LA system shows a potential positive correlation between user control and satisfaction/user experience. This confirms findings from studies on HCLA, indicating that learners need to be engaged in the decision-making if such systems are to be accepted and adopted 
\citep{buckingham2019human, ahn2019designing, dollinger2019working, west2020academics}.
This is also in line with several studies from the recommender systems and the human-AI interaction (HAII) communities, which have evidenced the positive effects of user control on satisfaction/user experience. For example, \citet{pu2011user} noted that user control significantly impacts the overall user experience with the recommender system. In a similar vein, \citet{sundar2020rise} stressed that the two key features of HAII, namely user awareness and user control, are the hallmarks of successful user experience with personalization services.
\section{Limitations}   \label{limitations}
Some limitations to this work need to be articulated. First, the \textit{Indicator Editor} was evaluated with students and teachers who know data analytics and visualization. While we reached a diverse user group from different countries and educational levels, a broader population, including less technical participants, would help assess the tool's effectiveness among users with more diverse backgrounds and experiences. Further, we conducted a qualitative user study with 15 participants. Therefore, the study's results should be interpreted cautiously and cannot be generalized. A quantitative user study with a larger sample would probably have yielded more significant and reliable results.
\section{Conclusion and Future Work}    \label{conclusion}
This paper contributes to developing transparent and trustworthy learning analytics (LA). We suggest that supporting user interaction and providing user control in the indicator development process can improve transparency and trust in LA systems. Following a transparency through exploration approach, we presented the systematic design, implementation, and evaluation details of the \textit{Indicator Editor}, which aims to support self-service LA (SSLA) by actively involving stakeholders with data analytics and visual analytics knowledge in the indicator implementation process. Applying a qualitative approach, we found first evidence that handing over control to users to steer the indicator implementation process can positively affect different aspects of LA, namely transparency, trust, satisfaction, and acceptance. While our results cannot be generalized, we are confident they pose valuable anchor points for SSLA systems' current and future design. Future work will focus on (a) improving the user experience with the \textit{Indicator Editor}, (b) helping users develop complex indicators beyond simple statistics, e.g., based on machine learning algorithms for clustering, classification, or prediction, and (c) adding configuration to support users with different levels of expertise, e.g., by providing basic and advanced modes of the \textit{Indicator Editor}. Moreover, we plan to validate our findings through quantitative research to investigate the effects of a transparency through exploration approach on users' perceptions of SSLA. 

\bibliographystyle{ACM-Reference-Format}
\bibliography{sample-base}

@article{conati2018ai,
  title   = {AI in Education needs interpretable machine learning: Lessons from Open Learner Modelling},
  author  = {Conati, Cristina and Porayska-Pomsta, Kaska and Mavrikis, Manolis},
  journal = {arXiv preprint arXiv:1807.00154},
  year    = {2018}
}

@book{shneiderman2022human,
  title     = {Human-Centered AI},
  author    = {Shneiderman, Ben},
  year      = {2022},
  publisher = {Oxford University Press}
}

@article{amershi2014power,
  title   = {Power to the people: The role of humans in interactive machine learning},
  author  = {Amershi, Saleema and Cakmak, Maya and Knox, William Bradley and Kulesza, Todd},
  journal = {Ai Magazine},
  volume  = {35},
  number  = {4},
  pages   = {105--120},
  year    = {2014}
}

@article{braun2006using,
  title     = {Using thematic analysis in psychology},
  author    = {Braun, Virginia and Clarke, Victoria},
  journal   = {Qualitative research in psychology},
  volume    = {3},
  number    = {2},
  pages     = {77--101},
  year      = {2006},
  publisher = {Taylor \& Francis}
}

@article{buckingham2019human,
  title   = {Human-centred learning analytics},
  author  = {Buckingham Shum, Simon and Ferguson, Rebecca and Martinez-Maldonado, Roberto},
  journal = {Journal of Learning Analytics},
  volume  = {6},
  number  = {2},
  pages   = {1--9},
  year    = {2019}
}

@article{carter2005utilization,
  title     = {The utilization of e-government services: citizen trust, innovation and acceptance factors},
  author    = {Carter, Lemuria and B{\'e}langer, France},
  journal   = {Information systems journal},
  volume    = {15},
  number    = {1},
  pages     = {5--25},
  year      = {2005},
  publisher = {Wiley Online Library}
}

@article{rehrey2019engaging,
  title     = {Engaging Faculty in Learning Analytics: Agents of Institutional Culture Change.},
  author    = {Rehrey, George and Shepard, Linda and Hostetter, Carol and Reynolds, Amberly and Groth, Dennis},
  journal   = {Journal of Learning Analytics},
  volume    = {6},
  number    = {2},
  pages     = {86--94},
  year      = {2019},
  publisher = {ERIC}
}

@inproceedings{chen2019towards,
  title     = {Towards value-sensitive learning analytics design},
  author    = {Chen, Bodong and Zhu, Haiyi},
  booktitle = {Proceedings of the 9th international conference on learning analytics \& knowledge},
  pages     = {343--352},
  year      = {2019}
}

@inproceedings{chatti2020design,
  title        = {How to design effective learning analytics indicators? A human-centered design approach},
  author       = {Chatti, Mohamed Amine and Muslim, Arham and Guesmi, Mouadh and Richtscheid, Florian and Nasimi, Dawood and Shahin, Amin and Damera, Ritesh},
  booktitle    = {European conference on technology enhanced learning},
  pages        = {303--317},
  year         = {2020},
  organization = {Springer}
}

@incollection{chatti2020lava,
  title     = {The LAVA model: Learning analytics meets visual analytics},
  author    = {Chatti, Mohamed Amine and Muslim, Arham and Guliani, Manpriya and Guesmi, Mouadh},
  booktitle = {Adoption of data analytics in higher education learning and teaching},
  pages     = {71--93},
  year      = {2020},
  publisher = {Springer}
}

@inproceedings{jivet2021quantum,
  title     = {Quantum of Choice: How learners’ feedback monitoring decisions, goals and self-regulated learning skills are related},
  author    = {Jivet, Ioana and Wong, Jacqueline and Scheffel, Maren and Valle Torre, Manuel and Specht, Marcus and Drachsler, Hendrik},
  booktitle = {LAK21: 11th international learning analytics and knowledge conference},
  pages     = {416--427},
  year      = {2021}
}

@inproceedings{whitelock2017students,
  title     = {What do students want? Towards an instrument for students' evaluation of quality of learning analytics services},
  author    = {Whitelock-Wainwright, Alexander and Ga{\v{s}}evi{\'c}, Dragan and Tejeiro, Ricardo},
  booktitle = {Proceedings of the seventh international learning analytics \& knowledge conference},
  pages     = {368--372},
  year      = {2017}
}

@article{roberts2017give,
  title     = {Give me a customizable dashboard: Personalized learning analytics dashboards in higher education},
  author    = {Roberts, Lynne D and Howell, Joel A and Seaman, Kristen},
  journal   = {Technology, Knowledge and Learning},
  volume    = {22},
  pages     = {317--333},
  year      = {2017},
  publisher = {Springer}
}

@article{chatti2019perla,
  title     = {The PERLA framework: Blending personalization and learning analytics},
  author    = {Chatti, Mohamed Amine and Muslim, Arham},
  journal   = {International review of research in open and distributed learning},
  volume    = {20},
  number    = {1},
  year      = {2019},
  publisher = {Athabasca University Press (AU Press)}
}

@article{davis1989perceived,
  title     = {Perceived usefulness, perceived ease of use, and user acceptance of information technology},
  author    = {Davis, Fred D},
  journal   = {MIS quarterly},
  pages     = {319--340},
  year      = {1989},
  publisher = {JSTOR}
}

@article{spinner2019explainer,
  title     = {explAIner: A visual analytics framework for interactive and explainable machine learning},
  author    = {Spinner, Thilo and Schlegel, Udo and Sch{\"a}fer, Hanna and El-Assady, Mennatallah},
  journal   = {IEEE transactions on visualization and computer graphics},
  volume    = {26},
  number    = {1},
  pages     = {1064--1074},
  year      = {2019},
  publisher = {IEEE}
}

@inproceedings{drachsler2016privacy,
  title     = {Privacy and analytics: it's a DELICATE issue a checklist for trusted learning analytics},
  author    = {Drachsler, Hendrik and Greller, Wolfgang},
  booktitle = {Proceedings of the sixth international conference on learning analytics \& knowledge},
  pages     = {89--98},
  year      = {2016}
}

@article{dollinger2019working,
  title     = {Working together in learning analytics towards the co-creation of value.},
  author    = {Dollinger, Mollie and Liu, Danny and Arthars, Natasha and Lodge, Jason M},
  journal   = {Journal of Learning Analytics},
  volume    = {6},
  number    = {2},
  pages     = {10--26},
  year      = {2019},
  publisher = {ERIC}
}

@article{bennett2019four,
  title     = {Four design principles for learner dashboards that support student agency and empowerment},
  author    = {Bennett, Liz and Folley, Sue},
  journal   = {Journal of Applied Research in Higher Education},
  volume    = {12},
  number    = {1},
  pages     = {15--26},
  year      = {2019},
  publisher = {Emerald Publishing Limited}
}

@article{dyckhoff2012design,
  title     = {Design and implementation of a learning analytics toolkit for teachers},
  author    = {Dyckhoff, Anna Lea and Zielke, Dennis and B{\"u}ltmann, Mareike and Chatti, Mohamed Amine and Schroeder, Ulrik},
  journal   = {Journal of Educational Technology \& Society},
  volume    = {15},
  number    = {3},
  pages     = {58--76},
  year      = {2012},
  publisher = {JSTOR}
}

@inproceedings{dollinger2018co,
  title     = {Co-creation strategies for learning analytics},
  author    = {Dollinger, Mollie and Lodge, Jason M},
  booktitle = {Proceedings of the 8th international conference on learning analytics and knowledge},
  pages     = {97--101},
  year      = {2018}
}

@article{du2020techniques,
  title     = {Techniques for interpretable machine learning},
  author    = {Du, Mengnan and Liu, Ninghao and Hu, Xia},
  journal   = {Communications of the ACM},
  volume    = {63},
  number    = {1},
  pages     = {68--77},
  year      = {2020},
  publisher = {ACM New York, NY, USA}
}

@article{gefen2003trust,
  title     = {Trust and TAM in online shopping: An integrated model},
  author    = {Gefen, David and Karahanna, Elena and Straub, Detmar W},
  journal   = {MIS quarterly},
  pages     = {51--90},
  year      = {2003},
  publisher = {JSTOR}
}

@article{he2016interactive,
  title     = {Interactive recommender systems: A survey of the state of the art and future research challenges and opportunities},
  author    = {He, Chen and Parra, Denis and Verbert, Katrien},
  journal   = {Expert Systems with Applications},
  volume    = {56},
  pages     = {9--27},
  year      = {2016},
  publisher = {Elsevier}
}

@article{hellmann2022development,
  title   = {Development of an Instrument for Measuring Users’ Perception of Transparency in Recommender Systems},
  author  = {Hellmann, Marco and Hernandez-Bocanegra, Diana C and Ziegler, J{\"u}rgen},
  journal = {system},
  volume  = {12},
  pages   = {7},
  year    = {2022}
}

@inproceedings{de2019student,
  title     = {Student centred design of a learning analytics system},
  author    = {de Quincey, Ed and Briggs, Chris and Kyriacou, Theocharis and Waller, Richard},
  booktitle = {Proceedings of the 9th international conference on learning analytics \& knowledge},
  pages     = {353--362},
  year      = {2019}
}

@article{holstein2019co,
  title   = {Co-designing a real-time classroom orchestration tool to support teacher--AI complementarity},
  author  = {Holstein, Kenneth and McLaren, Bruce M and Aleven, Vincent},
  journal = {Journal of Learning Analytics},
  volume  = {6},
  number  = {2},
  year    = {2019}
}

@article{jugovac2017interacting,
  title     = {Interacting with recommenders—overview and research directions},
  author    = {Jugovac, Michael and Jannach, Dietmar},
  journal   = {ACM Transactions on Interactive Intelligent Systems (TiiS)},
  volume    = {7},
  number    = {3},
  pages     = {1--46},
  year      = {2017},
  publisher = {ACM New York, NY, USA}
}

@inproceedings{keim2006challenges,
  title        = {Challenges in visual data analysis},
  author       = {Keim, Daniel A and Mansmann, Florian and Schneidewind, J{\"o}rn and Ziegler, Hartmut},
  booktitle    = {Tenth International Conference on Information Visualisation (IV'06)},
  pages        = {9--16},
  year         = {2006},
  organization = {IEEE}
}

@inproceedings{muslim2017goal,
  title     = {The Goal-Question-Indicator Approach for Personalized Learning Analytics-.},
  author    = {Muslim, Arham and Chatti, Mohamed Amine and Mughal, Memoona and Schroeder, Ulrik},
  booktitle = {CSEDU (1)},
  pages     = {371--378},
  year      = {2017}
}

@book{norman2013design,
  title     = {The design of everyday things: Revised and expanded edition},
  author    = {Norman, Don},
  year      = {2013},
  publisher = {Basic books}
}

@article{nunes2017systematic,
  title     = {A systematic review and taxonomy of explanations in decision support and recommender systems},
  author    = {Nunes, Ingrid and Jannach, Dietmar},
  journal   = {User Modeling and User-Adapted Interaction},
  volume    = {27},
  number    = {3},
  pages     = {393--444},
  year      = {2017},
  publisher = {Springer}
}

@article{pardo2014ethical,
  title     = {Ethical and privacy principles for learning analytics},
  author    = {Pardo, Abelardo and Siemens, George},
  journal   = {British Journal of Educational Technology},
  volume    = {45},
  number    = {3},
  pages     = {438--450},
  year      = {2014},
  publisher = {Wiley Online Library}
}

@inproceedings{alvarez2020deck,
  title     = {LA-DECK: A card-based learning analytics co-design tool},
  author    = {Alvarez, Carlos Prieto and Martinez-Maldonado, Roberto and Shum, Simon Buckingham},
  booktitle = {Proceedings of the tenth international conference on learning analytics \& knowledge},
  pages     = {63--72},
  year      = {2020}
}

@incollection{prieto2018co,
  title     = {Co-designing learning analytics tools with learners},
  author    = {Prieto-Alvarez, Carlos G and Martinez-Maldonado, Roberto and Anderson, Theresa Dirndorfer},
  booktitle = {Learning Analytics in the Classroom},
  pages     = {93--110},
  year      = {2018},
  publisher = {Routledge}
}

@inproceedings{pu2011user,
  title     = {A user-centric evaluation framework for recommender systems},
  author    = {Pu, Pearl and Chen, Li and Hu, Rong},
  booktitle = {Proceedings of the fifth ACM conference on Recommender systems},
  pages     = {157--164},
  year      = {2011}
}

@article{venkatesh2003user,
  title     = {User acceptance of information technology: Toward a unified view},
  author    = {Venkatesh, Viswanath and Morris, Michael G and Davis, Gordon B and Davis, Fred D},
  journal   = {MIS quarterly},
  pages     = {425--478},
  year      = {2003},
  publisher = {JSTOR}
}

@article{shneiderman2020bridging,
  title     = {Bridging the gap between ethics and practice: guidelines for reliable, safe, and trustworthy human-centered AI systems},
  author    = {Shneiderman, Ben},
  journal   = {ACM Transactions on Interactive Intelligent Systems (TiiS)},
  volume    = {10},
  number    = {4},
  pages     = {1--31},
  year      = {2020},
  publisher = {ACM New York, NY, USA}
}

@incollection{tintarev2015explaining,
  title     = {Explaining recommendations: Design and evaluation},
  author    = {Tintarev, Nava and Masthoff, Judith},
  booktitle = {Recommender systems handbook},
  pages     = {353--382},
  year      = {2015},
  publisher = {Springer}
}

@inproceedings{tsai2017providing,
  title     = {Providing control and transparency in a social recommender system for academic conferences},
  author    = {Tsai, Chun-Hua and Brusilovsky, Peter},
  booktitle = {Proceedings of the 25th Conference on User Modeling, Adaptation and Personalization},
  pages     = {313--317},
  year      = {2017}
}

@inproceedings{verbert2013visualizing,
  title     = {Visualizing recommendations to support exploration, transparency and controllability},
  author    = {Verbert, Katrien and Parra, Denis and Brusilovsky, Peter and Duval, Erik},
  booktitle = {Proceedings of the 2013 international conference on Intelligent user interfaces},
  pages     = {351--362},
  year      = {2013}
}

@inproceedings{verbert2020xla,
  title     = {XLA: Explainable Learning Analytics},
  author    = {Verbert, Katrien and De Laet, Tinne and Millecamp, Martijn and Broos, Tom and Chatti, Mohamed Amine and Muslim, Arham},
  booktitle = {Companion Proceedings 10th International Conference on Learning Analytics and Knowledge (LAK20)},
  pages     = {477--479},
  year      = {2020}
}

@article{gedikli2014,
  title     = {How should I explain? A comparison of different explanation types for recommender systems},
  author    = {Gedikli, Fatih and Jannach, Dietmar and Ge, Mouzhi},
  journal   = {International Journal of Human-Computer Studies},
  volume    = {72},
  number    = {4},
  pages     = {367--382},
  year      = {2014},
  publisher = {Elsevier}
}

@article{khosravi2022explainable,
  title     = {Explainable artificial intelligence in education},
  author    = {Khosravi, Hassan and Shum, Simon Buckingham and Chen, Guanliang and Conati, Cristina and Tsai, Yi-Shan and Kay, Judy and Knight, Simon and Martinez-Maldonado, Roberto and Sadiq, Shazia and Ga{\v{s}}evi{\'c}, Dragan},
  journal   = {Computers and Education: Artificial Intelligence},
  volume    = {3},
  pages     = {100074},
  year      = {2022},
  publisher = {Elsevier}
}

@inproceedings{zhao2019users,
  author    = {Zhao, Ruijing and Benbasat, Izak and Cavusoglu, Hasan},
  booktitle = {Proceedings of the 27th European Conference on Information Systems},
  title     = {Do users always want to know more? Investigating the relationship between system transparency and users’ trust in advice-giving systems},
  year      = {2019}
}

@article{west2020academics,
  title   = {Do academics and university administrators really know better? The ethics of positioning student perspectives in learning analytics},
  author  = {West, Deborah and Luzeckyj, Ann and Toohey, Danny and Vanderlelie, Jessica and Searle, Bill},
  journal = {Australasian Journal of Educational Technology},
  volume  = {36},
  number  = {2},
  pages   = {60--70},
  year    = {2020}
}

@inproceedings{hilliger2020learners,
  title        = {For learners, with learners: Identifying indicators for an academic advising dashboard for students},
  author       = {Hilliger, Isabel and De Laet, Tinne and Henr{\'\i}quez, Valeria and Guerra, Julio and Ortiz-Rojas, Margarita and Zu{\~n}iga, Miguel {\'A}ngel and Baier, Jorge and P{\'e}rez-Sanagust{\'\i}n, Mar},
  booktitle    = {Addressing Global Challenges and Quality Education: 15th European Conference on Technology Enhanced Learning, EC-TEL 2020, Heidelberg, Germany, September 14--18, 2020, Proceedings 15},
  pages        = {117--130},
  year         = {2020},
  organization = {Springer}
}

@article{teasley2017student,
  title     = {Student facing dashboards: One size fits all?},
  author    = {Teasley, Stephanie D},
  journal   = {Technology, Knowledge and Learning},
  volume    = {22},
  number    = {3},
  pages     = {377--384},
  year      = {2017},
  publisher = {Springer}
}

@inproceedings{jin2018effects,
  title     = {Effects of personal characteristics on music recommender systems with different levels of controllability},
  author    = {Jin, Yucheng and Tintarev, Nava and Verbert, Katrien},
  booktitle = {Proceedings of the 12th ACM Conference on Recommender Systems},
  pages     = {13--21},
  year      = {2018}
}

@inproceedings{andjelkovic2016mood,
  title     = {Moodplay: Interactive mood-based music discovery and recommendation},
  author    = {Andjelkovic, Ivana and Parra, Denis and O'Donovan, John},
  booktitle = {Proceedings of the 2016 conference on user modeling adaptation and personalization},
  pages     = {275--279},
  year      = {2016}
}

@article{jiang2019recent,
  title     = {Recent research advances on interactive machine learning},
  author    = {Jiang, Liu and Liu, Shixia and Chen, Changjian},
  journal   = {Journal of Visualization},
  volume    = {22},
  pages     = {401--417},
  year      = {2019},
  publisher = {Springer}
}

@article{hakami2020learning,
  title     = {How are learning analytics considering the societal values of fairness, accountability, transparency and human well-being?: A literature review},
  author    = {Hakami, Eyad and Hern{\'a}ndez Leo, Davinia},
  journal   = {Mart{\'\i}nez-Mon{\'e}s A, {\'A}lvarez A, Caeiro-Rodr{\'\i}guez M, Dimitriadis Y, editors. LASI-SPAIN 2020: Learning Analytics Summer Institute Spain 2020: Learning Analytics. Time for Adoption?; 2020 Jun 15-16; Valladolid, Spain. Aachen: CEUR; 2020. p. 121-41},
  year      = {2020},
  publisher = {CEUR Workshop Proceedings}
}

@inproceedings{pozdniakov2022question,
  title     = {The question-driven dashboard: how can we design analytics interfaces aligned to teachers’ inquiry?},
  author    = {Pozdniakov, Stanislav and Martinez-Maldonado, Roberto and Tsai, Yi-Shan and Cukurova, Mutlu and Bartindale, Tom and Chen, Peter and Marshall, Harrison and Richardson, Dan and Gasevic, Dragan},
  booktitle = {LAK22: 12th International Learning Analytics and Knowledge Conference},
  pages     = {175--185},
  year      = {2022}
}

@inproceedings{prinsloo2015student,
  title     = {Student privacy self-management: Implications for learning analytics},
  author    = {Prinsloo, Paul and Slade, Sharon},
  booktitle = {Proceedings of the fifth international conference on learning analytics and knowledge},
  pages     = {83--92},
  year      = {2015}
}

@inproceedings{slade2019learning,
  title     = {Learning analytics at the intersections of student trust, disclosure and benefit},
  author    = {Slade, Sharon and Prinsloo, Paul and Khalil, Mohammad},
  booktitle = {Proceedings of the 9th International Conference on learning analytics \& knowledge},
  pages     = {235--244},
  year      = {2019}
}

@inproceedings{shibani2019contextualizable,
  title     = {Contextualizable learning analytics design: A generic model and writing analytics evaluations},
  author    = {Shibani, Antonette and Knight, Simon and Buckingham Shum, Simon},
  booktitle = {Proceedings of the 9th international conference on learning analytics \& knowledge},
  pages     = {210--219},
  year      = {2019}
}

@inproceedings{tsai2020privacy,
  title     = {The privacy paradox and its implications for learning analytics},
  author    = {Tsai, Yi-Shan and Whitelock-Wainwright, Alexander and Ga{\v{s}}evi{\'c}, Dragan},
  booktitle = {Proceedings of the tenth international conference on learning analytics \& knowledge},
  pages     = {230--239},
  year      = {2020}
}

@article{khalil2023fairness,
  title   = {Fairness, Trust, Transparency, Equity, and Responsibility in Learning Analytics},
  author  = {Khalil, Mohammad and Prinsloo, Paul and Slade, Sharon},
  journal = {Journal of Learning Analytics},
  volume  = {10},
  number  = {1},
  pages   = {1--7},
  year    = {2023}
}

@article{slade2013learning,
  title     = {Learning analytics: Ethical issues and dilemmas},
  author    = {Slade, Sharon and Prinsloo, Paul},
  journal   = {American Behavioral Scientist},
  volume    = {57},
  number    = {10},
  pages     = {1510--1529},
  year      = {2013},
  publisher = {Sage Publications Sage CA: Los Angeles, CA}
}

@inproceedings{li2019impact,
  title     = {The impact of student opt-out on educational predictive models},
  author    = {Li, Warren and Brooks, Christopher and Schaub, Florian},
  booktitle = {Proceedings of the 9th international conference on learning analytics \& knowledge},
  pages     = {411--420},
  year      = {2019}
}

@inproceedings{clow2012learning,
  title     = {The learning analytics cycle: closing the loop effectively},
  author    = {Clow, Doug},
  booktitle = {Proceedings of the 2nd international conference on learning analytics and knowledge},
  pages     = {134--138},
  year      = {2012}
}

@inproceedings{barria2019explaining,
  title     = {Explaining need-based educational recommendations using interactive open learner models},
  author    = {Barria-Pineda, Jordan and Akhuseyinoglu, Kamil and Brusilovsky, Peter},
  booktitle = {Adjunct Publication of the 27th Conference on User Modeling, Adaptation and Personalization},
  pages     = {273--277},
  year      = {2019}
}

@inproceedings{barria2019making,
  title     = {Making educational recommendations transparent through a fine-grained open learner model},
  author    = {Barria Pineda, Jordan and Brusilovsky, Peter},
  booktitle = {Proceedings of Workshop on Intelligent User Interfaces for Algorithmic Transparency in Emerging Technologies at the 24th ACM Conference on Intelligent User Interfaces, IUI 2019, Los Angeles, USA, March 20, 2019},
  volume    = {2327},
  year      = {2019}
}

@inproceedings{cukurova2020modelling,
  title     = {Modelling collaborative problem-solving competence with transparent learning analytics: is video data enough?},
  author    = {Cukurova, Mutlu and Zhou, Qi and Spikol, Daniel and Landolfi, Lorenzo},
  booktitle = {Proceedings of the tenth international conference on learning analytics \& knowledge},
  pages     = {270--275},
  year      = {2020}
}

@inproceedings{abdi2020complementing,
  title     = {Complementing educational recommender systems with open learner models},
  author    = {Abdi, Solmaz and Khosravi, Hassan and Sadiq, Shazia and Gasevic, Dragan},
  booktitle = {Proceedings of the tenth international conference on learning analytics \& knowledge},
  pages     = {360--365},
  year      = {2020}
}

@inproceedings{bodily2018open,
  title     = {Open learner models and learning analytics dashboards: a systematic review},
  author    = {Bodily, Robert and Kay, Judy and Aleven, Vincent and Jivet, Ioana and Davis, Dan and Xhakaj, Franceska and Verbert, Katrien},
  booktitle = {Proceedings of the 8th international conference on learning analytics and knowledge},
  pages     = {41--50},
  year      = {2018}
}

@inproceedings{sarmiento2022participatory,
  title     = {Participatory and co-design of learning analytics: an initial review of the literature},
  author    = {Sarmiento, Juan Pablo and Wise, Alyssa Friend},
  booktitle = {LAK22: 12th International Learning Analytics and Knowledge Conference},
  pages     = {535--541},
  year      = {2022}
}

@article{martinez2015latux,
  title     = {LATUX: An Iterative Workflow for Designing, Validating, and Deploying Learning Analytics Visualizations.},
  author    = {Martinez-Maldonado, Roberto and Pardo, Abelardo and Mirriahi, Negin and Yacef, Kalina and Kay, Judy and Clayphan, Andrew},
  journal   = {Journal of Learning Analytics},
  volume    = {2},
  number    = {3},
  pages     = {9--39},
  year      = {2015},
  publisher = {ERIC}
}

@inproceedings{haythornthwaite2017information,
  title     = {An information policy perspective on learning analytics},
  author    = {Haythornthwaite, Caroline},
  booktitle = {Proceedings of the seventh international learning analytics \& knowledge conference},
  pages     = {253--256},
  year      = {2017}
}

@inproceedings{lang2018complexities,
  title     = {The complexities of developing a personal code of ethics for learning analytics practitioners: Implications for institutions and the field},
  author    = {Lang, Charles and Macfadyen, Leah P and Slade, Sharon and Prinsloo, Paul and Sclater, Niall},
  booktitle = {Proceedings of the 8th international conference on learning analytics and knowledge},
  pages     = {436--440},
  year      = {2018}
}

@inproceedings{swenson2014establishing,
  title     = {Establishing an ethical literacy for learning analytics},
  author    = {Swenson, Jenni},
  booktitle = {Proceedings of the Fourth International Conference on Learning Analytics and Knowledge},
  pages     = {246--250},
  year      = {2014}
}

@inproceedings{scheffel2015developing,
  title     = {Developing an evaluation framework of quality indicators for learning analytics},
  author    = {Scheffel, Maren and Drachsler, Hendrik and Specht, Marcus},
  booktitle = {Proceedings of the Fifth International Conference on Learning Analytics and Knowledge},
  pages     = {16--20},
  year      = {2015}
}

@article{ahn2019designing,
  title     = {Designing in Context: Reaching beyond Usability in Learning Analytics Dashboard Design.},
  author    = {Ahn, June and Campos, Fabio and Hays, Maria and DiGiacomo, Daniela},
  journal   = {Journal of Learning Analytics},
  volume    = {6},
  number    = {2},
  pages     = {70--85},
  year      = {2019},
  publisher = {ERIC}
}

@article{tsai2020empowering,
  title     = {Empowering learners with personalised learning approaches? Agency, equity and transparency in the context of learning analytics},
  author    = {Tsai, Yi-Shan and Perrotta, Carlo and Ga{\v{s}}evi{\'c}, Dragan},
  journal   = {Assessment \& Evaluation in Higher Education},
  volume    = {45},
  number    = {4},
  pages     = {554--567},
  year      = {2020},
  publisher = {Taylor \& Francis}
}

@inproceedings{tsai2017learning,
  title     = {Learning analytics in higher education---challenges and policies: a review of eight learning analytics policies},
  author    = {Tsai, Yi-Shan and Gasevic, Dragan},
  booktitle = {Proceedings of the seventh international learning analytics \& knowledge conference},
  pages     = {233--242},
  year      = {2017}
}

@inproceedings{prinsloo2013evaluation,
  title     = {An evaluation of policy frameworks for addressing ethical considerations in learning analytics},
  author    = {Prinsloo, Paul and Slade, Sharon},
  booktitle = {Proceedings of the third international conference on learning analytics and knowledge},
  pages     = {240--244},
  year      = {2013}
}

@inproceedings{hoel2017influence,
  title     = {The influence of data protection and privacy frameworks on the design of learning analytics systems},
  author    = {Hoel, Tore and Griffiths, Dai and Chen, Weiqin},
  booktitle = {Proceedings of the seventh international learning analytics \& knowledge conference},
  pages     = {243--252},
  year      = {2017}
}

@article{sundar2020rise,
  title     = {Rise of machine agency: A framework for studying the psychology of human--AI interaction (HAII)},
  author    = {Sundar, S Shyam},
  journal   = {Journal of Computer-Mediated Communication},
  volume    = {25},
  number    = {1},
  pages     = {74--88},
  year      = {2020},
  publisher = {Oxford University Press}
}

@article{dudley2018review,
  title     = {A review of user interface design for interactive machine learning},
  author    = {Dudley, John J and Kristensson, Per Ola},
  journal   = {ACM Transactions on Interactive Intelligent Systems (TiiS)},
  volume    = {8},
  number    = {2},
  pages     = {1--37},
  year      = {2018},
  publisher = {ACM New York, NY, USA}
}

@inproceedings{knijnenburg2011each,
  title     = {Each to his own: how different users call for different interaction methods in recommender systems},
  author    = {Knijnenburg, Bart P and Reijmer, Niels JM and Willemsen, Martijn C},
  booktitle = {Proceedings of the fifth ACM conference on Recommender systems},
  pages     = {141--148},
  year      = {2011}
}

@article{chatti2012reference,
  title={A reference model for learning analytics},
  author={Chatti, Mohamed Amine and Dyckhoff, Anna Lea and Schroeder, Ulrik and Th{\"u}s, Hendrik},
  journal={International journal of Technology Enhanced learning},
  volume={4},
  number={5-6},
  pages={318--331},
  year={2012},
  publisher={Inderscience Publishers}
}

@misc{elias2011learning,
  title={Learning analytics: Definitions, processes and potential},
  author={Elias, Tanya},
  year={2011}
}

@article{verbert2013learning,
  title={Learning analytics dashboard applications},
  author={Verbert, Katrien and Duval, Erik and Klerkx, Joris and Govaerts, Sten and Santos, Jos{\'e} Luis},
  journal={American Behavioral Scientist},
  volume={57},
  number={10},
  pages={1500--1509},
  year={2013},
  publisher={Sage Publications Sage CA: Los Angeles, CA}
}

@article{siepmann2023trust,
  title={Trust and transparency in recommender systems},
  author={Siepmann, Clara and Chatti, Mohamed Amine},
  journal={Human-Centered Explainable AI Workshop at the ACM CHI'23 conference},
  year={2023}
}

@article{kaur2022trustworthy,
  title={Trustworthy artificial intelligence: a review},
  author={Kaur, Davinder and Uslu, Suleyman and Rittichier, Kaley J and Durresi, Arjan},
  journal={ACM computing surveys (CSUR)},
  volume={55},
  number={2},
  pages={1--38},
  year={2022},
  publisher={ACM New York, NY}
}

@article{miller2022we,
  title={Are we measuring trust correctly in explainability, interpretability, and transparency research?},
  author={Miller, Tim},
  journal={arXiv preprint arXiv:2209.00651},
  year={2022}
}

@article{mcknight2002developing,
  title={Developing and validating trust measures for e-commerce: An integrative typology},
  author={McKnight, D Harrison and Choudhury, Vivek and Kacmar, Charles},
  journal={Information systems research},
  volume={13},
  number={3},
  pages={334--359},
  year={2002},
  publisher={INFORMS}
}

@inproceedings{yang2020visual,
  title={How do visual explanations foster end users' appropriate trust in machine learning?},
  author={Yang, Fumeng and Huang, Zhuanyi and Scholtz, Jean and Arendt, Dustin L},
  booktitle={Proceedings of the 25th international conference on intelligent user interfaces},
  pages={189--201},
  year={2020}
}

@article{knijnenburg2012explaining,
  title={Explaining the user experience of recommender systems},
  author={Knijnenburg, Bart P and Willemsen, Martijn C and Gantner, Zeno and Soncu, Hakan and Newell, Chris},
  journal={User modeling and user-adapted interaction},
  volume={22},
  number={4},
  pages={441--504},
  year={2012},
  publisher={Springer}
}

@inproceedings{ngo2020exploring,
  title={Exploring mental models for transparent and controllable recommender systems: a qualitative study},
  author={Ngo, Thao and Kunkel, Johannes and Ziegler, J{\"u}rgen},
  booktitle={Proceedings of the 28th ACM Conference on User Modeling, Adaptation and Personalization},
  pages={183--191},
  year={2020}
}

@article{ananny2018seeing,
  title={Seeing without knowing: Limitations of the transparency ideal and its application to algorithmic accountability},
  author={Ananny, Mike and Crawford, Kate},
  journal={new media \& society},
  volume={20},
  number={3},
  pages={973--989},
  year={2018},
  publisher={SAGE Publications Sage UK: London, England}
}

@article{hosseini2018four,
  title={Four reference models for transparency requirements in information systems},
  author={Hosseini, Mahmood and Shahri, Alimohammad and Phalp, Keith and Ali, Raian},
  journal={Requirements Engineering},
  volume={23},
  number={2},
  pages={251--275},
  year={2018},
  publisher={Springer}
}

@article{cramer2008effects,
  title={The effects of transparency on trust in and acceptance of a content-based art recommender},
  author={Cramer, Henriette and Evers, Vanessa and Ramlal, Satyan and Van Someren, Maarten and Rutledge, Lloyd and Stash, Natalia and Aroyo, Lora and Wielinga, Bob},
  journal={User Modeling and User-adapted interaction},
  volume={18},
  number={5},
  pages={455--496},
  year={2008},
  publisher={Springer}
}

@inproceedings{lang2023learning,
  title={Learning Analytics and Stakeholder Inclusion: What do We Mean When We Say" Human-Centered"?},
  author={Lang, Charles and Davis, Laura},
  booktitle={LAK23: 13th International Learning Analytics and Knowledge Conference},
  pages={411--417},
  year={2023}
}

@incollection{viberg2023designing,
  title={Designing culturally aware learning analytics: A value sensitive perspective},
  author={Viberg, Olga and Jivet, Ioana and Scheffel, Maren},
  booktitle={Practicable learning analytics},
  pages={177--192},
  year={2023},
  publisher={Springer}
}

@article{alfredo2024human,
  title={Human-centred learning analytics and AI in education: A systematic literature review},
  author={Alfredo, Riordan and Echeverria, Vanessa and Jin, Yueqiao and Yan, Lixiang and Swiecki, Zachari and Ga{\v{s}}evi{\'c}, Dragan and Martinez-Maldonado, Roberto},
  journal={Computers and Education: Artificial Intelligence},
  volume={6},
  pages={100215},
  year={2024},
  publisher={Elsevier}
}

@article{topali2025designing,
  title={Designing human-centered learning analytics and artificial intelligence in education solutions: a systematic literature review},
  author={Topali, Paraskevi and Ortega-Arranz, Alejandro and Rodr{\'\i}guez-Triana, Mar{\'\i}a Jes{\'u}s and Er, Erkan and Khalil, Mohammad and Ak{\c{c}}ap{\i}nar, G{\"o}khan},
  journal={Behaviour \& Information Technology},
  volume={44},
  number={5},
  pages={1071--1098},
  year={2025},
  publisher={Taylor \& Francis}
}

@inproceedings{sarmiento2020engaging,
  title={Engaging students as co-designers of learning analytics},
  author={Sarmiento, Juan Pablo and Campos, Fabio and Wise, Alyssa},
  booktitle={Companion proceedings of the 10th international learning analytics and knowledge conference},
  pages={29--32},
  year={2020},
  organization={SoLAR Frankfurt}
}

@book{hanington2019universal,
  title={Universal methods of design expanded and revised: 125 Ways to research complex problems, develop innovative ideas, and design effective solutions},
  author={Hanington, Bruce and Martin, Bella},
  year={2019},
  publisher={Rockport publishers}
}

@misc{buckingham2024human,
  title={Human-centred learning analytics: 2019--24},
  author={Buckingham Shum, Simon and Mart{\'\i}nez-Maldonado, Roberto and Dimitriadis, Yannis and Santos, Patricia},
  journal={British Journal of Educational Technology},
  volume={55},
  number={3},
  pages={755--768},
  year={2024},
  publisher={Wiley Online Library}
}

@incollection{dimitriadis2021human, 
  title={Human-centered design principles for actionable learning analytics},
  author={Dimitriadis, Yannis and Mart{\'\i}nez-Maldonado, Roberto and Wiley, Korah},
  booktitle={Research on E-learning and ICT in education: Technological, pedagogical and instructional perspectives},
  pages={277--296},
  year={2021},
  publisher={Springer}
}

@article{martinez2023human,
  title={Human-centred learning analytics: Four challenges in realising the potential},
  author={Martinez-Maldonado, Roberto},
  journal={Learning Letters},
  volume={1},
  pages={6--6},
  year={2023}
}

@article{shreiner2022information,
  title={The information won't just sink in: Helping teachers provide technology-assisted data literacy instruction in social studies},
  author={Shreiner, Tamara L and Guzdial, Mark},
  journal={British Journal of Educational Technology},
  volume={53},
  number={5},
  pages={1134--1158},
  year={2022},
  publisher={Wiley Online Library}
}

@inproceedings{oliver2022adapting,
  title={Adapting learning analytics dashboards by and for university students},
  author={Oliver-Quelennec, Katia and Bouchet, Fran{\c{c}}ois and Carron, Thibault and Fronton Casalino, Kathy and Pin{\c{c}}on, Claire},
  booktitle={European Conference on Technology Enhanced Learning},
  pages={299--309},
  year={2022},
  organization={Springer}
}

@misc{joarder2025isccreatorhumancentereddesign,
  title={The ISC Creator: Human-Centered Design of Learning Analytics Interactive Indicator Specification Cards}, 
  author={Shoeb Joarder and Mohamed Amine Chatti},
  year={2025},
  eprint={2504.07811},
  archivePrefix={arXiv},
  primaryClass={cs.CY},
  url={https://arxiv.org/abs/2504.07811}, 
}

@article{tsai2021effects,
  title={The effects of controllability and explainability in a social recommender system},
  author={Tsai, Chun-Hua and Brusilovsky, Peter},
  journal={User Modeling and User-Adapted Interaction},
  volume={31},
  number={3},
  pages={591--627},
  year={2021},
  publisher={Springer}
}

@inproceedings{ma2022glancee,
  title={Glancee: An adaptable system for instructors to grasp student learning status in synchronous online classes},
  author={Ma, Shuai and Zhou, Taichang and Nie, Fei and Ma, Xiaojuan},
  booktitle={Proceedings of the 2022 CHI conference on human factors in computing systems},
  pages={1--25},
  year={2022}
}

@article{shute2021design,
  title={The design, development, and testing of learning supports for the physics playground game},
  author={Shute, Valerie J and Smith, Ginny and Kuba, Renata and Dai, Chih-Pu and Rahimi, Seyedahmad and Liu, Zhichun and Almond, Russell},
  journal={International Journal of Artificial Intelligence in Education},
  volume={31},
  number={3},
  pages={357--379},
  year={2021},
  publisher={Springer}
}

@inproceedings{ahn2021co,
  title={Co-designing for privacy, transparency, and trust in K-12 learning analytics},
  author={Ahn, June and Campos, Fabio and Nguyen, Ha and Hays, Maria and Morrison, Jan},
  booktitle={LAK21: 11th international learning analytics and knowledge conference},
  pages={55--65},
  year={2021}
}

@inproceedings{shibani2022questioning,
  title={Questioning learning analytics? Cultivating critical engagement as student automated feedback literacy},
  author={Shibani, Antonette and Knight, Simon and Buckingham Shum, Simon},
  booktitle={LAK22: 12th international learning analytics and knowledge conference},
  pages={326--335},
  year={2022}
}

@article{alzahrani2023teaching,
  title={Do teaching staff trust stakeholders and tools in learning analytics? A mixed methods study},
  author={Alzahrani, Asma Shannan and Tsai, Yi-Shan and Aljohani, Naif and Whitelock-Wainwright, Emma and Gasevic, Dragan},
  journal={Educational technology research and development},
  volume={71},
  number={4},
  pages={1471--1501},
  year={2023},
  publisher={Springer}
}

@inproceedings{usmani2023human,
  title={Human-centered artificial intelligence: Designing for user empowerment and ethical considerations},
  author={Usmani, Usman Ahmad and Happonen, Ari and Watada, Junzo},
  booktitle={2023 5th international congress on human-computer interaction, optimization and robotic applications (HORA)},
  year={2023},
  organization={IEEE}
}

@inproceedings{ooge2023steering,
  title={Steering recommendations and visualising its impact: effects on adolescents’ trust in e-learning platforms},
  author={Ooge, Jeroen and Dereu, Leen and Verbert, Katrien},
  booktitle={Proceedings of the 28th international conference on intelligent user interfaces},
  pages={156--170},
  year={2023}
}

@article{wilson2021automated,
  title={Automated feedback and automated scoring in the elementary grades: Usage, attitudes, and associations with writing outcomes in a districtwide implementation of MI Write},
  author={Wilson, Joshua and Huang, Yue and Palermo, Corey and Beard, Gaysha and MacArthur, Charles A},
  journal={International Journal of Artificial Intelligence in Education},
  volume={31},
  number={2},
  pages={234--276},
  year={2021},
  publisher={Springer}
}

@article{duan2024towards,
  title={Towards transparent and trustworthy prediction of student learning achievement by including instructors as co-designers: A case study},
  author={Duan, Xiaojing and Pei, Bo and Ambrose, G Alex and Hershkovitz, Arnon and Cheng, Ying and Wang, Chaoli},
  journal={Education and Information Technologies},
  volume={29},
  number={3},
  pages={3075--3096},
  year={2024},
  publisher={Springer}
}

@article{lawrence2024teachers,
  title={How teachers conceptualise shared control with an AI co-orchestration tool: A multiyear teacher-centred design process},
  author={Lawrence, LuEttaMae and Echeverria, Vanessa and Yang, Kexin and Aleven, Vincent and Rummel, Nikol},
  journal={British Journal of Educational Technology},
  volume={55},
  number={3},
  pages={823--844},
  year={2024},
  publisher={Wiley Online Library}
}

@article{conijn2023effects,
  title={The effects of explanations in automated essay scoring systems on student trust and motivation},
  author={Conijn, Rianne and Kahr, Patricia and Snijders, Chris CP},
  journal={Journal of Learning Analytics},
  volume={10},
  number={1},
  pages={37--53},
  year={2023},
  publisher={UTS ePress}
}

@article{gedrimiene2023transparency,
  title={Transparency and Trustworthiness in User Intentions to Follow Career Recommendations from a Learning Analytics Tool.},
  author={Gedrimiene, Egle and Celik, Ismail and M{\"a}kitalo, Kati and Muukkonen, Hanni},
  journal={Journal of Learning Analytics},
  volume={10},
  number={1},
  pages={54--70},
  year={2023},
  publisher={ERIC}
}

@article{hutchins2024co,
  title={Co-designing teacher support technology for problem-based learning in middle school science},
  author={Hutchins, Nicole M and Biswas, Gautam},
  journal={British Journal of Educational Technology},
  volume={55},
  number={3},
  pages={802--822},
  year={2024},
  publisher={Wiley Online Library}
}

@article{wiley2024human,
  title={A human-centred learning analytics approach for developing contextually scalable K-12 teacher dashboards},
  author={Wiley, Korah and Dimitriadis, Yannis and Linn, Marcia},
  journal={British Journal of Educational Technology},
  volume={55},
  number={3},
  pages={845--885},
  year={2024},
  publisher={Wiley Online Library}
}

@article{campos2024leveraging,
  title={Leveraging cultural forms in human-centred learning analytics design},
  author={Campos, Fabio and Nguyen, Ha and Ahn, June and Jackson, Kara},
  journal={British Journal of Educational Technology},
  volume={55},
  number={3},
  pages={769--784},
  year={2024},
  publisher={Wiley Online Library}
}

@article{hilliger2024curriculum,
  title={Curriculum analytics adoption in higher education: A multiple case study engaging stakeholders in different phases of design},
  author={Hilliger, Isabel and Miranda, Constanza and Celis, Sergio and P{\'e}rez-Sanagust{\'\i}n, Mar},
  journal={British Journal of Educational Technology},
  volume={55},
  number={3},
  pages={785--801},
  year={2024},
  publisher={Wiley Online Library}
}

\end{document}